\documentclass[published]{JHEP3}
\JHEP{06(2004)042}

\usepackage{epsfig,multirow,bbm}
\usepackage{subfigure,latexsym}

\skip\footins = 1\bigskipamount plus 2pt minus 4pt

\newcommand{\Tr}{\mathop{\rm Tr}\nolimits}
\newcommand{\UU}{\mathop{\rm {}U}\nolimits}
\newcommand{\lsim}{\mathop{\lsi}}
\newcommand{\Z}{\mathbb{Z}}
\newcommand{\R}{\mathbb{R}}
\def\lsi{\lesssim}

\title{Phase diagram and dispersion relation of the non-commutative
  $\lambda \phi^{4}$ model in $d=3$}

\author{Wolfgang Bietenholz and Frank Hofheinz\\
        Institut f\"{u}r Physik, Humboldt Universit\"{a}t zu Berlin\\
        Newtonstr.\ 15, D-12489 Berlin, Germany\\
        E-mail: \email{bietenho@physik.hu-berlin.de}, 
                \email{hofheinz@physik.hu-berlin.de}}

\author{Jun Nishimura\\
        High Energy Accelerator Research Organization (KEK)\\
        1-1 Oho, Tsukuba 305-0801, Japan\\
        E-mail: \email{jnishi@post.kek.jp}}

\abstract{We present a non-perturbative study of the $\lambda
  \phi^{4}$ model in a three dimensional euclidean space, where the
  two spatial coordinates are non-commutative. Our results are
  obtained from numerical simulations of the lattice model, after its
  mapping onto a dimensionally reduced, twisted hermitean matrix
  model.  In this way we first reveal the explicit phase diagram of
  the non-commutative $\lambda \phi^{4}$ lattice model. We observe
  that the ordered regime splits into a phase of uniform order and a
  phase of two stripes of opposite sign, and more complicated
  patterns.  Next we discuss the behavior of the spatial and temporal
  correlators.  From the latter we extract the dispersion relation,
  which allows us to introduce a dimensionful lattice spacing.  To
  extrapolate to zero lattice spacing and infinite volume we perform a
  double scaling limit, which keeps the non-commutativity tensor
  constant.  The dispersion relation in the disordered phase
  stabilizes in this limit, which represents a non-perturbative
  renormalization.  In particular this confirms the existence of a
  striped phase in the continuum limit, in accordance with a
  conjecture by Gubser and Sondhi.  The extrapolated dispersion
  relation also exhibits UV/IR mixing as a non-perturbative effect.
  Finally we add some observations about a Nambu-Goldstone mode in the
  striped phase, and about the corresponding model in $d=2$.}

\received{April 15, 2004}
\accepted{June 23, 2004}

\begin{document}

\section{Non-commutative field theory}\label{section1}

The idea of introducing non-commutative (NC) coordinate operators ---
or a spatial uncertainty --- dates back to private communications
involving Heisenberg, Peierls, Pauli and Oppenheimer. The coordinates
are then given by hermitean operators $\hat x_{\mu}$ obeying a
commutation relation of the form
\begin{equation}  \label{NC1}
[ \hat x_{\mu} , \hat x_{\nu} ] = i \Theta_{\mu \nu} \,.
\end{equation}
The first papers on this idea appeared in 1947 on quantum theory in
flat~\cite{Sny} and curved~\cite{Yang} NC spaces. However, it was only
at the end of the twentieth century that it attracted attention on a
large scale in particle physics. Meanwhile, the mathematical
foundation for quantum field theories on a NC space, \emph{NC field
  theories}, was worked out; for an overview see refs.~\cite{books}.
First applications using this type of space as a formalism --- rather
than a possible description of nature --- emerged from solid state
physics, see for instance refs.~\cite{solid}.  Nowadays also the
possible reality of a NC space is a subject of intensive research.

It was string theory that finally boosted this field by identifying
strings in certain low energy limits with NC field
theory~\cite{String}.  This was the main reason why NC field theory
became extremely fashionable over the last years, so that more than
$1300$ papers have been produced on it up to now~\cite{Schapo}.  We
also take string theory as a motivation for our study, but in this
work we investigate a NC field theory as such, without working out
possible connections to other branches of physics.

A deep qualitative difference from ordinary (commutative) field theory
is the occurrence of a \emph{non-locality} of the range $\sqrt{ \Vert
  \Theta \Vert }$. This property obviously implies conceptual
problems, but from the optimistic point of view it may just provide
the crucial link to string theory or to quantum gravity. In fact,
there are arguments that a conciliation of gravity and quantum theory
induces some sort of non-commutativity under quite general
assumptions~\cite{gravity}.  This line of thought would naturally
include the time into the non-commutativity
relation~(\ref{NC1}). However, in that setting the problems related to
causality and unitarity are especially bad~\cite{causal}.  According
to refs.~\cite{unitar} unitarity is on safe grounds for the Minkowski
signature, but it is not yet clarified if the transition from a
Euclidean to a Minkowski signature can be justified.  For such
reasons, much of the literature excludes time from the
non-commutativity.  This will also be our framework, since the use of
a Euclidean space is vital for our numerical techniques, which will be
explained below.

In the early days people hoped for yet another possible virtue due to
the non-locality, namely that it would weaken or even remove the UV
divergences of the commutative field theories~\cite{Sny}, and
therefore simplify the renormalization. This hope was badly
disappointed: first it turned out that only part of the UV divergences
are removed whereas others remain. In particular the UV divergences in
the planar diagrams tend to survive the introduction of
$\Theta$~\cite{UV}.  What makes the situation much worse, however, is
that the remaining commutative UV divergences do not just disappear,
but they are typically converted into some kind of IR divergences. So
one ends up with a troublesome mixing of divergences at both ends of
the spectrum, denoted as \emph{UV/IR mixing}~\cite{UVIR}. A simple
intuitive picture of this effect based on the uncertainty principle is
described for instance in ref.~\cite{Szabo}. We are going to
illustrate the structure of such ``mixed'' divergences with an example
in subsection~\ref{section1.1}.  Beyond one loop, perturbation theory does not yet
provide any systematic machinery to handle this type of divergences,
which is unknown in commutative field theory.\footnote{For
  completeness we should mention that there are also models known
  where UV/IR mixing occurs, but nevertheless renormalizability can be
  shown to all orders. Examples for this are the NC Wess-Zumino
  model~\cite{Brazil}, the photon self-energy in NC QED~\cite{self-E},
  and the real, duality-covariant 4d NC $\lambda \phi^4$
  model~\cite{dual}.}  Hence adopting a fully non-perturbative
approach is highly motivated, and this is the goal of the work
presented here.

One issue that immediately arises is the question if UV/IR mixing is a
technical problem of perturbation theory, or if it should rather be
considered as a fundamental property of many NC field theories.  Our
previous numerical investigation of 2d NC $U(1)$ gauge theory clearly
supported the latter point of view~\cite{2dU1}, in agreement with
theoretical arguments~\cite{AMNS}. Manifestations of UV/IR mixing can
also be observed beyond perturbation theory --- as the present work
will demonstrate again --- hence such mixing effects belong to the
very nature of the corresponding NC models.\footnote{UV/IR mixing
  effects have also been observed on the semi-classical
  level~\cite{semi}.}  This implies, for instance, that it is not
promising to try to avoid such effects by performing some non-standard
perturbative expansion (as it has been suggested occasionally in the
literature).

In this work we consider the simplest case of an \emph{NC plane} with
a constant non-commutativity tensor, i.e.\ we deal with the
non-commutativity of only two coordinates. Its extent is characterized
by the parameter $\vartheta$,
\begin{equation}  \label{NCplane}
[ \hat x_{i} , \hat x_{j} ] = i \, \vartheta \, \epsilon_{ij} \,,
\qquad i,j \in \{ 1,2 \} \,,
\end{equation}
where $\epsilon$ is the antisymmetric unit tensor.  In addition we
have a Euclidean time coordinate $t$. The whole 3d space is lattice
discretized with equidistant lattice spacings.  This is done by the
standard procedure for the time coordinate, while we follow the
instruction of refs.~\cite{AMNS} for the NC plane. There we impose the
operator equation
\begin{equation}
\exp \Big( i \frac{2\pi}{a} \hat x_{j} \Big) = \hat {\mathbbm 1}\,,
\end{equation}
where $a$ is the spatial lattice constant. Thus the spatial lattice
sites are somewhat fuzzy. In our formulation the momentum components
$p_{j}$ commute, and we have the usual periodicity over the Brillouin
zone,
\begin{equation}
\exp \left( i \left[ p_{j} + \frac{2\pi}{a} \right] \hat x_{j} \right)
= \exp (i p_{j} \hat x_{j} ) \,.
\end{equation}
Multiplication by a factor $\exp ( -i \sum_{\ell =1}^{2} p_{\ell} \hat
x_{\ell})$ leads to the condition
\begin{equation}  \label{intcon}
\frac{1}{2a} \, \vartheta \, p_{j} \in \Z \,.
\end{equation}
Hence our NC lattice at fixed $\vartheta$ is automatically
\emph{periodic}.

Let us assume now that we are dealing with a torus of lattice size $N
\times N$: then the momenta can take the discrete values
\begin{equation}
p^{(n)} = \frac{2\pi}{aN} \, n \,, \qquad n = (n_{1},n_{2}) \,, \qquad
n_{j} = 1, \dots , N \,.
\end{equation}
Together with relation~(\ref{intcon}) we identify the
non-commutativity parameter as
\begin{equation}  \label{double}
\vartheta = \frac{1}{\pi} N a^{2} \,.
\end{equation}
Taking the continuum limit of a NC theory requires to keep $\vartheta$
finite. By condition~(\ref{double}), the continuum limit $a \to 0$ and
the thermodynamic limit $N \to \infty$ are entangled; we will take
them simultaneously in such a way that the product $N a^{2}$ remains
constant. Then obviously also the physical volume diverges.  This kind
of limit is denoted as the \emph{double scaling limit}, and the
entanglement that it involves is related to the UV/IR mixing: based on
relation~(\ref{double}) we have to take the UV limit and the IR limit
simultaneously.

\subsection{The non-commutative $\lambda \phi^{4}$ model}\label{section1.1}

NC field theories can be formulated in a form which looks similar to
their commutative counterpart, if all field multiplications are
performed by the \emph{star product}~\cite{starprod}
\begin{equation}  \label{stern}
\phi (x) \star \psi (x) := \phi (x) \exp \left( \frac{i}{2}
\overleftarrow{\partial_{\mu}} \Theta_{\mu \nu}
\overrightarrow{\partial_{\nu}} \right) \psi (x) \,,
\end{equation}
as we recognize if the fields are decomposed into plane waves; a
derivation is sketched in appendix~\ref{maplatmat}.  Then we can work
with ordinary coordinates $x_{\mu}$, and the non-commutativity is
encoded in the star product.

In the specific case of \emph{bilinear} terms in the action, the star
product is equivalent to the ordinary product, as we see from
integration by parts and the antisymmetry of the tensor $\Theta$.
Equipped with these rules we can write down the action of the NC
$\lambda \phi^{4}$ model in the $d$-dimensional Euclidean space,
\begin{equation}  \label{cont-act}
S [ \phi ] = \int d^{d}x \, \left[ \frac{1}{2} \partial_{\mu} \phi(x)
  \partial_{\mu} \phi(x) + \frac{m^{2}}{2} \phi(x)^{2} +
  \frac{\lambda}{4} \phi(x) \star \phi(x) \star \phi(x) \star \phi(x)
  \right].
\end{equation}
We see that the strength of the self-interaction $\lambda$ also
determines the extent of the effects due to the non-commutativity.

To illustrate now the UV/IR mixing --- that we mentioned before --- we
consider the perturbative expansion of the one particle irreducible
two-point function,
\begin{equation}
\Gamma (p) = \langle \tilde \phi (p)^{*} \tilde \phi (p) \rangle_{1PI}
= \sum_{\ell =0}^{\infty} \lambda^{\ell} \Gamma^{(\ell )}(p) \,,
\end{equation}
where $\tilde \phi$ is the scalar field in momentum space.  To the
leading order $\ell = 0$ the action is bilinear, hence the star
product is not needed and we find for the free field the same result
as in the commutative case, $\Gamma^{(0)}(p) = p^{2} + m^{2}$.
However, if we move on to $\Gamma^{(1)}$ the commutative term splits
into a planar contribution, which is not affected by $\Theta$, plus a
non-planar term, which is altered by the non-commutativity,
$\Gamma^{(1)}(p) = \Gamma^{(1)}_{\rm p} + \Gamma^{(1)}_{\rm np}(p)$,
\begin{equation}  \label{pnp}
\Gamma^{(1)}_{\rm p} = 2 \int \frac{d^{d}q}{(2\pi )^{d}}
\frac{1}{q^{2}+m^{2}} \,, \quad \Gamma^{(1)}_{\rm np} (p) = \int
\frac{d^{d}q}{(2\pi )^{d}} \frac{\exp (iq_{\mu} \Theta_{\mu \nu}
  p_{\nu})}{q^{2}+m^{2}} \,.
\end{equation}
The planar term confirms that part of the commutative UV divergences
persist. It has been shown that this behavior holds for the planar
terms to all orders~\cite{UV,UVIR}.

The non-planar term can be evaluated for instance with the
Pauli-Villars regularization. In $d=4$ one obtains~\cite{UVIR,Szabo}
\begin{equation}
\Gamma^{(1)}_{\rm np} (p) = \frac{1}{96 \pi^{2}} \left[ \Lambda_{\rm
    eff}^{2} - m^{2} \log \left( \frac{\Lambda_{\rm eff}^{2}}{m^{2}}
  \right) + O(1) \right] \,, \qquad \Lambda_{\rm eff}^{2} =
\frac{1}{\frac{1}{\Lambda^{2}} - p_{\mu} \Theta_{\mu \nu} p_{\nu}} \,.
\label{Lambdaeff}
\end{equation}
$\Lambda$ is the usual momentum cutoff as it appears in the
commutative model and in the planar term of eq.~(\ref{pnp}).
Eq.~(\ref{Lambdaeff}) shows explicitly that $\Gamma^{(1)}_{\rm np}
(p)$ is UV finite (with respect to $\Lambda$) in the NC model at
finite external momentum $p$.  It diverges, however, if we take in
addition the IR limit $p \to 0$, \emph{or} the limit $\Theta \to 0$.

A step beyond standard perturbation theory in the discussion of the NC
$\lambda \phi^{4}$ model was undertaken by Gubser and Sondhi, who
performed a self-consistent Hartree-Fock type one-loop
calculation~\cite{GuSo}. This method would be exact for the $O(N)$
symmetric $\sigma$-model in the large $N$ limit. Gubser and Sondhi
conjectured its (qualitative) applicability also at $N=1$.  In
particular they derived a prediction for the qualitative feature of
the phase diagram. A strongly negative coefficient $m^{2}$ corresponds
to a very low temperature, such that some ordered structure is
enforced.

In dimensions $d=3$ and $4$, with two NC coordinates obeying
eq.~(\ref{NCplane}), ref.~\cite{GuSo} predicted that
\begin{itemize}
\item at small $\vartheta$, decreasing $m^{2}$ results in the usual
  Ising type uniform order, as in the commutative model
\item at larger values of $\vartheta$ the ordering favors the
  formation of \emph{stripes} in some direction, where $\phi (x)$
  takes opposite signs, or even the superposition of stripes in
  different directions (``checker board patterns'').
\end{itemize}

The phenomenon of a spontaneous stripe formation is also known in
solid state physics, see for instance ref.~\cite{solidref}.  However,
such a phase is unknown in the commutative $\lambda \phi^{4}$
model. Its predicted appearance can be understood as a consequence of
an IR divergence, which shifts the energy minimum to non-vanishing
momenta. At $m^{2} \ll 0$ the modes close to the energy minimum may
condense, which corresponds to the stabilization of such stripe
patterns.

The conjecture by Gubser and Sondhi was later supported by a study
using an effective action, which was treated approximately by the
Raileigh-Ritz method~\cite{CZ}.  The possible existence of this
striped phase was also discussed in the framework of a renormalization
group analysis in $4-\varepsilon$ dimensions~\cite{ChenWu}. For
generalities on the wilsonian renormalization group in NC field
theory, see refs.~\cite{RG}.

However, this new phase remained on the level of conjectures until the
system could be studied non-perturbatively by means of Monte Carlo
simulations. The explicit result for the phase diagram and for the
dispersion relation on the 3d lattice, including its limit to the
continuum and to infinite volume, will be presented here.  Some points
of this work have been anticipated in various proceeding
contributions~\cite{Procs} and in a Ph.D.\ thesis~\cite{Diss}.

We note that the occurrence of a striped phase implies the spontaneous
breaking of translation and rotation invariance. The properties of the
Nambu-Goldstone boson related to the broken translation symmetry will
be discussed in section~\ref{section8}.  Based on this symmetry
breaking, Gubser and Sondhi did not expect a striped phase in
$d=2$. However, it was observed numerically that a striped phase
\emph{does} exist in two dimensions~\cite{AC}, and we are going to
confirm that observation in appendix~\ref{2dmod}. As ref.~\cite{AC}
pointed out, this does not contradict the Mermin-Wagner
Theorem~\cite{MW}, since the proof for that Theorem assumes properties
like locality and a regular IR behavior, which do not hold in most NC
field theories.

\section{The matrix model formulation}\label{section2}

A lattice formulation of the model~(\ref{cont-act}) is obtained
from the discretization as described in section~\ref{section1}, where the derivatives
can be replaced (for instance) by differences between nearest neighbor
lattice sites. However, in such a formulation the field variables
on any two lattice sites are coupled through the star product.
This property would make a direct simulation very tedious.

A way out of this problem is the mapping of the lattice model onto a
dimensionally reduced, twisted matrix model. This procedure was
suggested in refs.~\cite{AMNS}, as a refinement of a previous work on
NC gauge theories in the continuum~\cite{IIKK}. Some aspects of this
map are summarized in appendix~\ref{maplatmat}.

For our scalar field $\phi (\vec x, t)$ on a periodic lattice of size
$N^{2} \times T$ and unit lattice spacing, this mapping leads to the
action
\begin{equation}
S [ \hat \phi ] \!=\! N  \Tr  \sum_{t=1}^{T} \left[ \frac{1}{2}
  \sum_{j=1}^{2} \left( \hat \Gamma_{j} \hat \phi (t) \hat
  \Gamma_{j}^{\dagger} - \hat \phi (t) \right)^{2} + \frac{1}{2}
  \left( \hat \phi (t +1) - \hat \phi (t) \right)^{2} +
  \frac{m^{2}}{2} \hat \phi (t)^{2} + \frac{\lambda}{4} \hat \phi
  (t)^{4} \right], \label{matact}
\end{equation}
where $\hat \phi (t)$ represents a hermitean $N \times N$ matrix,
living on one of the discrete time points $t = 1, \dots ,T$.

Since the time direction is commutative, its contribution to the
kinetic term can be discretized in the usual way. On the other hand,
in the spatial directions the shift by one lattice unit has to be
arranged for by some matrix transformation, in our case by the
multiplication with the \emph{twist eaters} $\hat \Gamma_{j}$. The
condition for them is that they obey the 't~Hooft-Weyl algebra
\begin{equation}
\hat \Gamma_{i} \hat \Gamma_{j} = {\cal Z}_{ji} \hat \Gamma_{j} \hat
\Gamma_{i} \,,
\end{equation}
where the phase factor ${\cal Z}_{ji} = {\cal Z}_{ij}^{*}$ is the
\emph{twist}.

In general a phase factor like ${\cal Z}_{12}$ in a twisted matrix
model can take the form~\cite{GAO}
\begin{equation}  \label{twist1}
{\cal Z}_{12} = e ^{2\pi i k /N} \,,  \quad k \in \Z \,, 
\end{equation}
since the twist originates from the boundary conditions before
compactification. In our formulation we set
\begin{equation}  \label{twist2}
k = \frac{N+1}{2} \,, 
\end{equation}
as we are going to explain in subsection~\ref{section2.1}. Thus our
twist differs from the conventional choice $k=1$. Of course, this
means that we have to use odd values for $N$. With this choice, the 3d
lattice model and the 1d twisted matrix model (where the twist appears
implicitly in the form of the twist eaters) can be rigorously
identified~\cite{AMNS} by means of \emph{Morita
  equivalence}~\cite{Morita}, which means that the algebras in both
formulations are identical.

Note that typical matrices $\hat \phi (t)$ are densely filled.  This
is in contrast to ordinary lattice field theory simulations, where the
matrices that appear upon integrating out some of the fields tend to
be sparse. Hence the matrix multiplications in action~(\ref{matact})
require a considerable computational effort (it grows as $O(N^{3})$,
i.e.\ faster than the system size of $O(N^2 )$), since the standard
lattice techniques for matrix operations cannot be applied.  For a
description of our numerical methods we refer to appendix
\ref{simu}. At this point we only remark that the property of densely
filled matrices can be considered as a problem inherited from the star
products in the lattice action, though the mapping onto the matrix
model still simplifies the simulation drastically.

\subsection{The twist}\label{section2.1}

Let us briefly discuss in this subsection our choice of the twist
factor.  It is instructive to return to a general spatial lattice
constant $a$ in this context.  First, since we are on a torus we
should not use the unbounded operators $\hat x_{j}$. Instead we
introduce the unitary matrices
\begin{equation}  \label{Zop}
\hat Z_{j} := \exp \left( \frac{2\pi}{Na} i \hat x_{j} \right), 
\end{equation}
which obey the commutation relation
\begin{equation}  \label{con1}
\hat Z_{i} \hat Z_{j} = \exp \left( -\frac{4\pi^{2}}{N^{2}a^{2}} i
\Theta_{ij} \right) \hat Z_{j} \hat Z_{i} = \exp \left(
-\frac{4\pi}{N} i \epsilon_{ij} \right) \hat Z_{j} \hat Z_{i} \,.
\end{equation} 
In the last step we have used the relations~(\ref{NCplane})
and~(\ref{double}).

In order to describe the shift by one lattice unit we need the
translation operator
\begin{equation}  \label{shiftop}
\hat D_{j} := e^{a \hat \partial_{j}} \,;
\end{equation}
the operator $\hat \partial_{j}$ is introduced in appendix
\ref{maplatmat}.  To perform this shift correctly, $\hat D_{j}$ has to
fulfill the relation
\begin{equation}  \label{con2}
\hat D_{j} \hat Z_{i} \hat D_{j}^{\dagger} = e^{2\pi i a \delta_{ij}
  /N} \hat Z_{i} \,.
\end{equation}

The issue is now to find a matrix solution for the
conditions~(\ref{con1}) and~(\ref{con2}). To this end we use the
unitary twist eaters
\begin{equation}
\hat \Gamma_{1} = \pmatrix{0 & 1 &   &   &   &   &   \cr
  & 0 & 1 &   &   &   &   \cr
& & . & . & & & \cr
& & & . & . & & \cr
& & & & . & . & \cr
& & & & & 0 & 1 \cr
1 &&&&&& 0}, 
\qquad 
\hat \Gamma_{2} = \pmatrix{1 &   &   &   &   &   &   \cr
  & {\cal Z}_{21} &   &   &   &   &   \cr
  &   & {\cal Z}_{21}^{2} &   &   &   &   \cr
  &   &   & {\cal Z}_{21}^{3} &   &   &   \cr
  &   &   &   & . &   &   \cr
  &   &   &   &   & . &   \cr
  &   &   &   &   &   & },
\end{equation}
where ${\cal Z}_{12} = {\cal Z}_{21}^{*}$ characterizes the twist.
Now we see that the ansatz
\begin{equation}
\hat D_{j} = \hat \Gamma_{j} \,, \qquad \hat Z_{1} = \hat
\Gamma_{2}^{(N+1)/2} \,, \qquad \hat Z_{2} = \hat \Gamma_{1}^{\dagger
  ~ (N+1)/2}
\end{equation}
does indeed solve the conditions~(\ref{con1}) and~(\ref{con2}) for odd
$N$, \emph{iff} the twist has the form anticipated in
eqs.~(\ref{twist1}) and~(\ref{twist2}).  Hence this form represents a
satisfactory solution to the two matrix conditions given above (this
solution is unique up to symmetries of the algebra).

\section{A momentum dependent order parameter}\label{section3}

We denote by $\tilde \phi (\vec p , t)$ the scalar field with the two
spatial components expressed in momentum space (the Fourier transform
can be carried out in the usual way, see appendix~\ref{maplatmat}).

Now we introduce the quantity
\begin{equation}  \label{Mkdef}
M(k) := \frac{1}{NT} \ ^{\rm ~~max}_{\frac{N}{2\pi} 
\, \vert \vec p
  \vert = k} \, \left\vert \sum_{t=1}^{T} \tilde \phi (\vec p ,t)
\right\vert \,.
\end{equation}
The field $\tilde \phi (\vec p , t)$ is averaged over the time
direction and then rotated such that its absolute value is maximal at
a specific value of $\vert \vec p \vert$. The expectation value
$\langle M(k) \rangle$ is our \emph{momentum dependent order
  parameter}.

Of course, $\langle M(0) \rangle$ is the standard order parameter for
the $Z_{2}$ symmetry (magnetization). For $k>0$ this order parameter
is sensitive to some kind of staggered order, as it can occur if
anti-ferromagnetic couplings over some distance are involved.  In
particular, $\langle M(1) \rangle$ detects the formation of exactly
two stripes parallel to one of the axes; more precisely it measures
the leading sine component of $\frac{1}{T} \sum_{t} \phi (\vec x , t)$
in such a pattern.  The value $k = \sqrt{2}$ captures the
corresponding case with diagonal stripes, and $\langle M(k \geq 2)
\rangle$ is suitable for the search for orders with higher modes of
$\frac{1}{T} \sum_{t} \phi (\vec x , t)$.

\FIGURE[b]{\epsfig{file=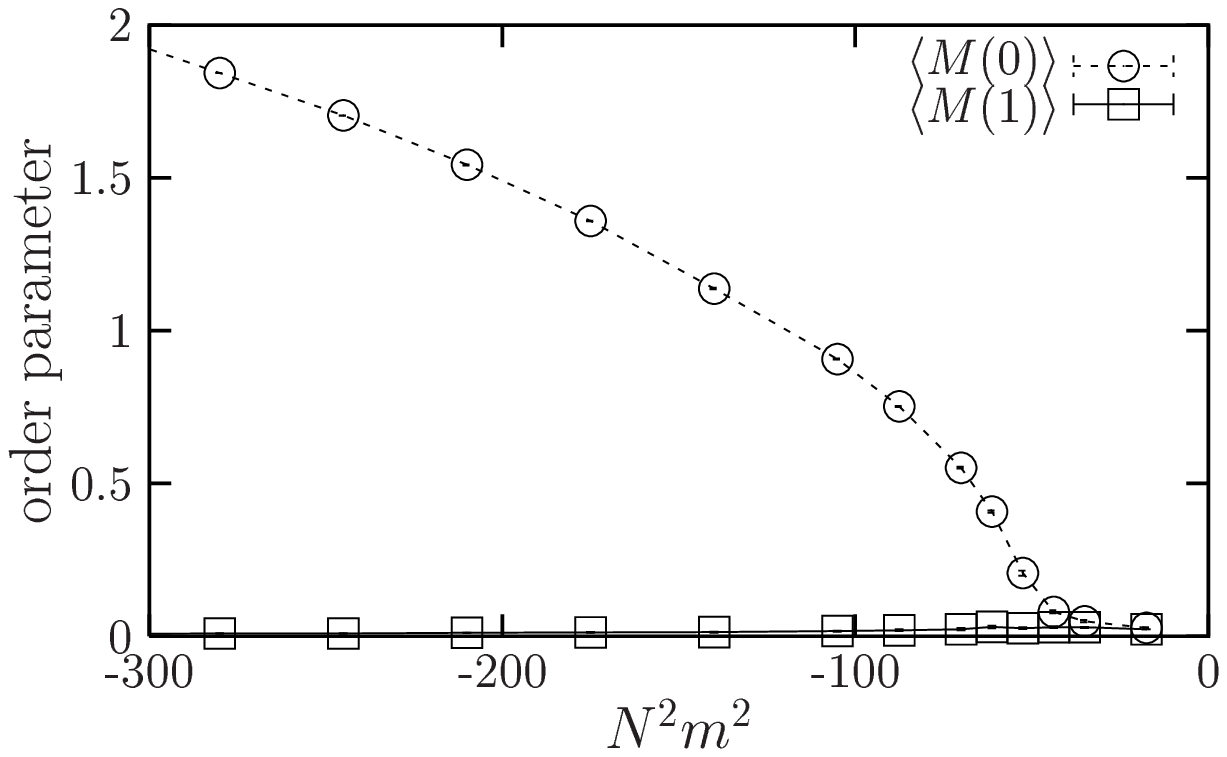,width=.48\linewidth,clip=} 
  \epsfig{file=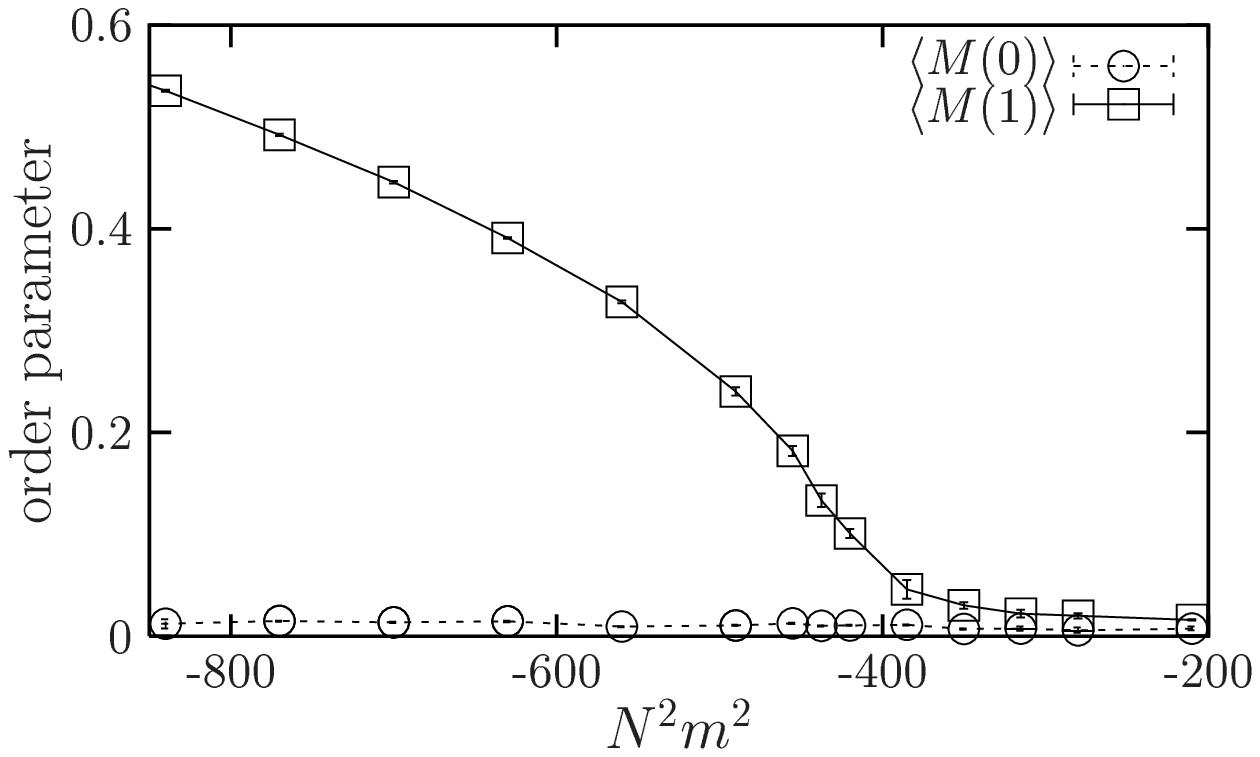,width=.50\linewidth,clip=}
  \caption{The momentum dependent order parameter $\langle
    M(k)\rangle$ (at $k=0,\, 1$) against $N^2 m^2$ at $N=35$. On the
    left we fixed $N^2\lambda = 70$, which leads to the uniform
    phase. On the right we set $N^2\lambda=350$, so that a strongly
    negative $m^{2}$ leads to the striped phase.\label{orderpar-fig}}}

As an example, figure~\ref{orderpar-fig} shows the results for
$\langle M(0) \rangle$ and $\langle M(1) \rangle$ at $N=T=35$ and
small $\lambda$ (on the left) resp.\ large $\lambda$ (on the
right). As $m^{2}$ decreases, the uniform resp.\ staggered order is
observed to set in unambiguously.

{\renewcommand\belowcaptionskip{-.6em}
\FIGURE{\epsfig{file=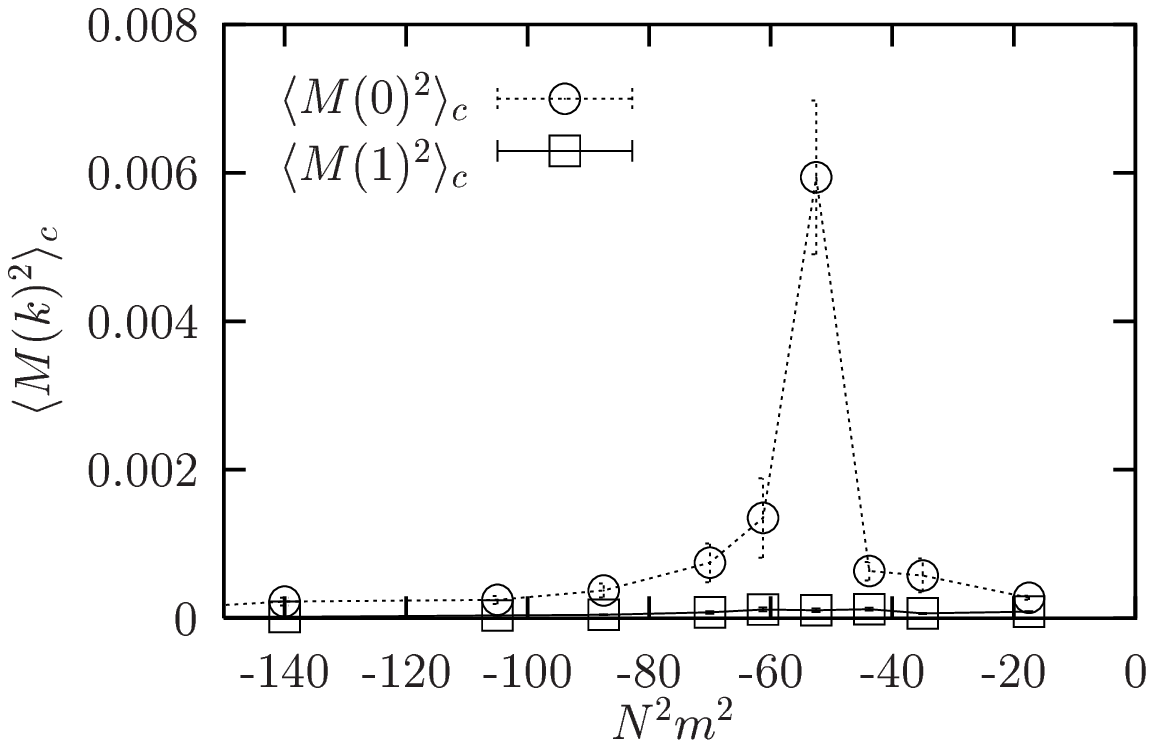,width=.48\linewidth,clip=}
  \epsfig{file=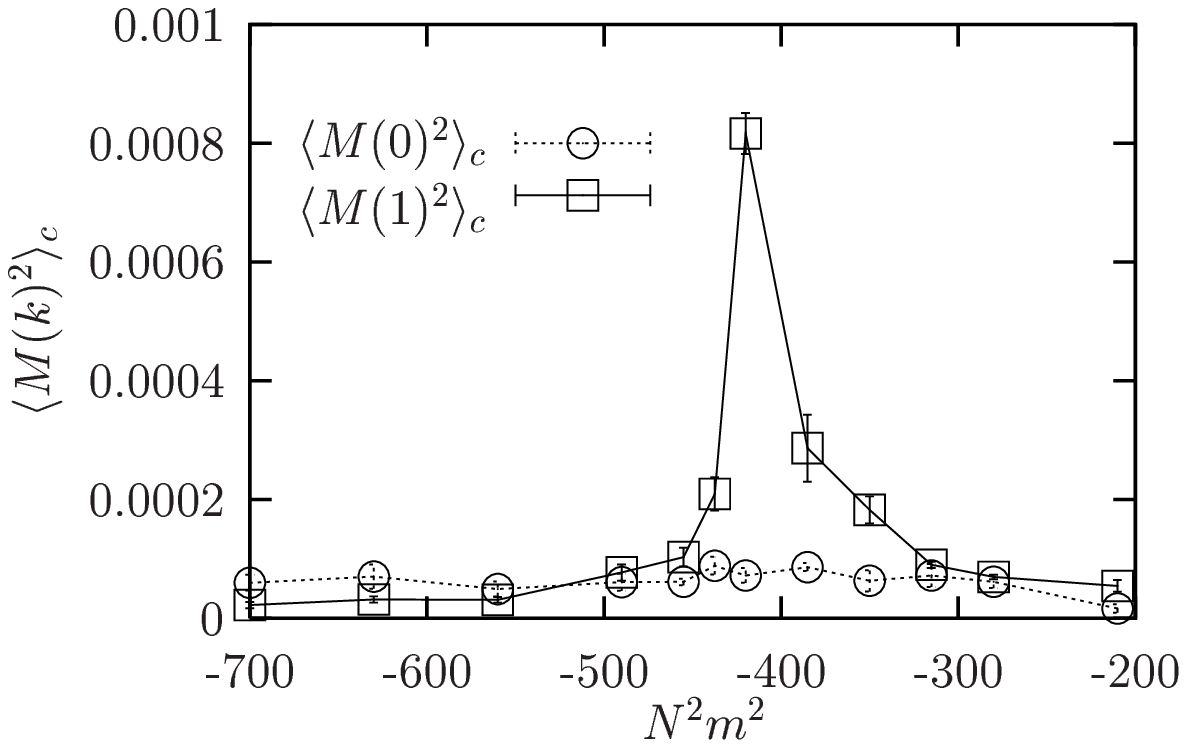,width=.50\linewidth,clip=}
  \caption{The connected two-point function of the momentum
  dependent order parameter, $\langle M(k)^{2} \rangle_{c}$
  against $N^2 m^2$ at $N=35$ and $N^{2}\lambda =70$ (on the left)
  resp.\ $N^{2} \lambda = 350$ (on the right).
  We observe marked peaks, which allow us
  to localize well the phase transitions that we saw before in
  figure~\ref{orderpar-fig}.\label{order2-fig}}}
}

However, from such plots the critical value of $m^{2}$ can only be read
off approximately. For a more precise localization of the disorder-order
phase transition we measured the connected two-point function
$\langle M(k)^{2} \rangle_{c}$, which has a peak at the phase transition
(for $N \to \infty$ it diverges at this point). In figure~\ref{order2-fig}
we show the results for the phase transitions illustrated before 
in figure~\ref{orderpar-fig}. Indeed we find marked
peaks which allow for an accurate determination of the transition from the
disordered phase to the uniformly ordered phase (on the left) or --- at 
larger $\lambda$ --- to a striped phase (on the right).

\pagebreak[3]

{\renewcommand\belowcaptionskip{-.6em}
\begin{figure}[t]
  \centering
\subfigure[\footnotesize{$N^2 \lambda =90,~N^2 m^2 = -22.5$}]
{\epsfig{figure=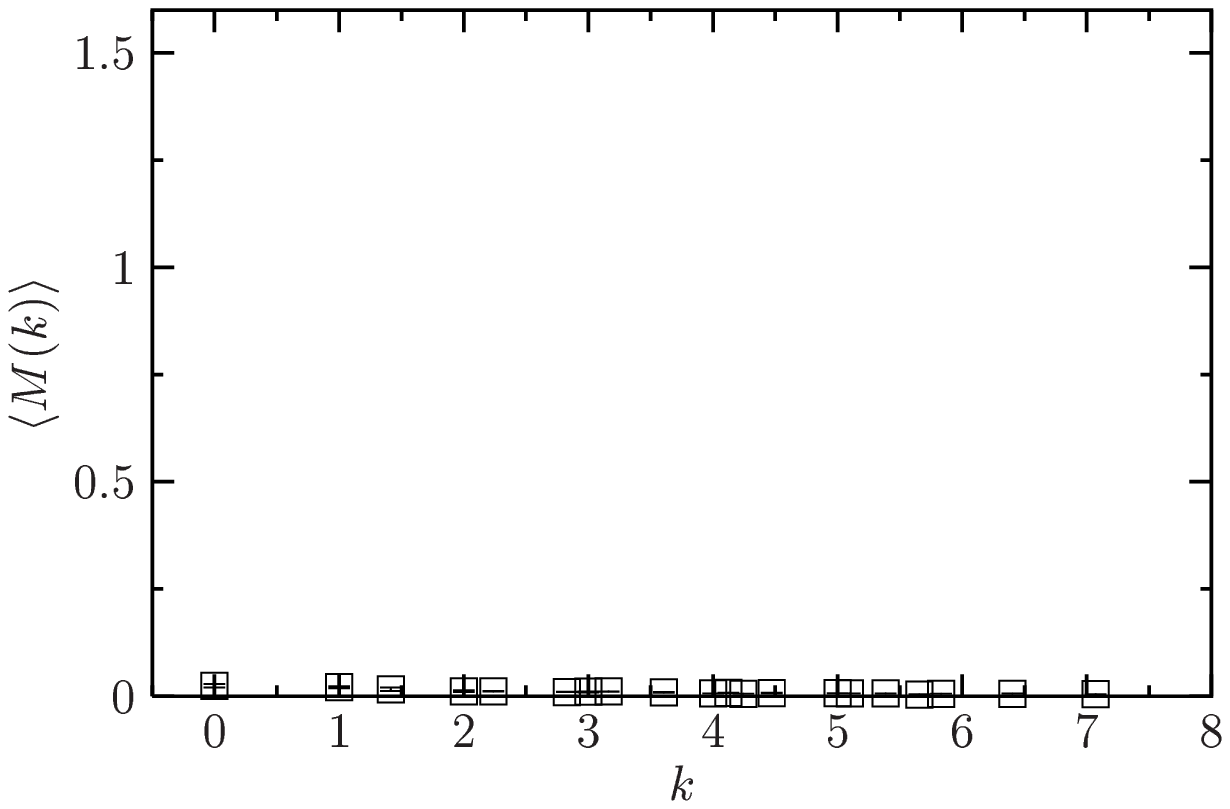,width=.4\linewidth}}%
\hspace*{1cm}
\subfigure[\footnotesize{$N^2\lambda = 900, ~N^2 m^2 =-360$}]
{\epsfig{figure=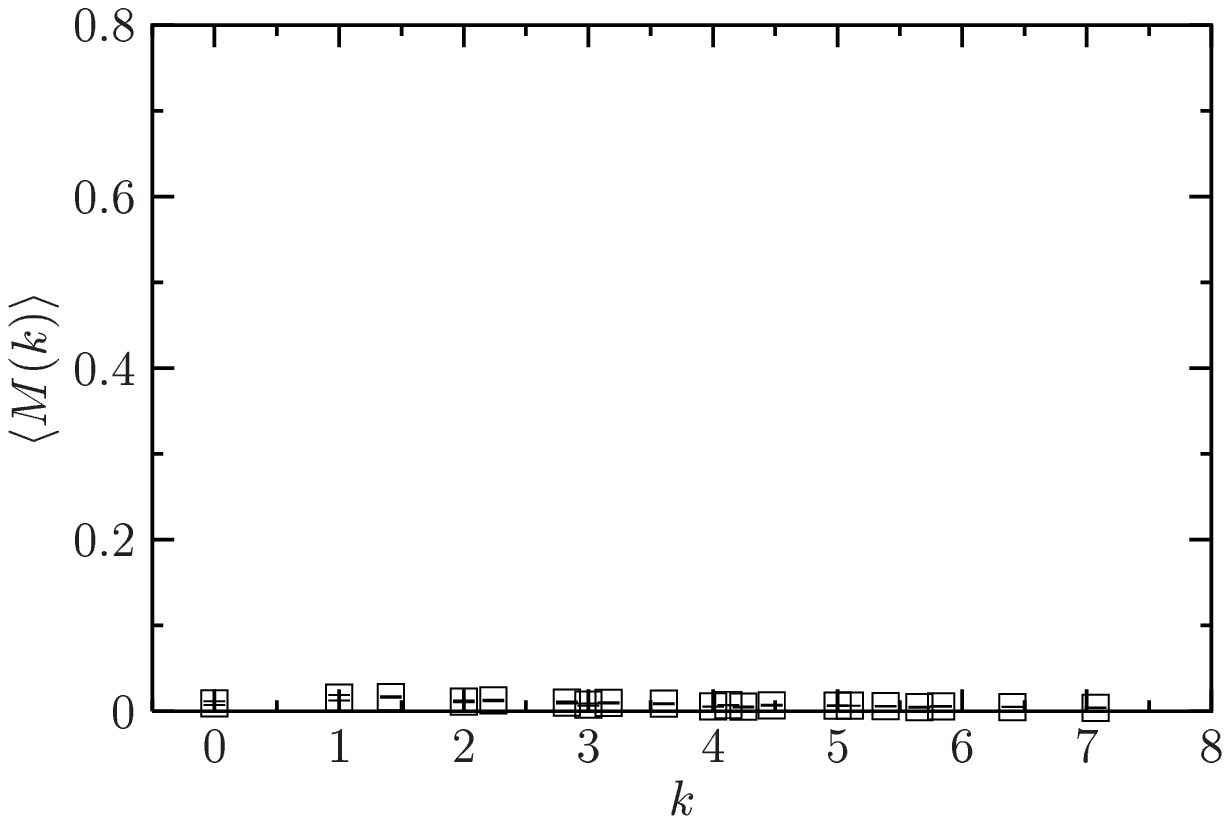,width=.4\linewidth}}
\\
\subfigure[\footnotesize{$N^2 \lambda = 90,~N^2 m^2 =-225$}]
{\epsfig{figure=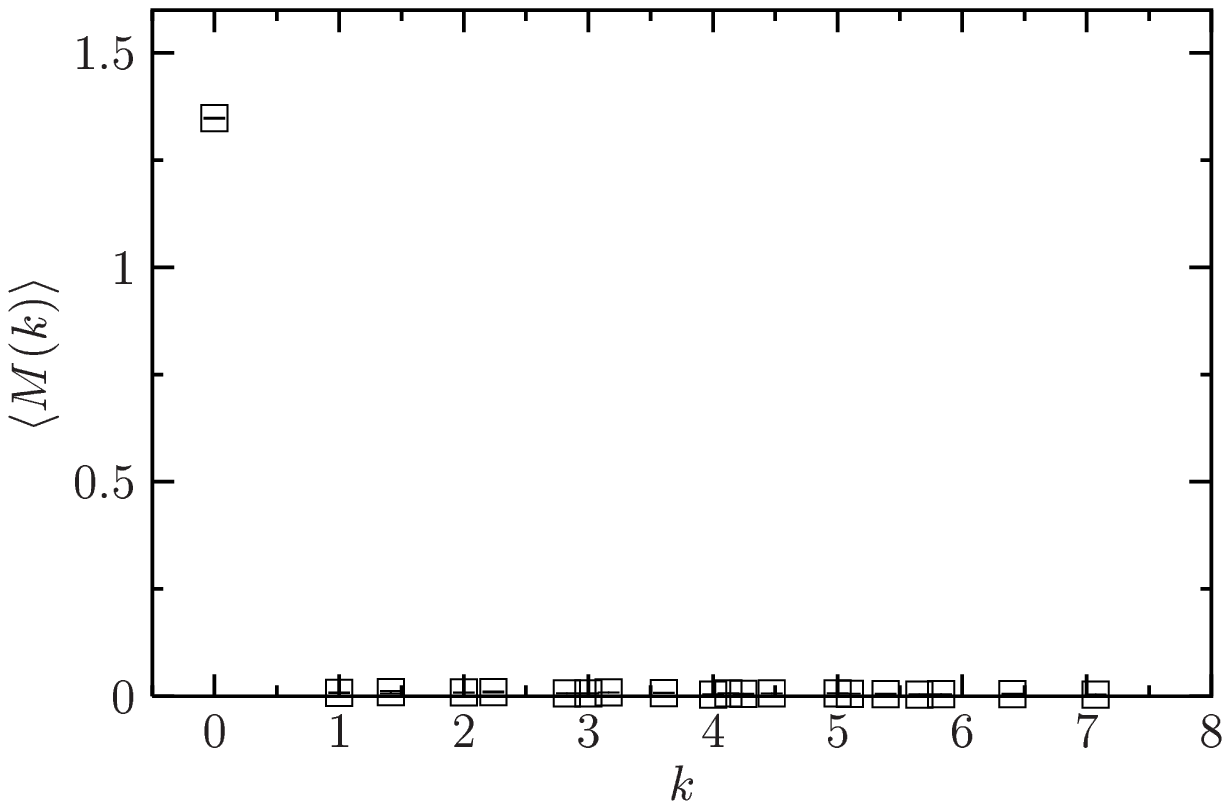,width=.4\linewidth}}%
\hspace*{1cm}
\subfigure[\footnotesize{$N^2 \lambda = 900,~N^2 m^2 = -1350$}]
{\epsfig{figure=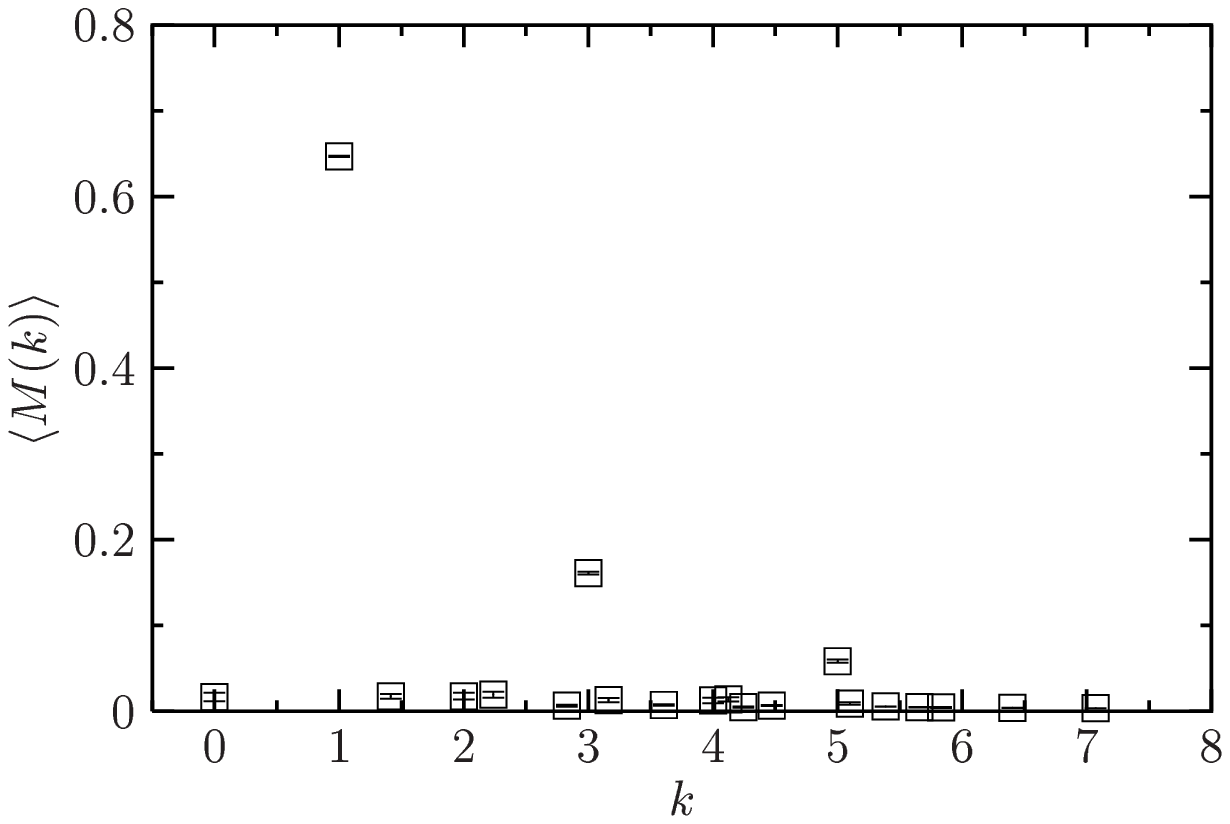,width=.4\linewidth}}
  \caption{The order parameter $\langle M(k) \rangle$ at different
  momenta $k$, at $N=T=45$.}
\label{order-k}
\end{figure}
}

In figure~\ref{order-k} we take a look at $\langle M(k) \rangle$ at
the (discrete) values of $k$ in the range $k = 0 \dots 7$.  As long as
$m^{2}$ is not strongly negative, no order can be found at any
frequency, so we are manifestly in the disordered phase (plots on
top).  Moving on to $m^{2} \ll 0$ we find at small $\lambda$ clearly
the uniform order, since only $\langle M(0) \rangle$ deviates from
zero (plot below on the left). Finally at larger $\lambda$ the Ising
order parameter $\langle M(0) \rangle$ drops to zero again, and we
observe a clear signal at $k=1$, i.e.\ a two-stripe pattern (plot
below on the right).  In this case, we also see some signals at $k=3$
and $5$. However, this should \emph{not} be interpreted as an
indication of an underlying multi-stripe structure. What happens is
that only close to the phase transition the two\pagebreak[3] stripes approximate a
sine shape; for even lower $m^{2}$ they approximate more and more a
step shape, and what we see here is simply the Fourier decomposition
of such a pattern.  This transition is illustrated in figure~\ref{profile}.

{\renewcommand\belowcaptionskip{-1em}
\FIGURE[t]{\epsfig{file=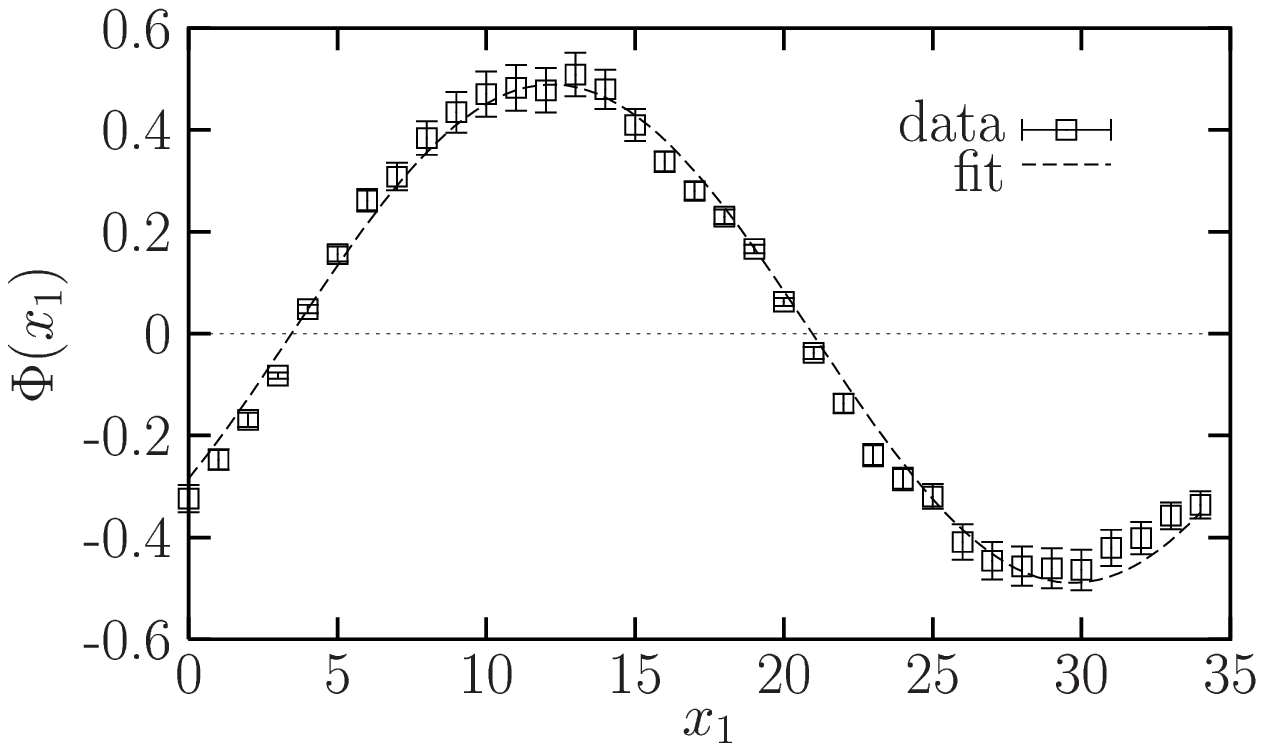,width=.48\linewidth,clip=}
  \epsfig{file=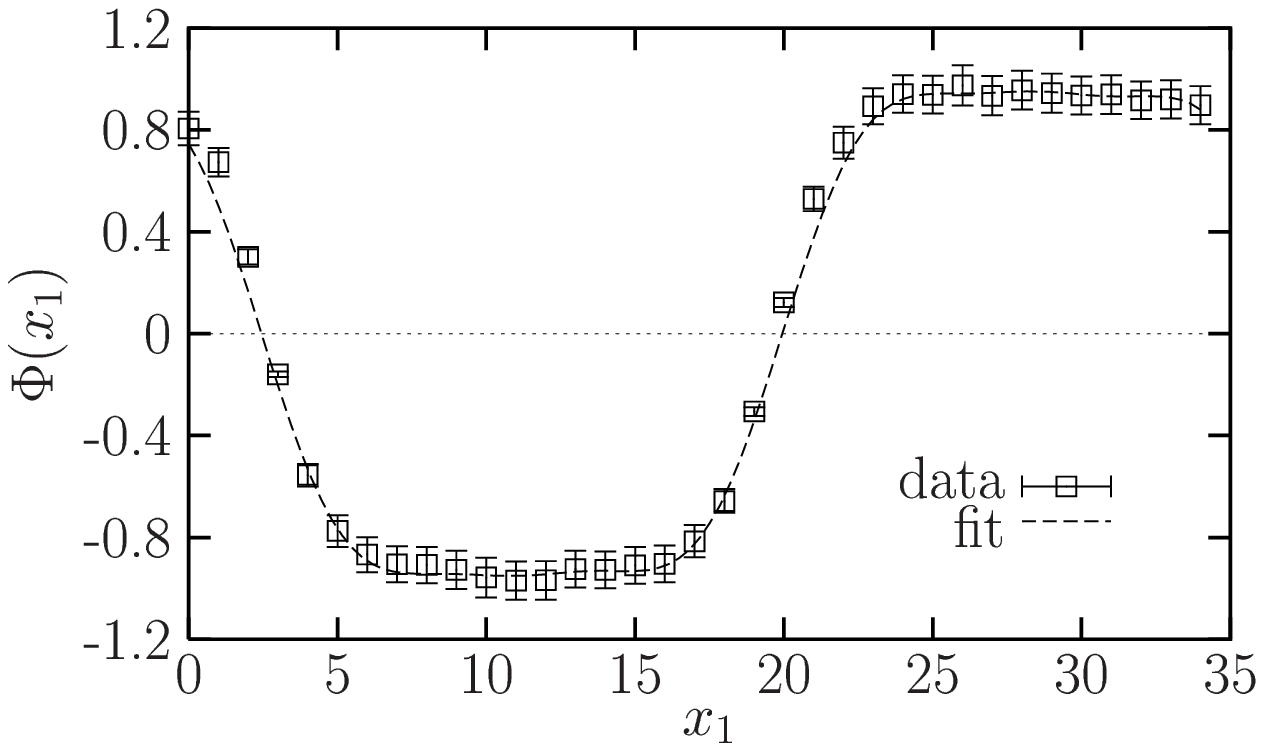,width=.50\linewidth,clip=}
  \caption{The profile $\Phi (x_1 ) = \frac{1}{NT} \sum_{x_{2},t}
  \phi(x)$ of a simple two-stripe pattern at $N=35$, $N^2 \lambda
  =350$: near the disordered phase a sine shape dominates (on the
  left, $N^2 m^2 = -250$), but at even lower $m^{2}$ the profile
  approximates a step shape (on the right, $N^2 m^2 = -620$).  (These
  two plots are based on single configurations, and the error bars
  refer to the sum in $\Phi (x_1 )$.)\label{profile}}}
}

A really stable multi-stripe pattern is hard to find --- as we will
discuss later --- and in that case we should see $\langle M(k \leq
\sqrt{2}) \rangle \simeq 0$ and find a signal only at higher $k$.

\vspace{-1pt minus 8pt}

\section{The phase diagram}\label{section4}

\vspace{-1pt minus 3pt}

In the previous section we explained our tools to explore the phase
diagram.  In particular, decreasing $m^{2}$ at fixed $\lambda$ we
could localize the disorder-order phase transition accurately, and
determine the type of order that emerged.

The next step is to repeat this procedure at various values of $N=T$
and search for suitable axes so that the phase diagram stabilizes for
increasing $N$. The result is shown in figure~\ref{phasedia}.  We
obtained an explicit phase diagram with the qualitative features
conjectured by Gubser and Sondhi.

It turns out that the axes $N^{2}m^{2}$ and $N^{2} \lambda$ are
suitable for a large $N$ extrapolation. The power of $N$ multiplying
the self-coupling is singled out by the transition inside the ordered
regime between the uniform and striped phase. That phase transition,
however, cannot be localized to the same precision as the
disorder-order transition. The reason for this property will be
clarified later on in the discussion of the dispersion relation
(section~\ref{section6}). Still the stabilization of this region in
$N$ is ultimately compelling.

Of course, the question about the \emph{order of the phase
  transitions} is of interest, but (as in many other models) it is
difficult to arrive at an absolutely safe answer. To get insight into
this question, we searched for a hysteresis behavior by crossing the
phase transition lines.  If we cross the transition between disorder
and order, we do \emph{not} see any hysteresis effect. We have checked
this behavior to a high precision for the transition from the
disordered phase to both, the uniform and the striped phase, see
figure~\ref{hyst}.  Hence we assume that transition to
be of \emph{second order}.

The situation with respect to the uniform-striped transition is
unclear: for increasing $\lambda$ in the ordered regime, stripes do
eventually show up. On the other hand, for decreasing $\lambda$ it
seems to be hardly possible to make the stripes disappear again.  This
behavior is illustrated in figure~\ref{hyst}.
\pagebreak[3]

{\renewcommand\belowcaptionskip{-1.2em}
\FIGURE[t]{\centerline{\epsfig{file=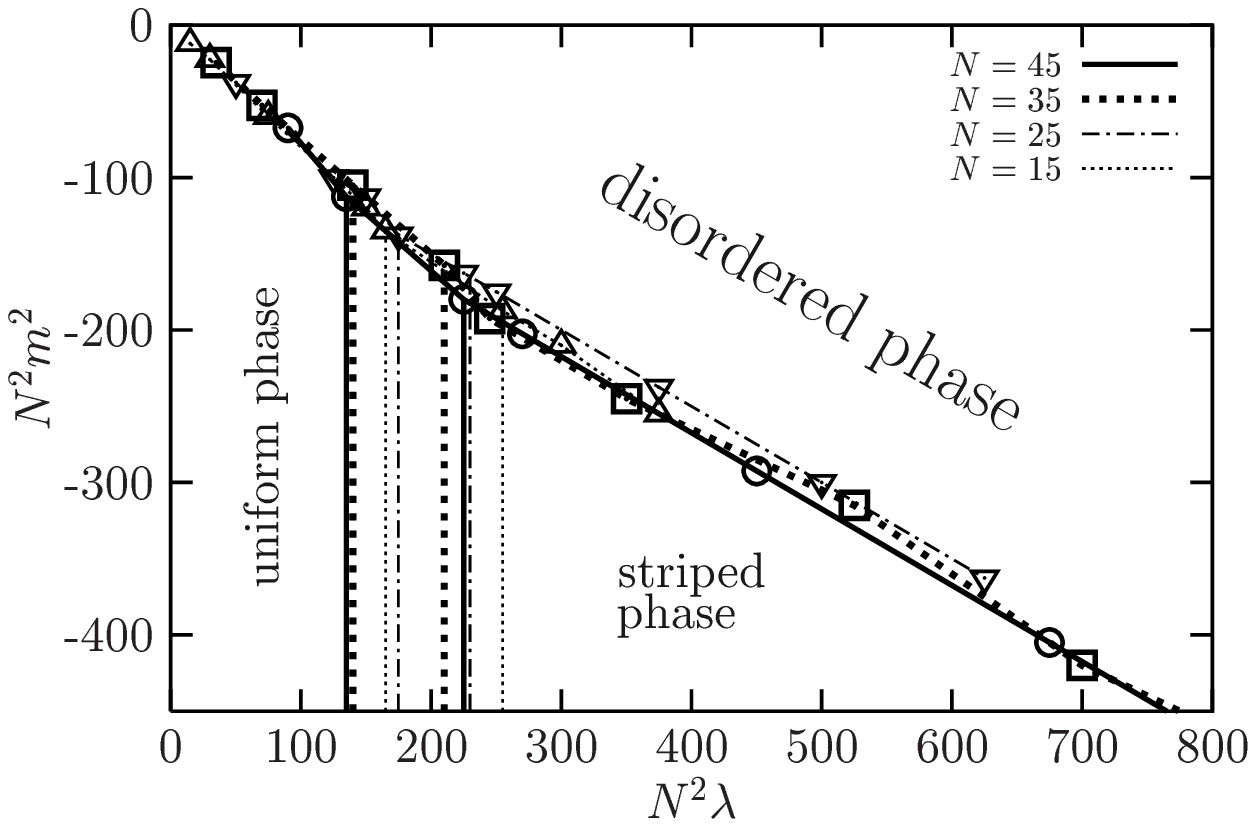,width=.75\linewidth,clip=}}%
  \caption{The phase diagram of the 3d $\lambda \phi^{4}$ model with
    two NC coordinates, identified from (indirect) lattice simulations
    with lattice size $N^{3}$.\label{phasedia}}}

\begin{figure}[t]\centering
\subfigure[\footnotesize{$N^2 \lambda =50$}]%
{\epsfig{figure=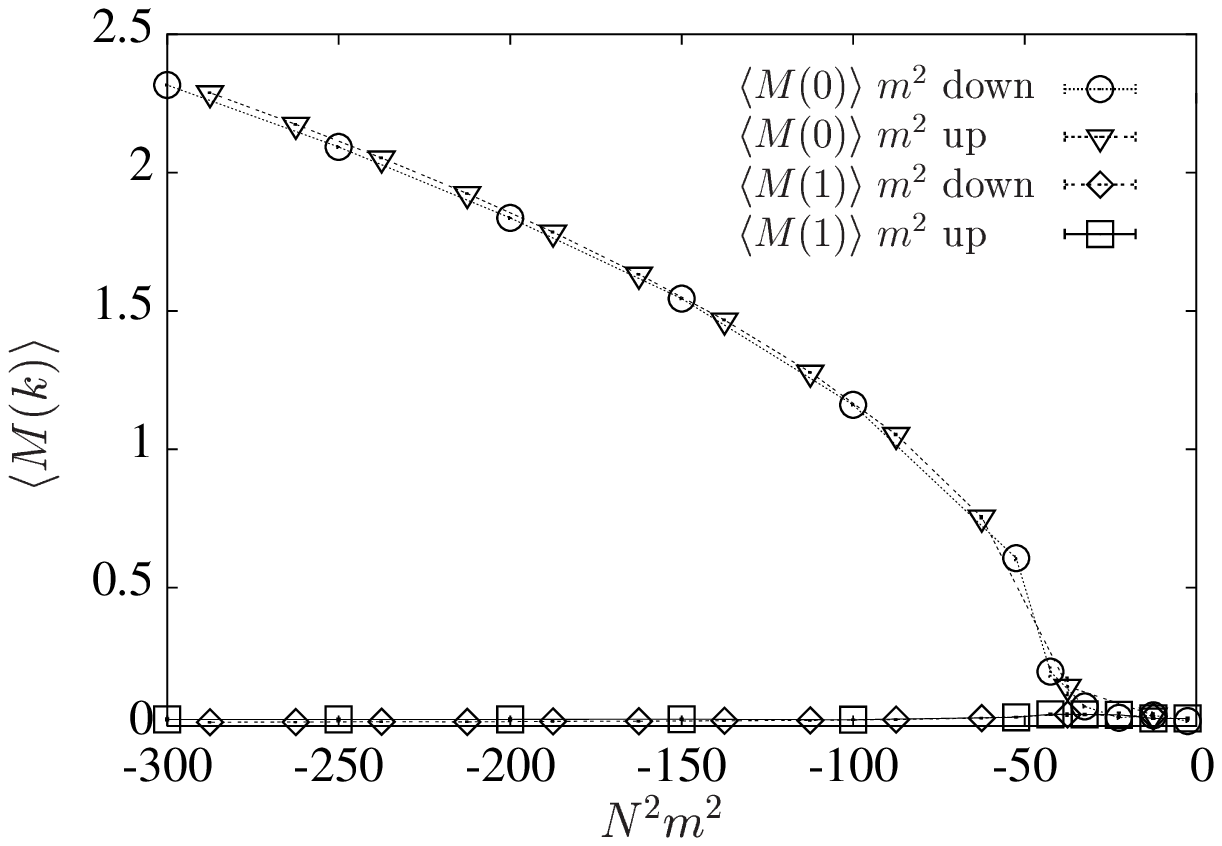,width=.45\linewidth}}%
\hspace*{1cm}
\subfigure[\footnotesize{$N^2\lambda = 500$}]%
{\epsfig{figure=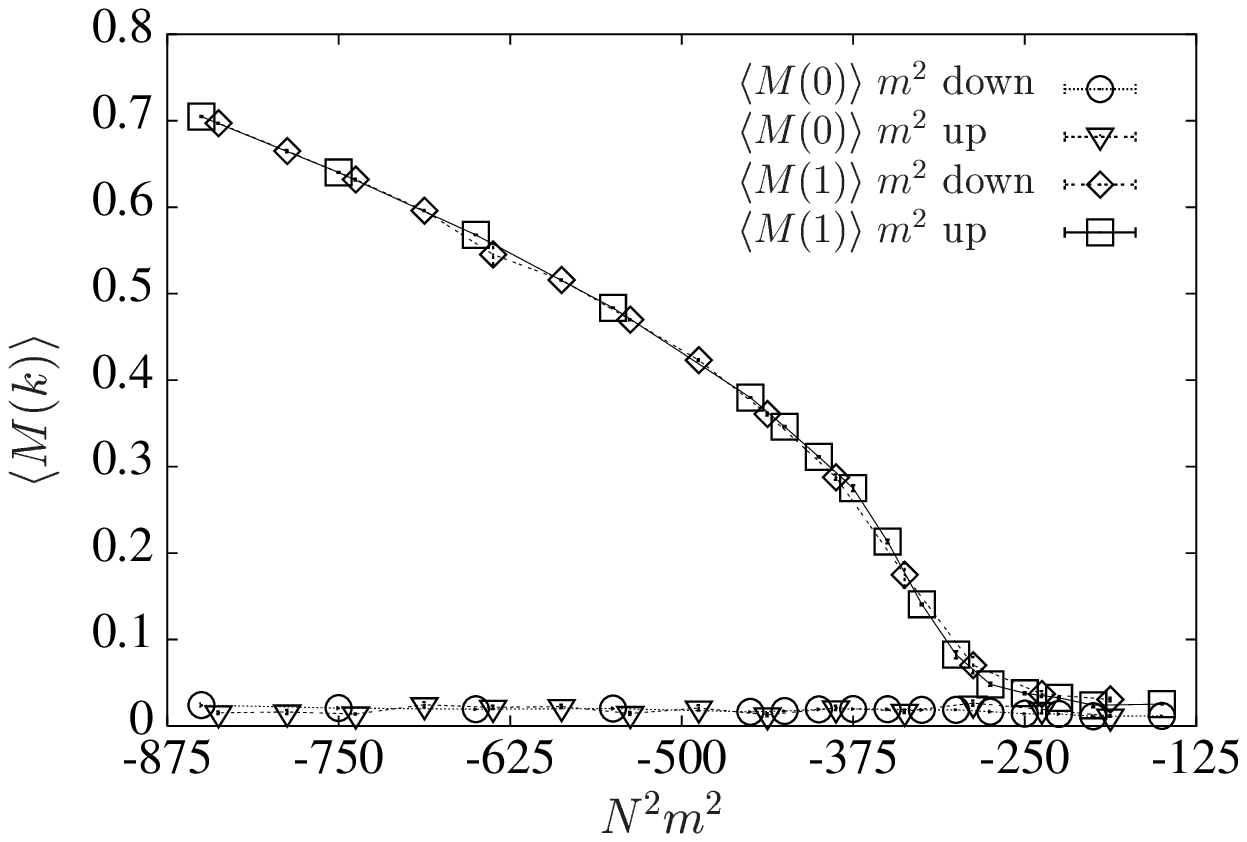,width=.45\linewidth}}
\\[-10pt]
\subfigure[\footnotesize{$N^2 m^2 =-250$}]%
{\epsfig{figure=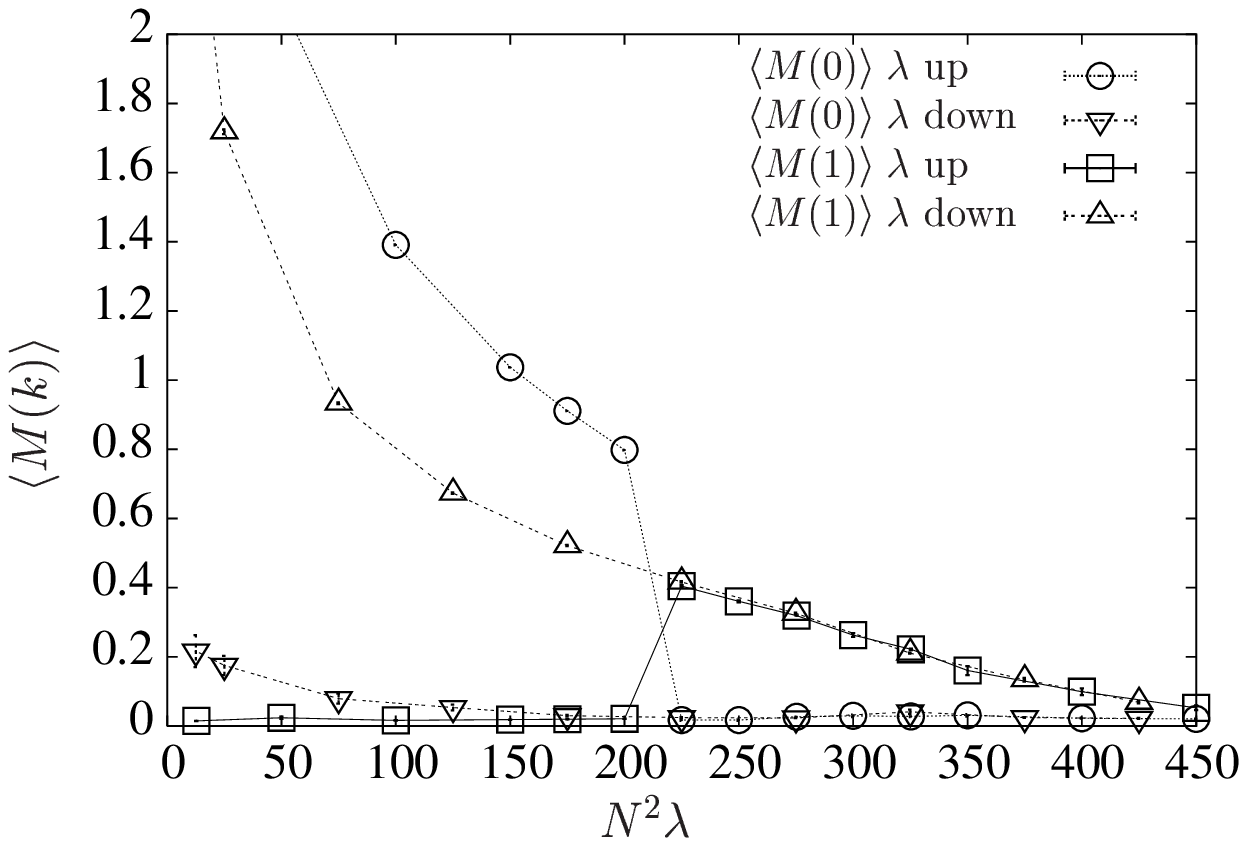,width=.45\linewidth}}%
\caption{An illustration of the hysteresis behavior described in the
  text: for the disorder-order phase transition, no hysteresis can be
  observed, as we show for the case that the uniform resp.\ striped
  order is involved (above, on the left resp.\ on the right).  Below
  we show the outcome for the uniform-striped transformation: once a
  stripe pattern is built, a return to the uniform order is hardly
  possible inside the ordered regime, hence the ``thermal cycle'' does
  not close.\label{hyst}}
\end{figure}
}

\clearpage

\begin{figure}[t]
  \centering
\subfigure[\footnotesize{$N^2 \lambda =70,~N^2 m^2 = -17.5$}]
{\epsfig{figure=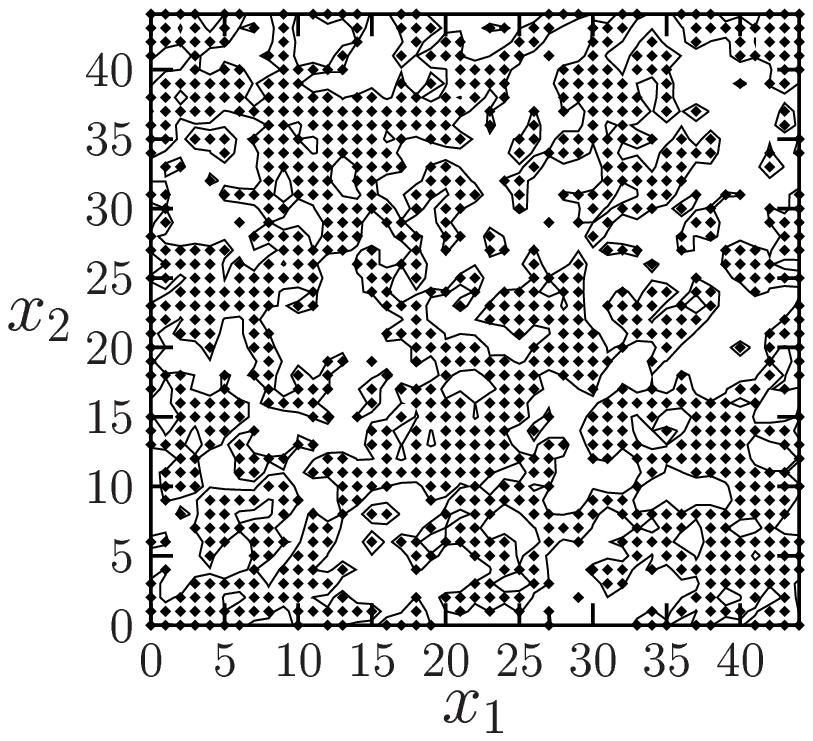,width=.35\linewidth}}%
\hspace*{1cm}
\subfigure[\footnotesize{$N^2\lambda = 700, ~N^2 m^2 =-350$}]
{\epsfig{figure=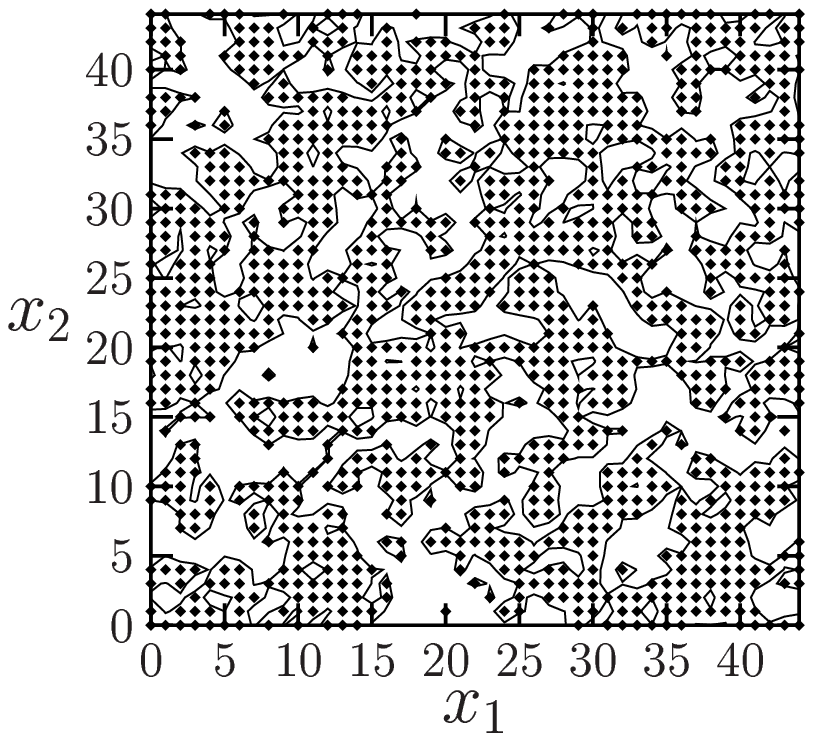,width=.35\linewidth}}
\\
\subfigure[\footnotesize{$N^2 \lambda = 70,~N^2 m^2 =-280$}]
{\epsfig{figure=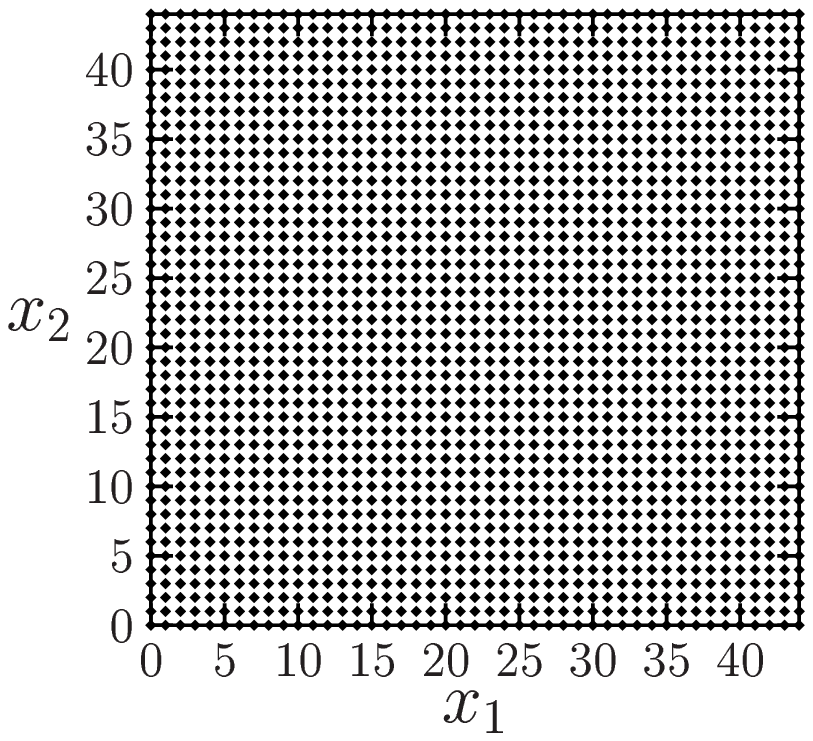,width=.35\linewidth}}%
\hspace*{1cm}
\subfigure[\footnotesize{$N^2 \lambda = 700,~N^2 m^2 = -910$}]
{\epsfig{figure=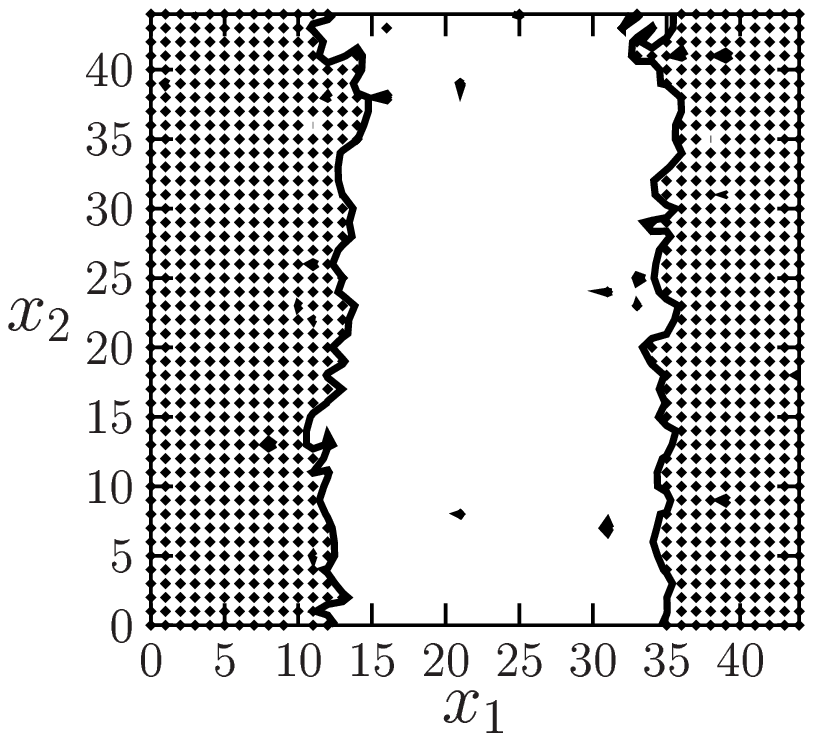,width=.35\linewidth}}%
\caption{Typical lattice configurations at $N = T = 45$ for low $-
  m^{2}$ (on top) and high $- m^{2}$ (below), and for weak $\lambda$
  (on the left) and strong $\lambda$ (on the right). The dotted and
  blank area represent the domains with different signs of $\phi (x)$.
  We recognize disorder in both cases on top, and a uniform resp.\
  striped order below.\label{snap}}
\end{figure}

It is instructive to visualize typical configurations in the different
phases. This can be achieved by mapping back the matrix configurations
to the lattice. The mapping and inverse mapping between lattice
configurations and matrices are briefly described in
appendix~\ref{maplatmat}.  For further details we refer to
ref.~\cite{Diss}.  Typical snapshots at low and at high values of
$\lambda$ and of $- m^{2}$ are shown in figure~\ref{snap}.

\begin{figure}[t]
  \centering
\subfigure[\footnotesize{$N =35$, $\lambda =10$, $m^2 = -4$}]
{\epsfig{figure=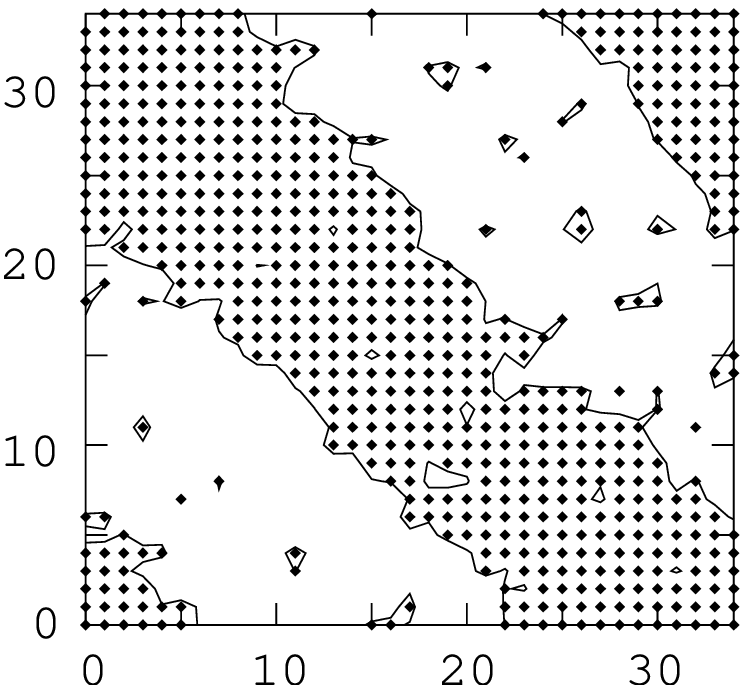,width=.3\linewidth}}%
\hspace*{1cm}
\subfigure[\footnotesize{$N =55$, $\lambda =50$, $m^2 = -22$}]
{\epsfig{figure=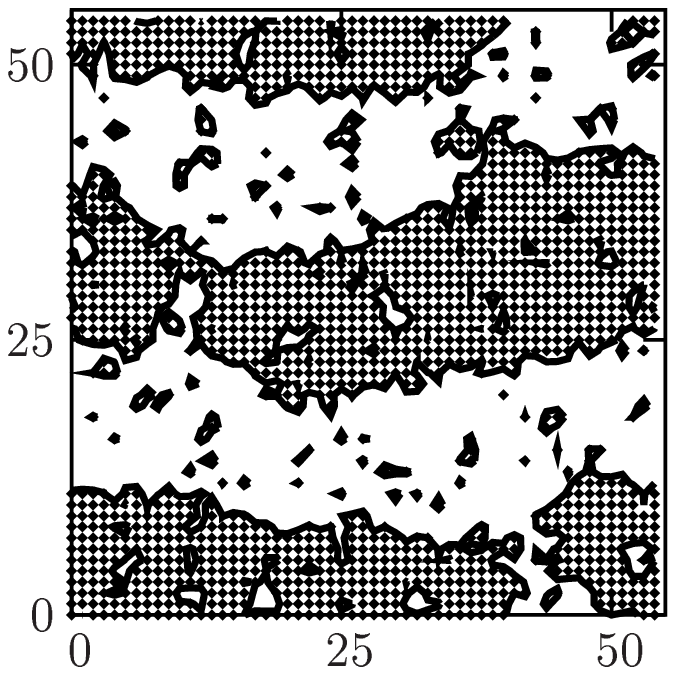,width=.285\linewidth}}
\caption{Examples for a pattern with two diagonal stripes (on the left)
and of four stripes parallel to an axis (on the right). They were
observed in the histories at $N=35$, $\lambda =10$, $m^2 = -4$, resp.\
$N=55$, $\lambda =50$, $m^2 = -22$. However, it is difficult to verify
the ultimate stability of such non-minimal stripe patterns.}
\label{multi}
\end{figure}

In particular we clearly see a two stripe pattern in the plot on
bottom on the right. The search for such obvious pictures also for
multi-stripe and checkerboard patterns --- which are safely stable in
the Monte Carlo history --- turned out be very difficult. Quite clear
hints for such patterns can be seen at relatively large $N$ and a
strong self-coupling $\lambda$; examples are shown in
figure~\ref{multi}.

However, if we just map the configurations back to the lattice and
illustrate it as in figures~\ref{snap} and~\ref{multi}, the emerging
pictures do not show a convincing dominance of one stable multi-stripe
pattern, unlike the case of two stripes parallel to one axis at lower
$\lambda$.

If we really want to confirm the Gubser-Sondhi conjecture we need
evidence that in the limit, which corresponds to the continuum and to
infinite volume, a finite stripe width dominates. Hence in this limit
we expect an infinite number of stripes in various directions, with a
finite width.  Since it is very difficult to find evidence for this
behavior by a direct illustration, we will search for it in a more
subtle manner in sections~\ref{section6} and~\ref{section7}.  That
investigation is based on correlation functions, so this is what we
want to discuss next.

\section{Correlation functions}\label{section5}

We first take a look at the \emph{spatial correlation
  function}\footnote{Theoretically one could also fix $\vec y$ and
  $t$, but the summation is useful in practice to enhance the
  statistics.}
\begin{equation}  \label{spacoreq}
C(\vec x ) = \frac{1}{N^2 T} \sum_{\vec y , t} \langle \phi (\vec x +
\vec y , t ) \phi (\vec y , t) \rangle \,.
\end{equation}
As an example we plot in figure~\ref{spacorfig1} the correlators
$C(x_{1},0)$ and $C(0, x_{2})$ at $N=T=45$ in the four sectors of the
phase diagram, which we also distinguished in figure~\ref{snap}. In
all the sectors we confirm the expected behavior: a fast decay in both
directions in the disordered phase, but hardly any decay in the
uniform phase. In the case of two stripes (parallel to an axis), we
observe a ferromagnetic behavior parallel to the stripes and an
anti-ferromagnetic behavior vertical to them.

{\renewcommand\belowcaptionskip{-.5em}
\begin{figure}[t]
\centering
\subfigure[\footnotesize{$N^2 \lambda =90,~N^2 m^2 = -22.5$}]
{\epsfig{figure=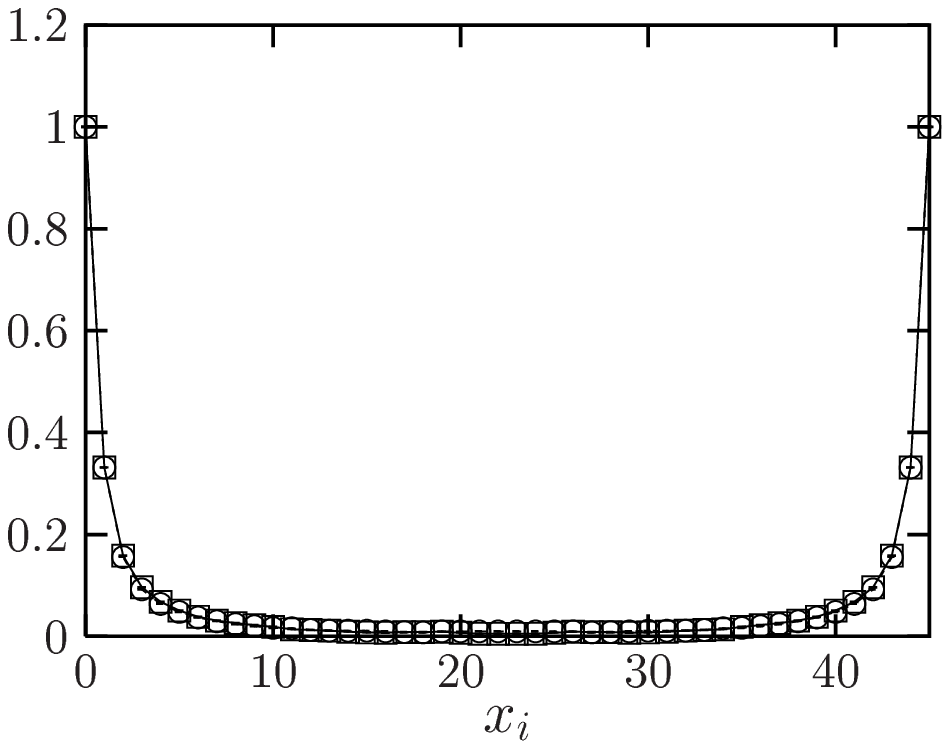,width=.35\linewidth}}%
\hspace*{1cm}
\subfigure[\footnotesize{$N^2\lambda = 900, ~N^2 m^2 =-360$}]
{\epsfig{figure=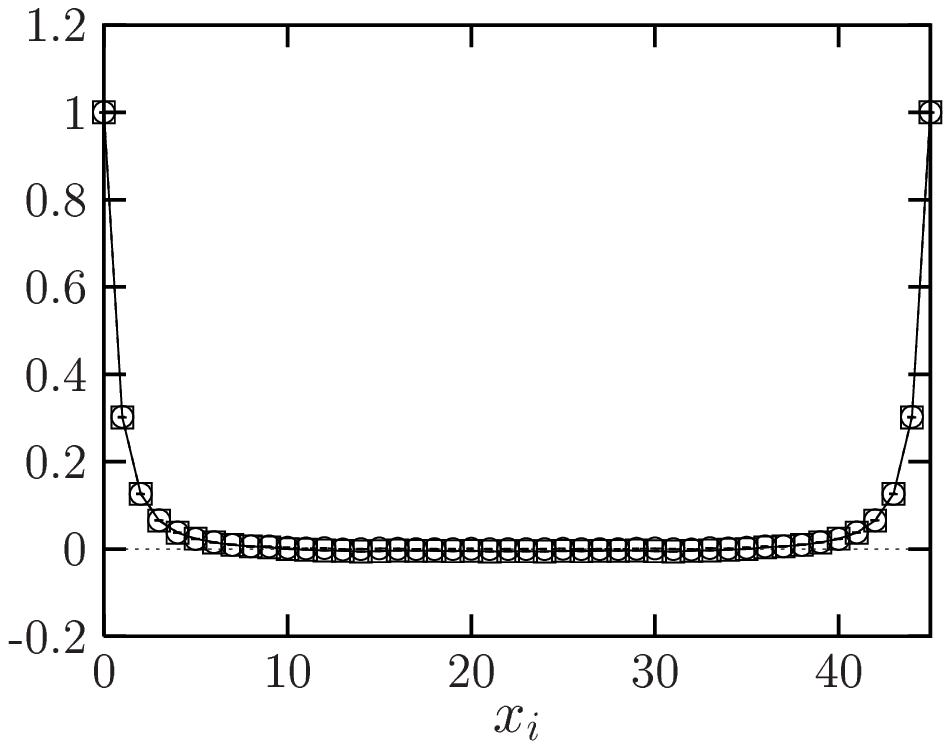,width=.35\linewidth}}
\\
\subfigure[\footnotesize{$N^2 \lambda = 90,~N^2 m^2 =-225$}]
{\epsfig{figure=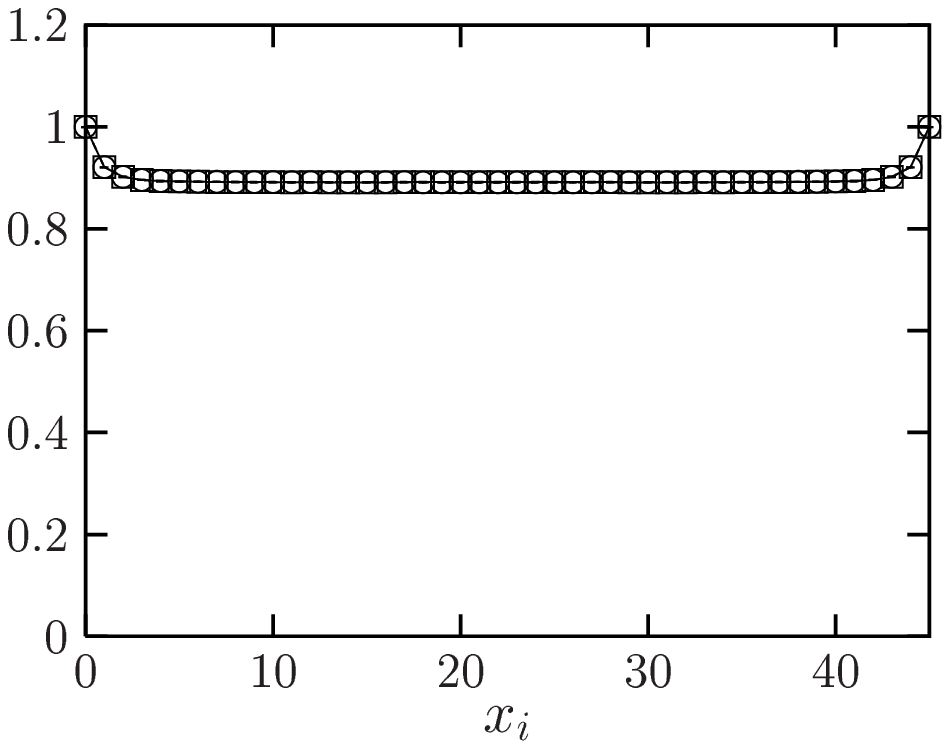,width=.35\linewidth}}%
\hspace*{1cm}
\subfigure[\footnotesize{$N^2 \lambda = 900,~N^2 m^2 = -945$}]
{\epsfig{figure=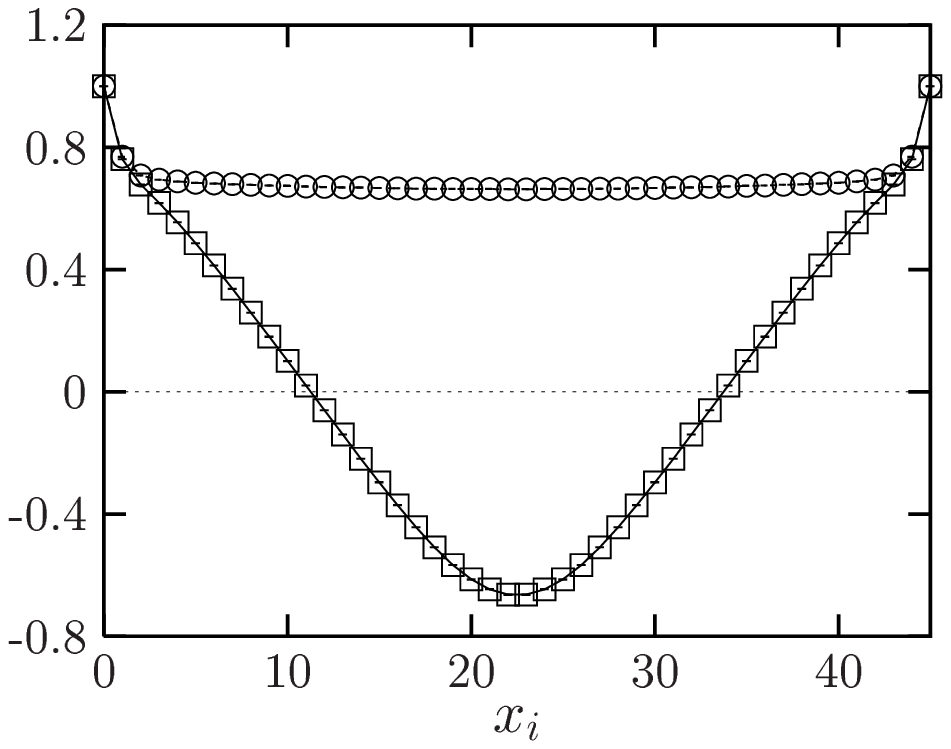,width=.35\linewidth}}
\caption{The spatial correlation function $C(\vec x )$, defined in
  eq.\ (\protect\ref{spacoreq}) (with one component of $\vec x $ being
  $0$) in the different sectors of the phase diagram in figure
  \protect\ref{phasedia}, at $N=T=45$.  In the disordered phase (the
  two plots above) we see a fast decay in both directions, both, for
  small and for large $\lambda$. In the uniform phase (below, left)
  there is hardly any decay, and in the case of two stripes (below,
  right) we observe a mixed ferromagnetic and anti-ferromagnetic
  behavior in the two directions.\label{spacorfig1}}
\end{figure}
}

Next we turn our interest to the fast decay in the disordered phase,
at some point close to the ordering phase
transition. Figure~\ref{spacorlog} shows logarithmic plots for these
correlators, one close to the uniform phase (on the left) and the
other one close to the striped phase (on the right). We see that the
decay is somehow irregular: it is faster than polynomial, but it does
not follow an exponential either. Of course the exponential decay is
standard in the commutative world, so here we see that this property
is distorted by the non-commutativity.

{\renewcommand\belowcaptionskip{-.5em}
\begin{figure}[t]
\centering
\subfigure[\footnotesize{$N^2 \lambda =70,~N^2 m^2 = -17.5$}]
{\epsfig{figure=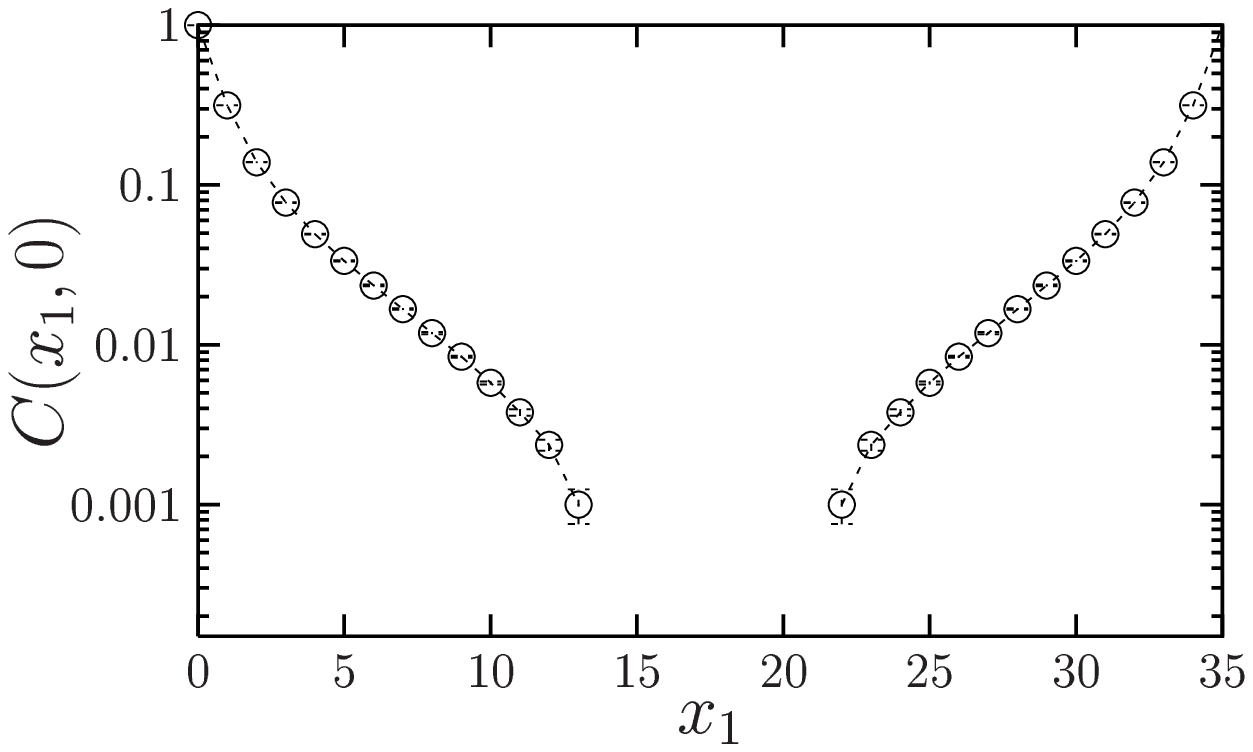,width=.35\linewidth}}%
\hspace*{1cm}
\subfigure[\footnotesize{$N^2\lambda = 350, ~N^2 m^2 =-140$}]
{\epsfig{figure=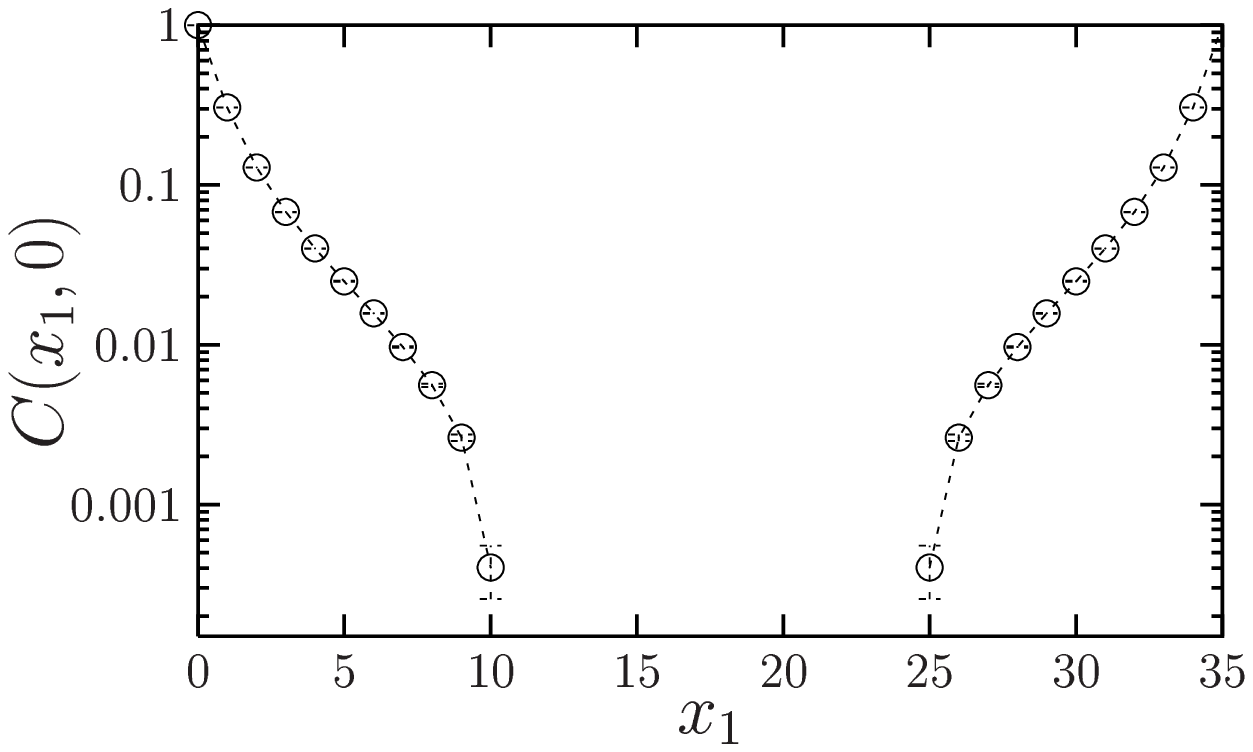,width=.35\linewidth}}
  \caption{The spatial correlator in the disordered phase, close to
the transition to the uniform phase (on the left) resp.\ to the
striped phase (on the right). The logarithmic plots show that the
decay is fast, but it does not follow an exponential.\label{spacorlog}}
\end{figure}
}

Now we turn our attention to another type of correlation function,
namely to the quantity
\begin{equation}  \label{Gtau}
G ( \tau ) := \frac{1}{T} \sum_{t} \langle \tilde \phi (\vec p = \vec 0 ,t)^* 
\tilde \phi (\vec p = \vec 0 , t + \tau ) \rangle \,.
\end{equation}
This function measures the correlation between two averaged spatial
layers with a temporal separation of $\tau$.  We denote $G(\tau )$ as
the \emph{temporal correlator}, but we should stress that the
difference from the spatial correlator is not only that we deal with a
separation in time.

\begin{figure}[t]
\centering
\subfigure[\footnotesize{$N^2 \lambda =70,~N^2 m^2 = -35$}]
{\epsfig{figure=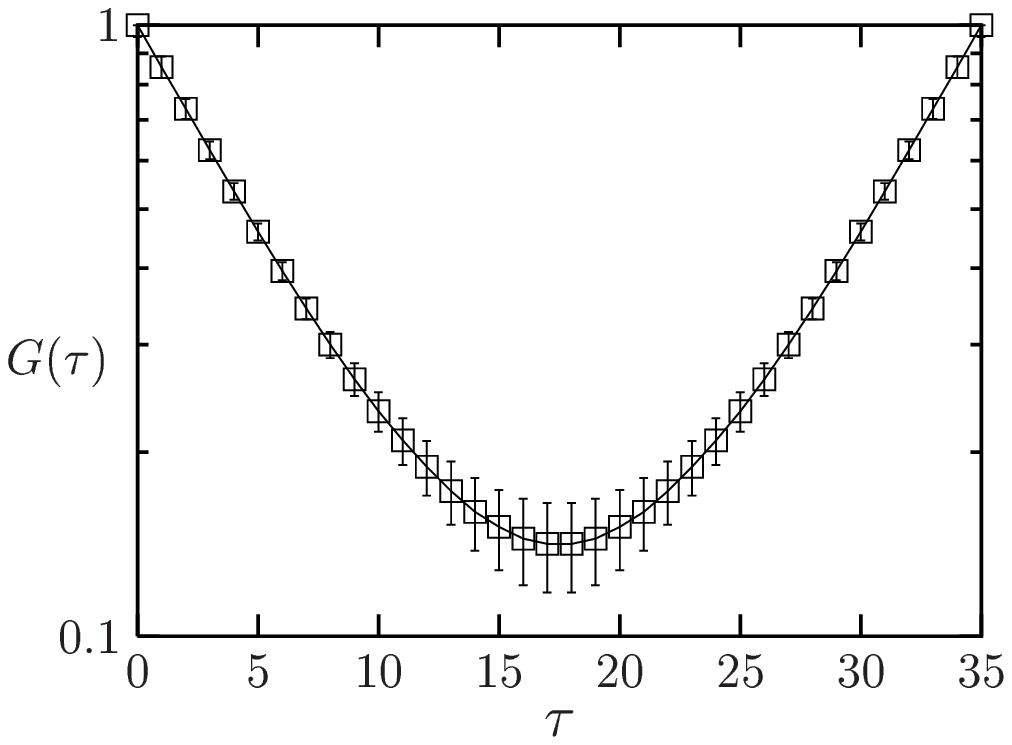,width=.35\linewidth}}%
\hspace*{1cm}
\subfigure[\footnotesize{$N^2\lambda = 525, ~N^2 m^2 =-140$}]
{\epsfig{figure=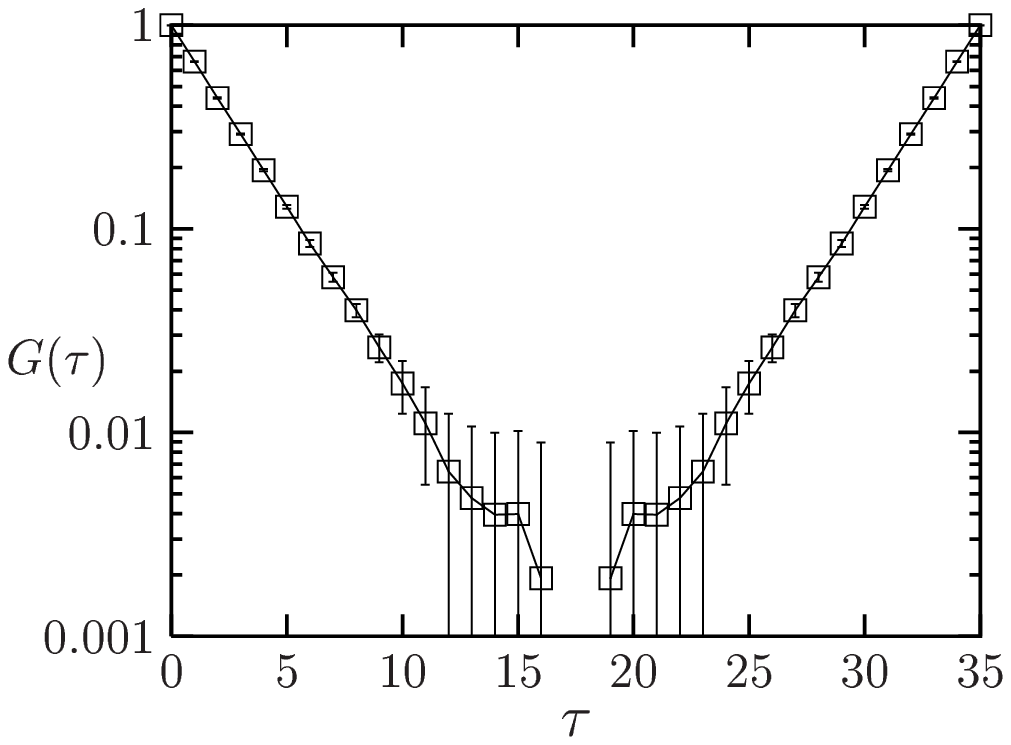,width=.35\linewidth}}
\\
\subfigure[\footnotesize{$N^2 \lambda = 70,~N^2 m^2 =-140$}]
{\epsfig{figure=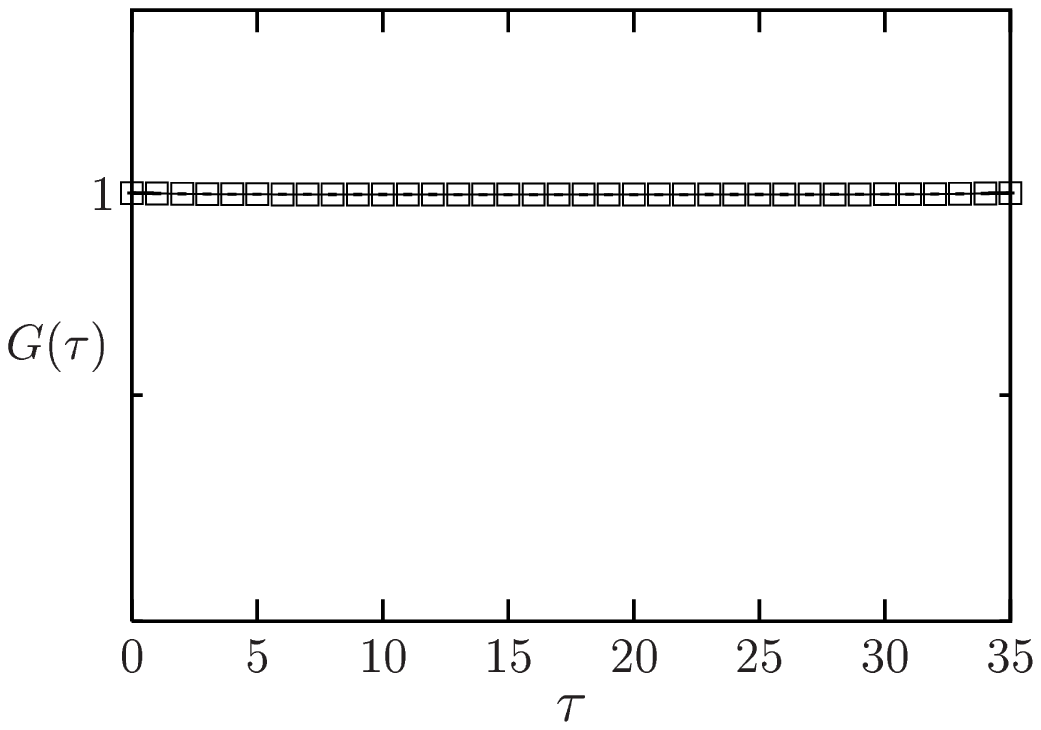,width=.35\linewidth}}%
\hspace*{1cm}
\subfigure[\footnotesize{$N^2 \lambda = 525,~N^2 m^2 = -560$}]
{\epsfig{figure=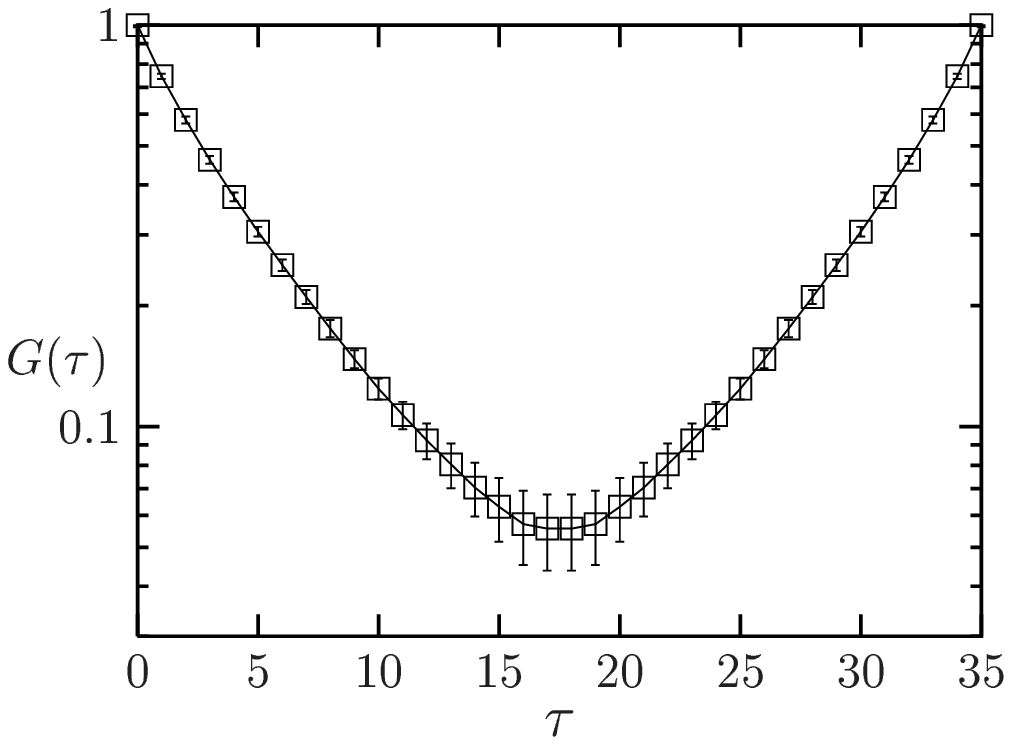,width=.35\linewidth}}
  \caption{The temporal correlator $G(\tau )$, defined
in eq.\ (\protect\ref{Gtau}), in the different sectors of the phase
diagram in figure \protect\ref{phasedia}, at $N=T=35$.
We see a fast decay, which now follows
well an exponential (resp.\  a $\cosh$ shape) in the disordered
phase (two plots above) and in the striped phase (below, right), 
whereas $G(\tau )$ is nearly constant
in the uniform phase (below, left).
}
\label{temcor}
\end{figure}

In figure~\ref{temcor} we show typical examples also for this
correlator in the four sectors of the phase diagram, in analogy to
figure~\ref{spacorfig1}.  Here we show directly logarithmic plots,
which illustrate that this decay does follow an exponential --- resp.\
a $\cosh$ function, i.e.\ an exponential with periodic boundary
conditions --- in the disordered phase and in the striped phase.  In
the uniform phase we obtain $G(\tau ) \simeq 1$ for all values of
$\tau$.

From the exponential decay in the disordered phase close to ordering
we can now extract the energy at momentum $\vec p = \vec 0$, i.e.\ the
\emph{rest energy} $E_{0}$.  We evaluate this quantity at some time
separation $\tau$ as
\begin{equation}  \label{E0}
E_{0} = - \ln \frac{G(\tau + 1)}{G(\tau )} \,.
\end{equation}
Such a term is sensible if this evaluation yields a plateau over some
range of $\tau$, which is not very close to the boundary nor to the
center $T/2$, so that the exponential decay dominates.
Figure~\ref{temcorE0} shows an example of the corresponding decay and
for the neat plateau that one obtains for $E_{0}$ according to
eq.~(\ref{E0}).

\FIGURE[t]{\epsfig{file=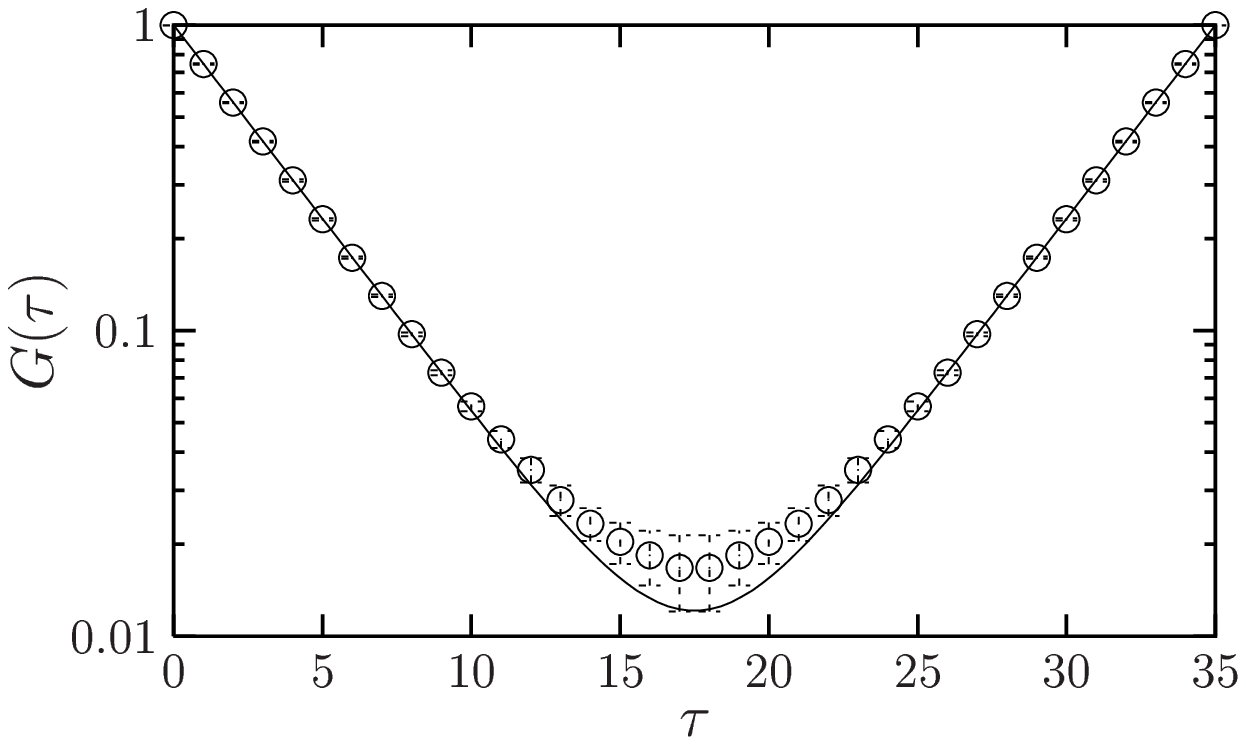,width=.45\linewidth,clip=} 
  \epsfig{file=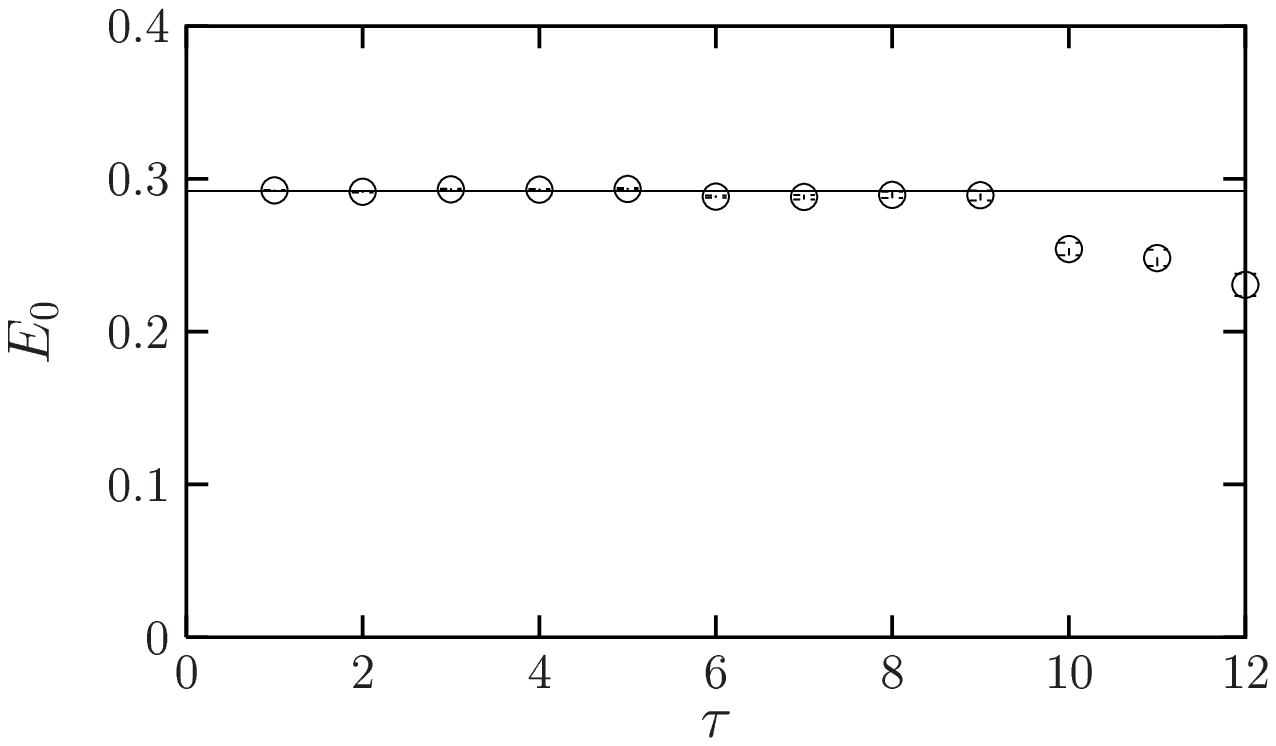,width=.45\linewidth,clip=}
  \caption{On the left: the temporal correlator $G(\tau )$ in the
    disordered phase, close to the transition to the striped phase (at
    $N=T=35$, $N^{2} \lambda = 350$, $N^{2}m^{2} = -140$).  For
    comparison, the line represents a {\tt cosh} function.  From the
    decay we extract the rest energy $E_0$ at different values of
    $\tau$, according to eq.\ (\protect\ref{E0}).  The plot on the
    right shows that this provides a stable plateau, up to the region
    around $T/2$ where the finite $T$ effects are
    significant.\label{temcorE0}}}

The same procedure to extract the effective energy from an exponential
decay can be repeated also for finite momenta $\vec p$.  This leads to
the full dispersion relation, which is the subject of the next
section~\ref{section6}. For more general aspects of dispersion
relations in NC field theory we refer to ref.~\cite{dispersion}.

\section{The dispersion relation}\label{section6}

We now generalize the correlation function $G(\tau )$ of
eq.~(\ref{Gtau}) and study the decay of the expectation value
\begin{equation}
\frac{1}{T} \sum_{t} \langle \tilde \phi (\vec p , t)^* \tilde \phi
(\vec p , t + \tau ) \rangle
\end{equation}
at various momenta $\vec p$. In the disordered phase close to the
ordering transition we always find an exponential decay in $\tau$,
which allows us to extract the energy $E(\vec p )$.  In
figure~\ref{disp35} we show the resulting dispersion relation at
$N=T=35$ for a small, a moderate and a large value of $\lambda$.  In
the first case --- near the uniform phase --- $E^{2}(\vec p^{\, 2})$
follows closely the usual linear form, which confirms again that at
small $\lambda$ this model looks like its commutative counterpart.

\FIGURE[t]{\centering  \hspace*{10mm}
  \includegraphics[width=.45\linewidth]{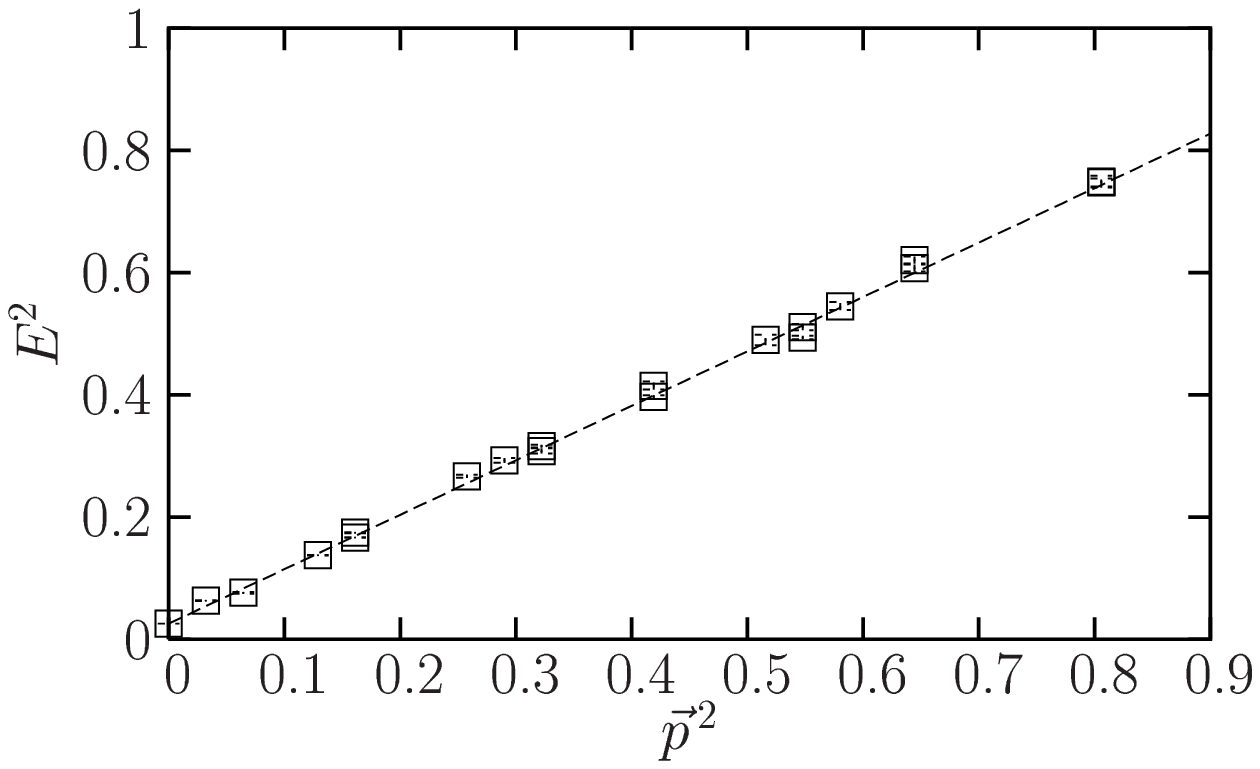} \vspace*{3mm} \newline
  \hspace*{0mm}
  \includegraphics[width=.45\linewidth]{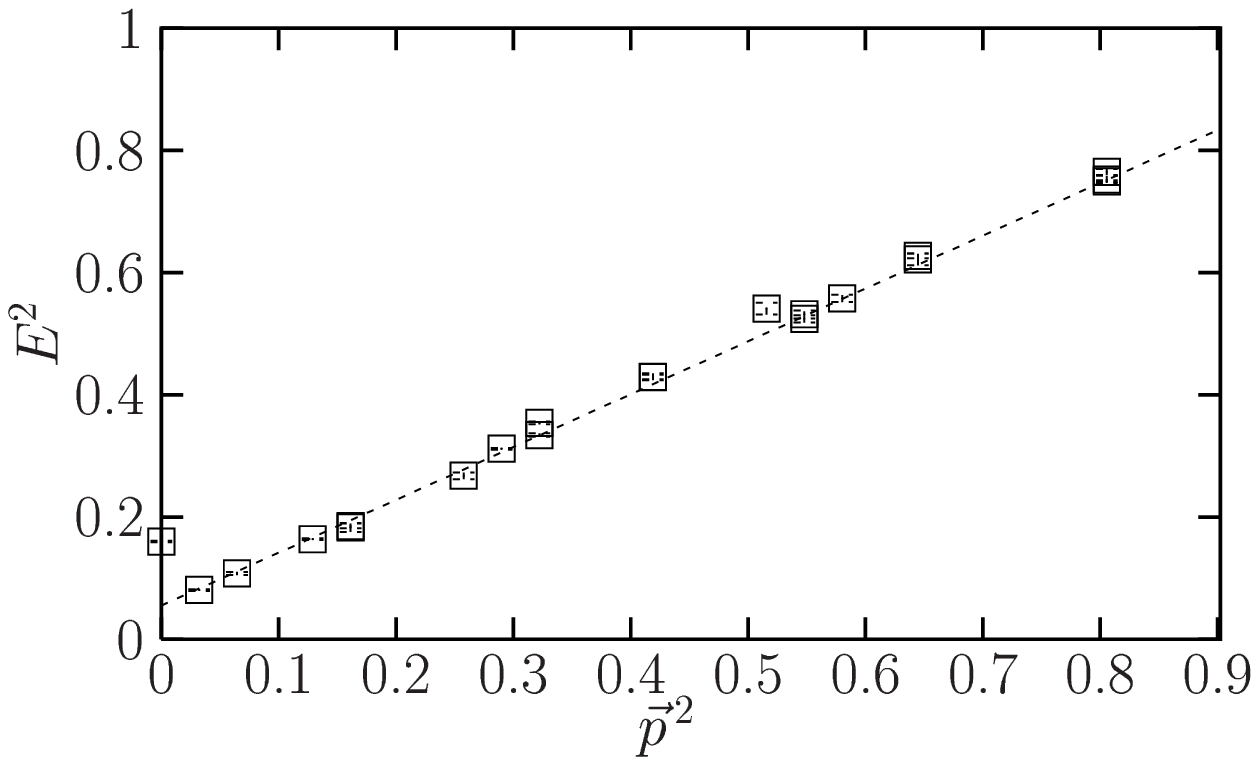} \hspace*{4mm}
  \includegraphics[width=.45\linewidth]{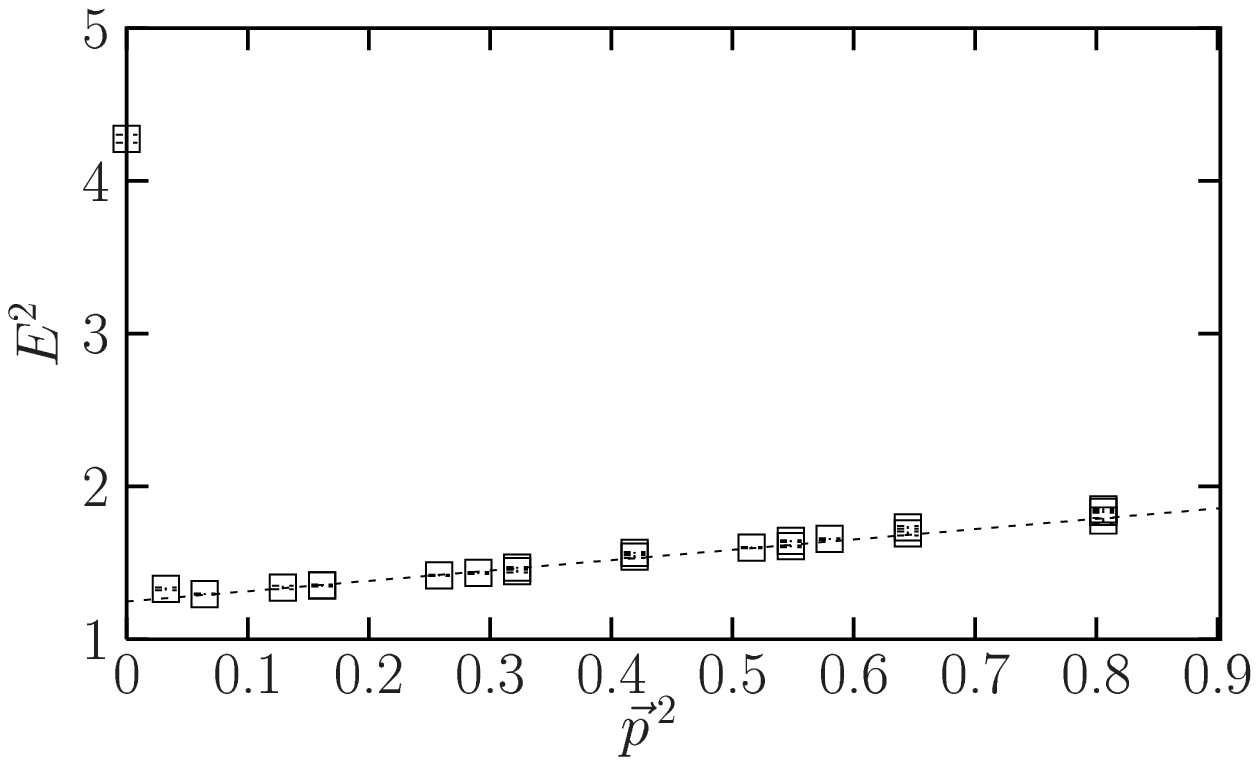} 
  \caption{The dispersion relations measured at $N=35$ in the disordered
phase, close to the ordering transition. We proceed from small $\lambda$
(on top: $\lambda = 0.06$, $m^2 = -0.01$) to moderate $\lambda$ 
(on the left: $\lambda = 0.6$, $m^2 = - 0.23$) and finally to large
$\lambda$ (on the right: $\lambda = 100$, $m^2 = - 7$). 
In the first case, which is close to
the uniform phase, we see the standard linear dispersion relation.
As $\lambda$ increases we are close to the striped phase, and the energy
minimum moves to a finite momentum $\vert \vec p \vert$.
The location of the minimum and the rest energy grow for
raising values of $\lambda$.}
\label{disp35}}

As $\lambda$ increases we see an amplified rest energy (at $\vec p =
\vec 0$), followed by a sharp dip at low momenta and again the linear
curve asymptotically at larger momenta.  Of course, the observation
$E_{0} > 0$ agrees with the UV/IR mixing. Our result confirms that
this mixing also occurs non-perturbatively, as we announced in
section~\ref{section1}.  The question of an IR divergence, i.e.\ a
divergence of $E_0$ in the double scaling limit of zero lattice
spacing and infinite volume, will be addressed in the next section.

The fact that the energy minimum drives to finite momenta is the basis
for the striped phase that we illustrated before in
section~\ref{section4}. If such momenta condense they manifest
themselves in the stabilization of some stripe patterns. For instance,
in the second plot of figure~\ref{disp35} (at $\lambda =0.6$) the
leading non-zero momentum which occurs in this volume is clearly the
location of the energy minimum.  This momentum corresponds to $k =1$
in the notation of eq.~(\ref{Mkdef}), hence in this case a
condensation of the mode with minimal energy leads to two stripes
parallel to one of the axes.

In the last example in figure~\ref{disp35} there are several non-zero
momenta which are close to the minimum. In such situations various
types of stripe patterns may condense. The transition between them
does not come about easily, so they all appear stable even if they do
not correspond to the exact minimum.  In such cases we could see
multi-stripe patterns as in figure~\ref{multi}, but it was difficult
to figure out which pattern is ultimately most stable.  The dispersion
relation now clarifies the situation as a coexistence of qualitatively
different, practically stable patterns in the vicinity of the energy
minimum. If we start from such a point in the disordered phase and
lower $m^2$, the configuration takes one of the stripe patterns which
correspond to the $k$ values near the minimum. Once this pattern is
built up, it is hardly possible to change it again as long as we are
in the ordered regime (obviously this would require a large
deformation in the structure of the configuration).  For the same
reason the transition between the uniform and the striped phase could
not be determined accurately based on a direct illustration, as we
described in section~\ref{section4}.  However, it is the very nature
of the system that such transitions happen gradually over a broad
interval in~$\lambda$.

As a further example, we show a dispersion relation for $N=T=55$ in
figure~\ref{disp55}. Here we have a finer resolution of the momenta.
Now the momentum that corresponds to $k=1$ is clearly \emph{not} the
minimum any more, but the minimal vicinity involves the cases $k =
\sqrt{2}$, $2$ and $\sqrt{5}$. From the phase diagram in
figure~\ref{phasedia} it is clear that for the given parameters $m^2 =
-15$, $\lambda = 50$ we are near the striped phase, so here we have at
last an evidence for the existence of stable multi-stripe patterns.

\FIGURE[t]{\centerline{\epsfig{file=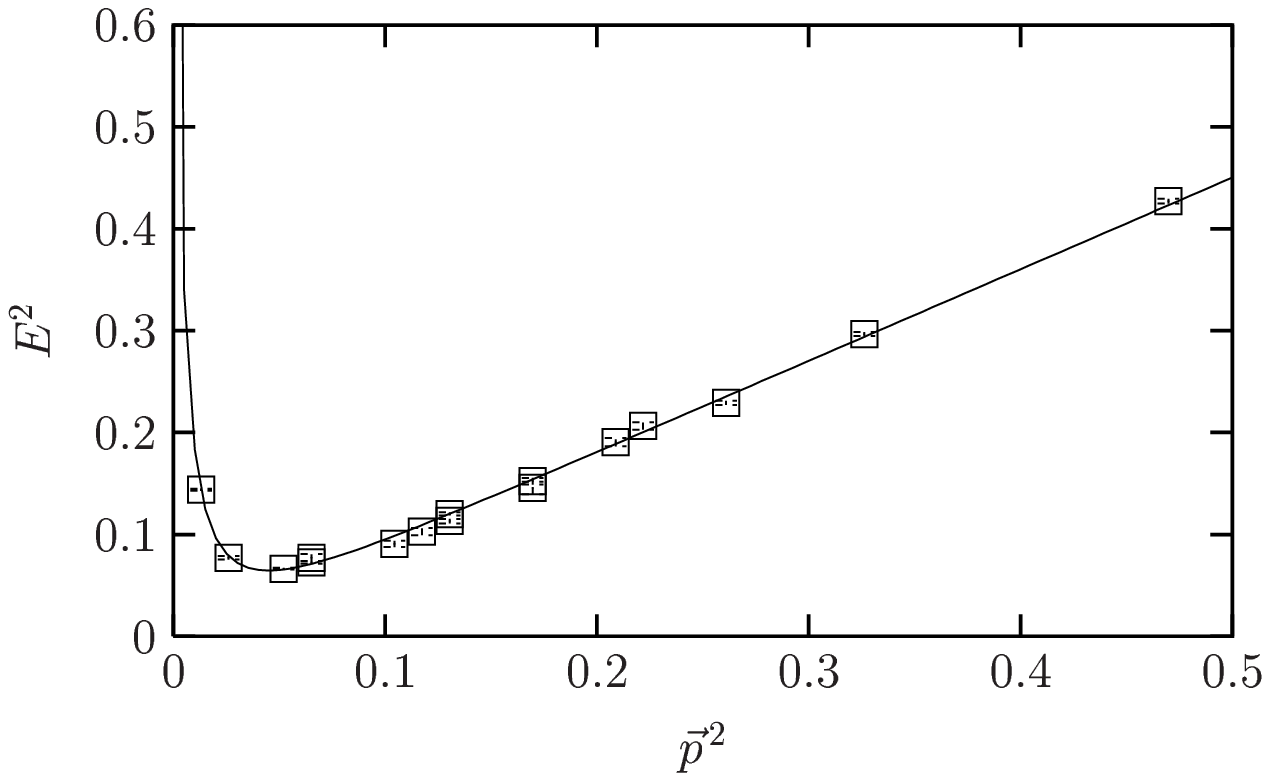,width=.55\linewidth,clip=}}%
  \caption{The dispersion relation determined at $N=55$, $m^{2} =
    -15$, $\lambda =50$. The symbols correspond to the discrete values
    of $k = N \vert \vec p \vert / (2\pi )$ that occur, i.e.\ $k= 1$,
    $\sqrt{2}$, $2$, $\sqrt{5}$, $\sqrt{8}$, $\dots$
    \label{disp55}}}

The dispersion relation observed in figure~\ref{disp55} can be
parameterized to a good accuracy~as
\begin{eqnarray}
E^{2}(\vec p^{\, 2}) &=& c_{0} \, \vec p ^{\, 2} + m^{2} +
\frac{c_{1}}{\sqrt{\vec p^{\, 2} +\bar m^{2}}} \exp \left(
- c_{2} \sqrt{\vec p^{\, 2} + \bar m^{2}} \right) 
\nonumber \\
\hbox{with} \quad c_{0} &=& 0.919(4)\,,  \quad c_{1} = 0.07(1) \,,  \quad
c_{2} = 13(1) \,,  \quad \bar m^{2} = 0.0004(1) \,,  \label{fitfun}
\end{eqnarray}
which is illustrated by the curve in figure~\ref{disp55}.  (We assume
the slight deviation of $c_0$ from its expected value $1$ to be a
finite $N$ artifact.)  The square root in this formula is
characteristic for the three dimensional case, and an obvious ansatz
for a linear IR divergence in $E^{2}$ (regularized by the term $\bar
m^2$ and suppressed at larger $\vec p^{\, 2}$).  We remark that in
$d=4$ one would expect the corresponding formula with the square root
replaced both times by $(\vec p^{\, 2} + \bar m^{2})$. In $d=3$ and
$4$ this form can be related to $\Gamma^{(1)}_{\rm np}$ given in
eq.~(\ref{pnp})~\cite{Diss}. In our 3d results the fits work in fact
much better with eq.~(\ref{fitfun}) than with the expected 4d formula.

\section{The continuum limit}\label{section7}

In order to study the extrapolation of our results to a continuous,
non-commutative space, we first have to identify a dimensionful
lattice spacing. We perform this identification in the \emph{planar
  limit}, where one sends $N \to \infty$ at fixed parameters $m^{2}$
and $\lambda$.

We saw in section~\ref{section6} that the dispersion relation $E^2
(\vec p^{\,2} )$ becomes linear for relatively large momenta. In
figure~\ref{plandec} we study this behavior in the planar limit: we
see that the linear regime is not altered any more as we approach this
limit, and that the dip for the minimum is squeezed towards zero
momentum.  This implies that the behavior at relatively large momenta
is dominated by the planar contribution.  Hence we can extrapolate a
dispersion relation to the planar limit by just extending the linear
slope down to $\vec p^{\,2} = 0$.

This is analogous to our study of the 2d NC $\UU(1)$ gauge theory,
where we saw an area law for Wilson loops at a \emph{small}
area~\cite{2dU1}. Thus it coincided with the analytic result by Gross
and Witten for (commutative) $\UU(N)$ Yang-Mills theory in the planar
limit~\cite{GW}.

\FIGURE[t]{\centerline{\epsfig{file=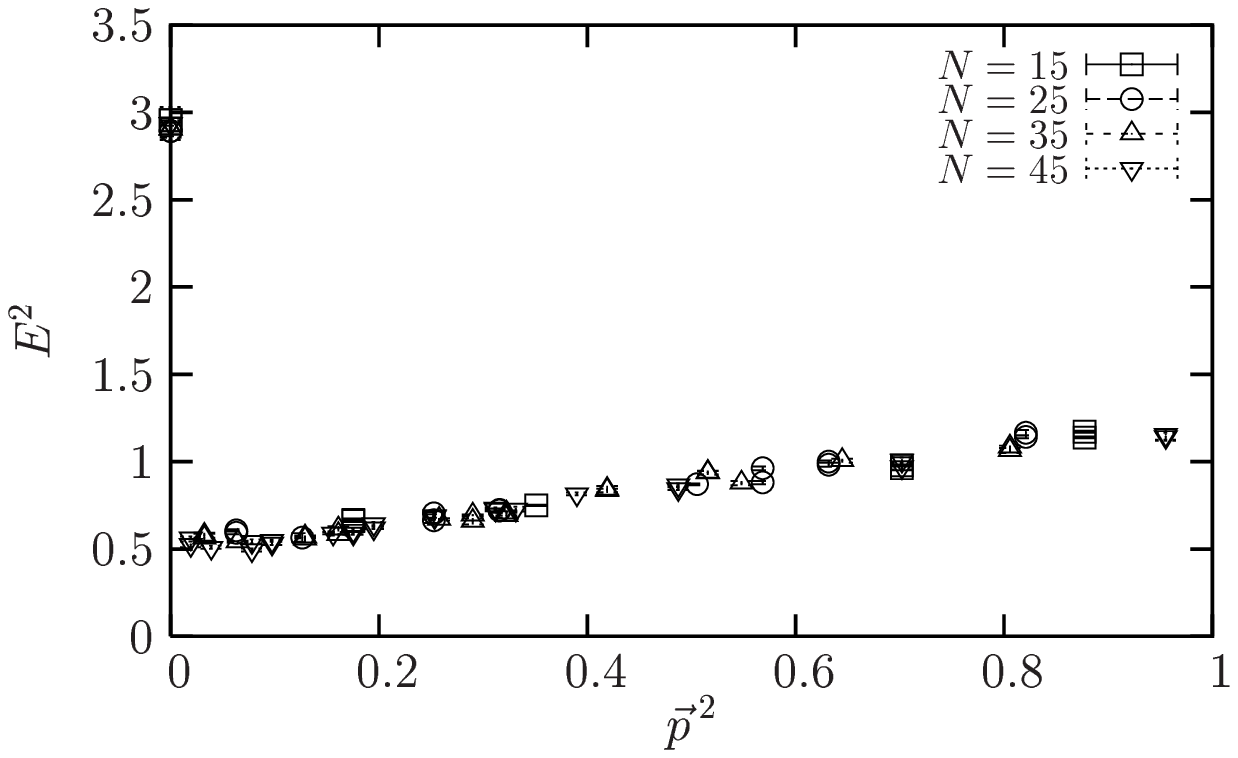,width=.46\linewidth,clip=}}%
  \caption{The dispersion relation (at $\lambda =50$, $m^{2} = -13.2$)
    as we approach the planar limit.  The linear regime is stable, and
    the minimum moves close to $0$.\label{plandec}}}

Making use of this property, we determine an \emph{effective mass}
$M_{\rm eff}$ in the planar limit. We only consider the ``linear
dispersion regime'' where the momentum $\vert \vec p \vert$ is large
enough to follow the linear dispersion behavior
\begin{equation}
E^{2} \simeq M_{\rm eff}^{2} + \vec p^{\, 2} \,.
\end{equation}
If the momenta are not too small, this relation is fulfilled to a high
precision --- as we saw from figures~\ref{disp35} and~\ref{disp55} ---
hence we can evaluate an accurate value for $M_{\rm eff}^{2}$ in
lattice units. In particular, we fixed again $\lambda =50$ and
considered various values of $m^{2}$; in each case we searched for the
planar limit, i.e.\ a stabilization of the resulting effective mass as
$N$ increases.

\FIGURE[t]{\centerline{\epsfig{file=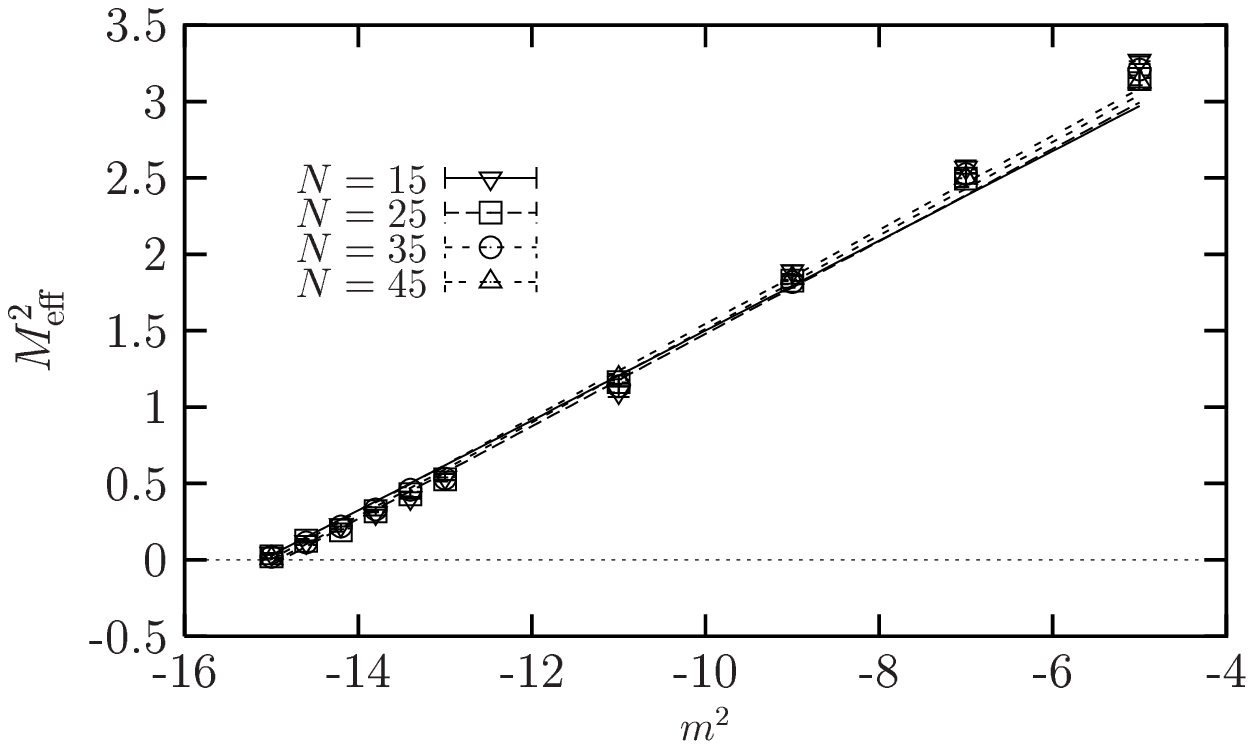,width=.55\linewidth,clip=}}%
  \caption{The effective mass squared as a function of the bare mass squared
in the disordered phase. We observe a linear behavior, which stabilizes 
accurately in the planar large $N$ limit.\label{figMeff}}}

This is indeed the case in the range $N=15 \dots 45$.
Figure~\ref{figMeff} shows that the results for $M^{2}_{\rm eff}$
depend linearly on $m^{2}$, as long as we are in the disordered phase.
We parameterize this linear curve as
\begin{equation}  \label{c0c1}
M^{2}_{\rm eff} = \mu^2 + \gamma \, m^{2} \,,
\end{equation}
and we show in figure~\ref{figc0c1} that the values we obtain
for $\mu^2$ and $\gamma$ are stable in $N$.

\FIGURE[t]{\hfil\epsfig{file=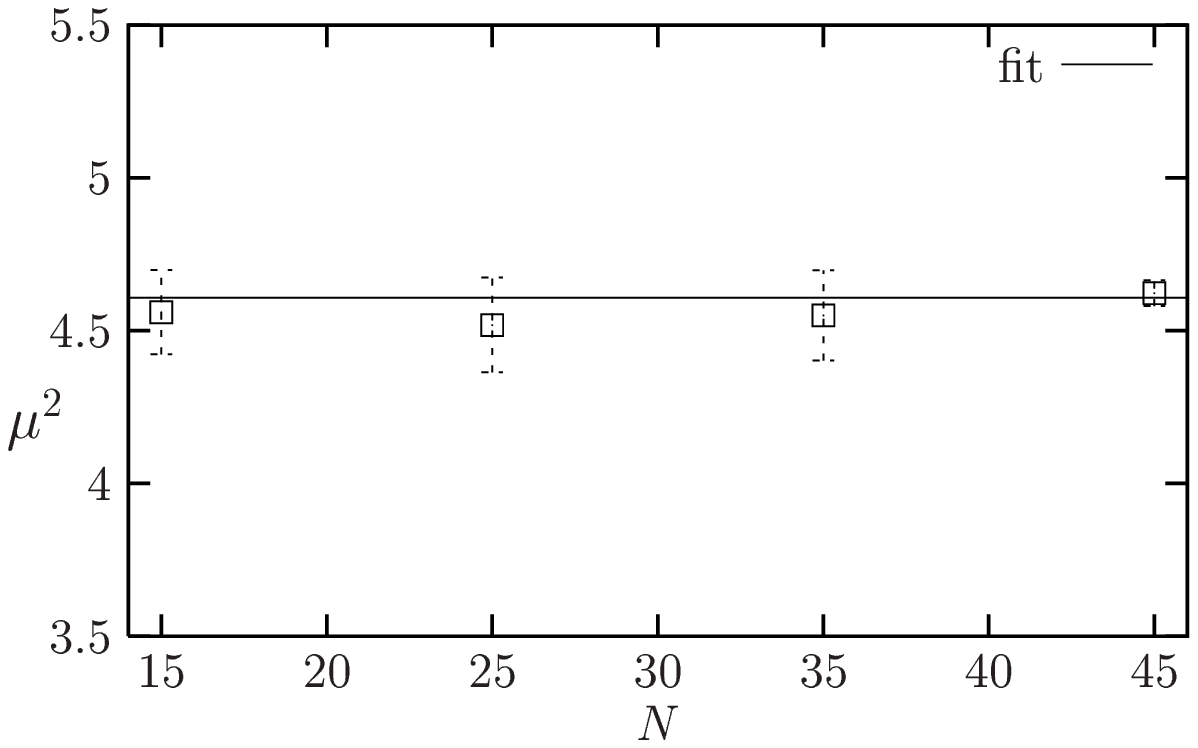,width=.45\linewidth}\hfil
  \epsfig{file=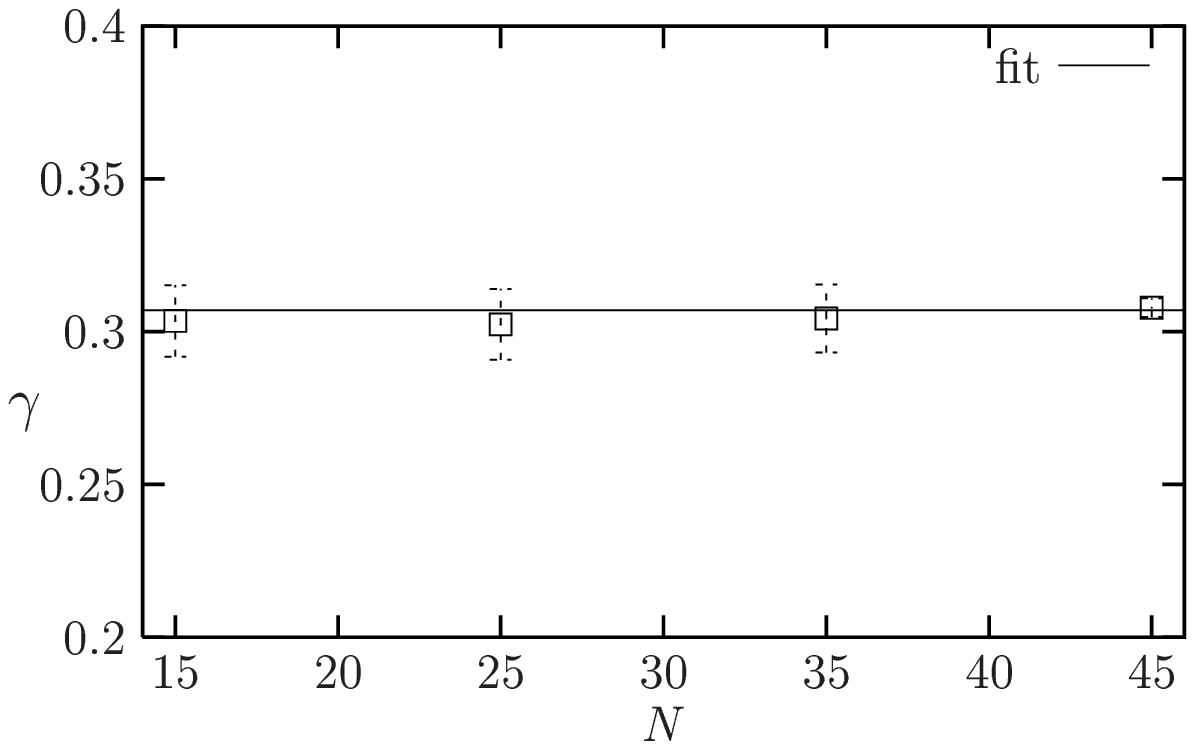,width=.45\linewidth}\hfil
  \caption{The terms $\mu^2$ and $\gamma$ in the
    parameterization~(\ref{c0c1}), at $\lambda = 50$. They are found
    to be constant in $N$ to a good approximation.\label{figc0c1}}}

Our condition for the continuum limit is now that we keep the
\emph{effective mass in physical units}, $M_{\rm eff}/ a$, constant in
the planar limit, where $a$ is the lattice spacing.  Without loss of
generality we simply set $M_{\rm eff}/ a = 1$, which means that we
measure every quantity which has the dimension of a length in units of
$\, a / M_{\rm eff}$.  This corresponds to identifying the lattice
spacing as
\begin{equation}
a^2 = \mu^2 + \gamma \, m^2 \,.
\end{equation}
The ratio $m_c^2 = - \mu^2 / \gamma$ represents the squared critical
bare mass, and according to our result illustrated in
figure~\ref{figc0c1} we obtain
\begin{equation}
m_{c}^{2} = -15.01(8) \,, 
\end{equation}
which is of course also in agreement with figure~\ref{figMeff}.

Therefore, in the \emph{double scaling limit} --- which keeps the
non-commutativity parameter $\vartheta$ in eq.~(\ref{double}) constant
--- we have to fix $N(\mu^2 + \gamma \, m^2 )$, while taking the
limits $N \to \infty$ and $m^2 \to m_{c}^{2}$. This limit involves a
free constant, which fixes the value of the non-commutativity
parameter $\vartheta$. We choose it as
\begin{eqnarray}
N a^2 = 100 \gamma \quad \Rightarrow &&
N ( m^2 - m_{c}^2 ) = 100 \,,  \nonumber \\
&& \vartheta = \frac{100 \gamma}{\pi} = 9.77(6) \,. \label{Na2}
\end{eqnarray}
The value for $N ( m^2 - m_{c}^2 )$ is taken rather large, so that we
do not get too close to the critical mass for the $N$ values that we
are using (otherwise we would risk numerical problems due to large
fluctuations).\footnote{The question if the dimensionless
  non-commutativity parameter is rational or not is not an issue here,
  because we hand it over to the computer which makes it rational
  anyhow.}

\FIGURE[t]{\centerline{\epsfig{file=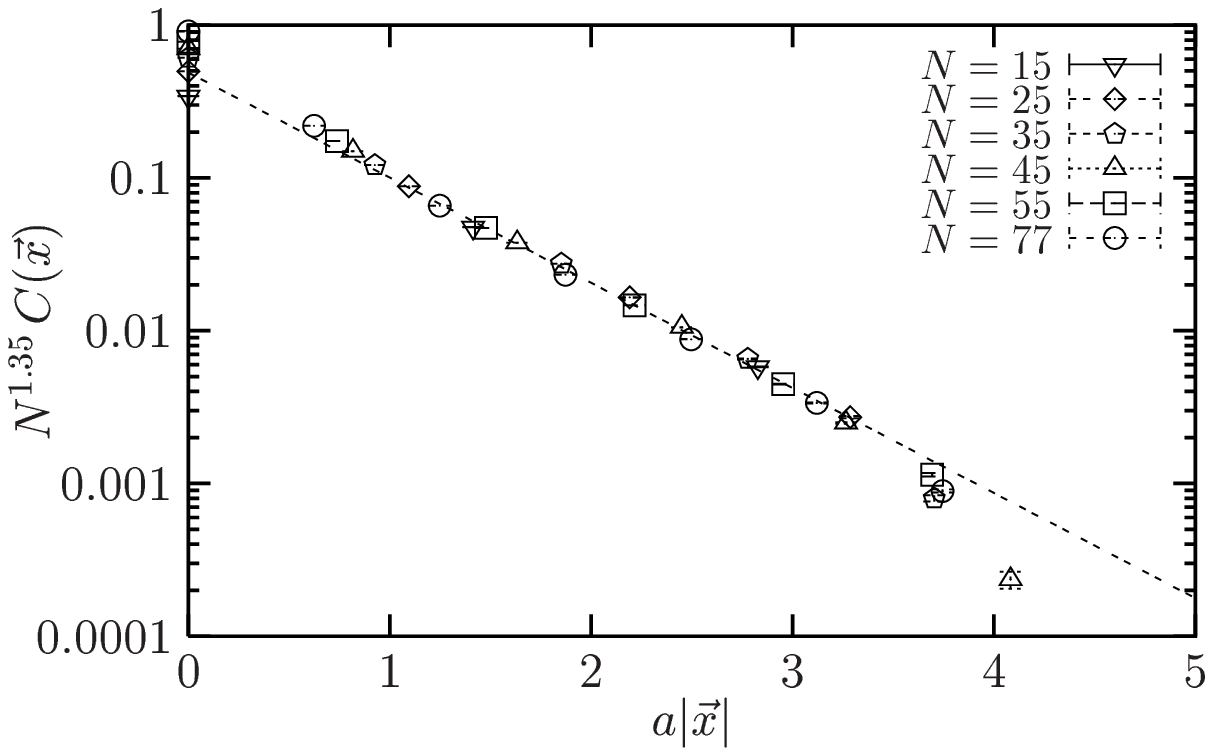,width=.55\linewidth,clip=}}%
  \caption{The spatial correlator $C(\vec x )$ defined in
  eq.~(\ref{spacoreq}), amplified by a wave function renormalization
  factor $N^{1.35}$, against the physical distance $a \vert \vec x
  \vert$. We see a double scaling regime, which corresponds to the
  non-commutative continuum limit.  (The dashed line marks the linear
  regime of this double scaling limit.)
\label{spacorfig2}}}

In this framework, we first measure the spatial correlator, which was
defined before in eq.~(\ref{spacoreq}).  A double scaling behavior can
be observed if the spatial correlator is multiplied by a wave function
renormalization factor $N^{\alpha}$, where the suitable power is found
to be $\alpha = 1.35$. Then a scaling region shows up, as
figure~\ref{spacorfig2} shows.\footnote{The correlator shown here is
  \emph{not} normalized.  This is in contrast to
  section~\ref{section6}, where we set $C(\vec 0 ) =1$.}

\looseness=1Finally we are equipped to approach the question, which represents the
real challenge in this context: we now want to verify if the striped
phase persists in the continuum limit.

To this end, we consider the double scaling limit of the dispersion
relation. The crucial question is whether or not the dispersion
minimum stabilizes at a finite momentum. If such a minimum survives
the double scaling limit, it indicates the existence of a striped
phase in the continuum, where the value of the minimal momentum
characterizes the dominating stripe width in the infinite,
non-commutative space.

\FIGURE[t]{\epsfig{file=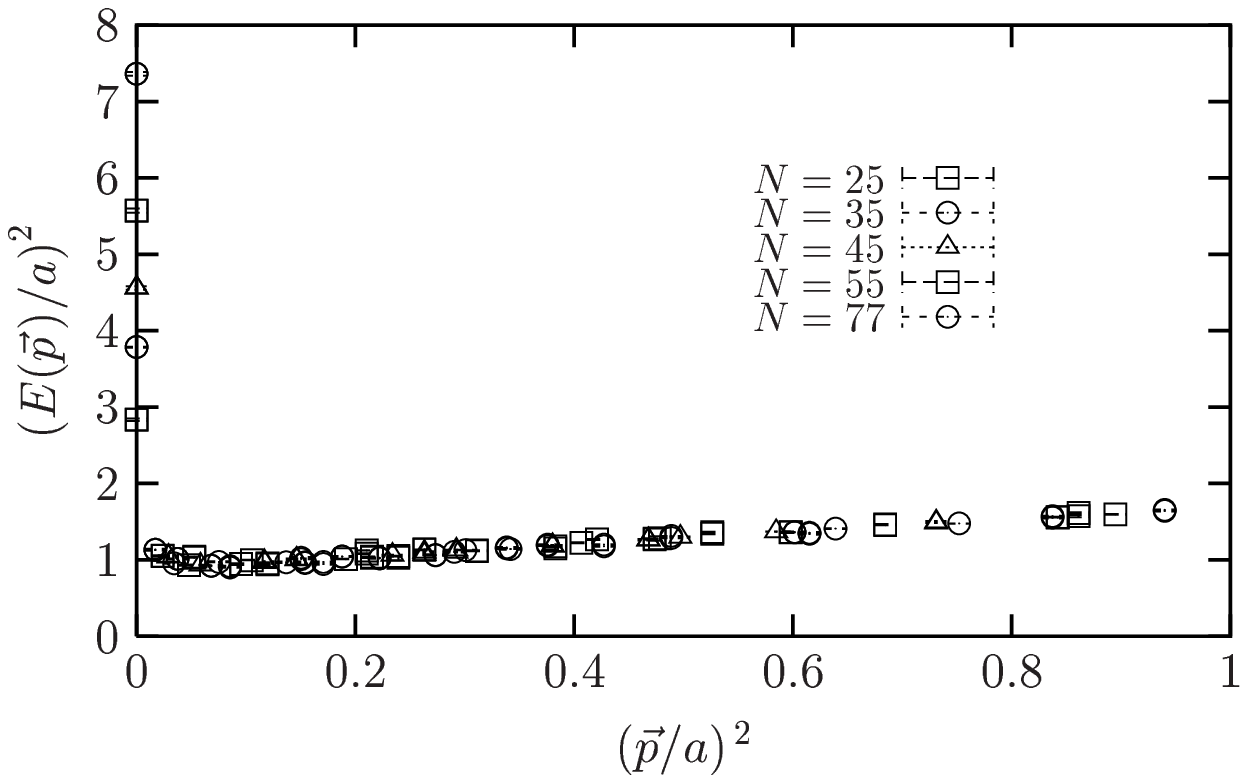,width=.55\linewidth,clip=}
  \epsfig{file=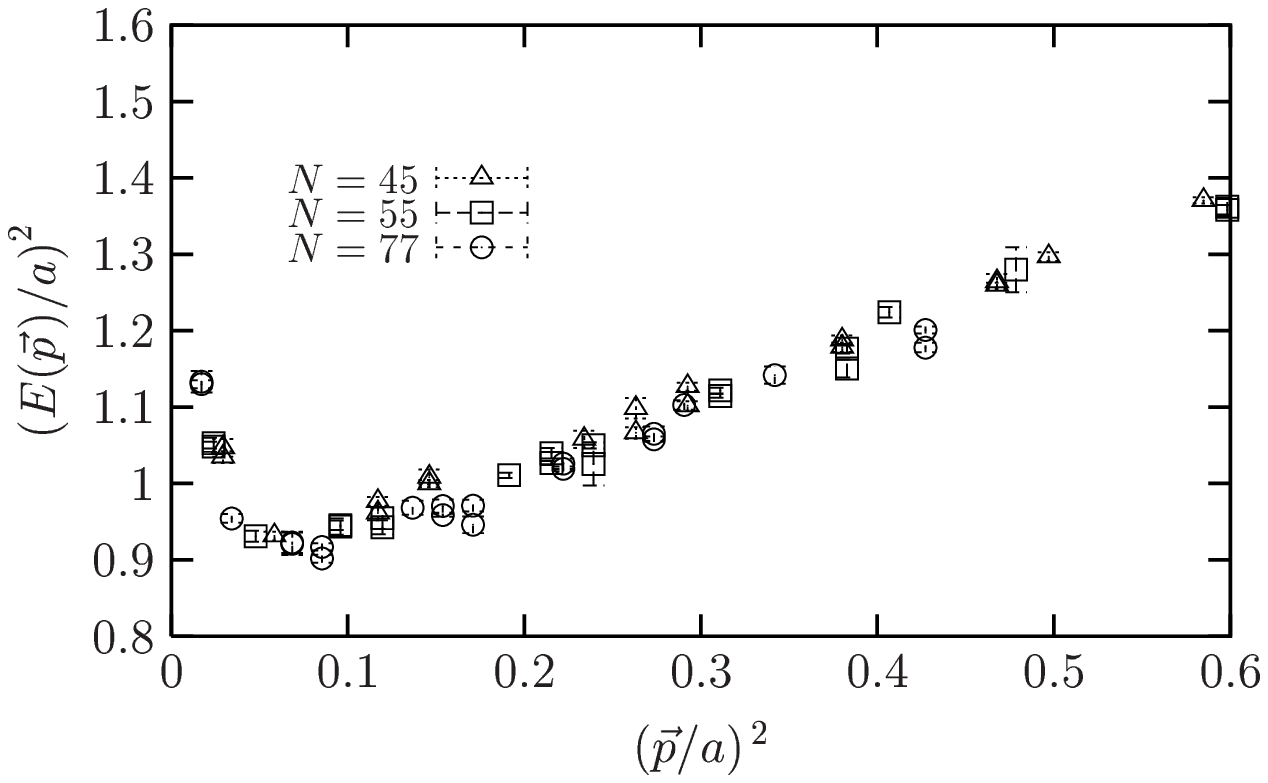,width=.58\linewidth,clip=}%
  \caption{The dispersion relation at $\lambda =50$ and $\vartheta
    \simeq 9.77$ (in physical units), with energy and momentum
    expressed in physical units. We see the linear regime starting at
    moderate momenta, a minimum around $(\vec p /a)^{2} \lsim 0.1$,
    and a rest energy $E_0$, which increases rapidly with $N$, see
    also figure~\ref{E0scal}. The plot on top provides an overview,
    and below we zoom the region around the minimum. This result
    confirms that the dispersion relation stabilizes in the double
    scaling limit to the continuum and infinite volume, which
    corresponds to large $N$ in this figure.\label{dispcont}}}

The plots in figures~\ref{dispcont} demonstrate that at finite
momentum the dispersion relation does indeed stabilize in the double
scaling limit. For $N = 25 \dots 77$, a universal minimum is found
around $(\vec p /a)^{2}_{\rm min} \lsim 0.1$.\footnote{We omit $N=15$
  because in that case the correlator $G(\tau )$ decays too fast to
  extract the energy.  Note that also $N=25$ is problematic in this
  respect.}  Of course, this number refers to the specific values of
$\vartheta$ and $\lambda$ that we have chosen.  We recall that here we
are working in the disordered phase, which also holds for
figure~\ref{dispcont}.  It would be very difficult to extrapolate to
the double scaling limit inside the striped phase, because there the
finite $N$ boundary conditions have a very strong impact.  Hence our
strategy is to work in the disordered phase (where the finite $N$
effects are rather harmless, as figures~\ref{spacorfig1}
and~\ref{spacorlog} show) close to the striped phase, and extract from
there information about the striped phase in the continuum, by
identifying the modes that will condense at lower $m^2$.

\looseness=1In particular this result shows that the model is
\emph{non-perturbatively renormalizable}. In fact the way we
identified the lattice spacing amounts to choosing it in such a way
that the high momentum region of the dispersion relation scales in the
double scaling limit.  What is really non-trivial, therefore, is that
the large $N$ scaling extends to the low momentum region, where it
deviates from the planar linear behavior observed at higher momenta.
This corresponds to the continuum limit of the non-commutative field
theory at \emph{finite}~$\vartheta$.

We remark that in this model also the renormalizability of the
perturbative series may seem less problematic than in other cases,
since the commutative $\lambda \phi^4$ model is super-renormalizable.
However, to the best of our knowledge the perturbative
renormalizability of the 3d NC $\lambda \phi^4$ model has not been
shown in the literature.

At last we take a look at the double scaling of the rest energy
$E_0$. Figure~\ref{E0scal} shows that it diverges linearly in
$\sqrt{N} \propto 1/a$. Therefore we do find an IR divergence in the
continuum limit, in full agreement with the concept of UV/IR mixing
--- we recall that also the UV divergence in this model is linear, as
eq.~(\ref{pnp}) shows.  This IR divergence confirms that the
non-planar terms survive the continuum limit that we take by the
double scaling described here, and they strongly affect the continuum
behavior.

{\renewcommand\belowcaptionskip{-1em}
\FIGURE[t]{\centerline{\epsfig{file=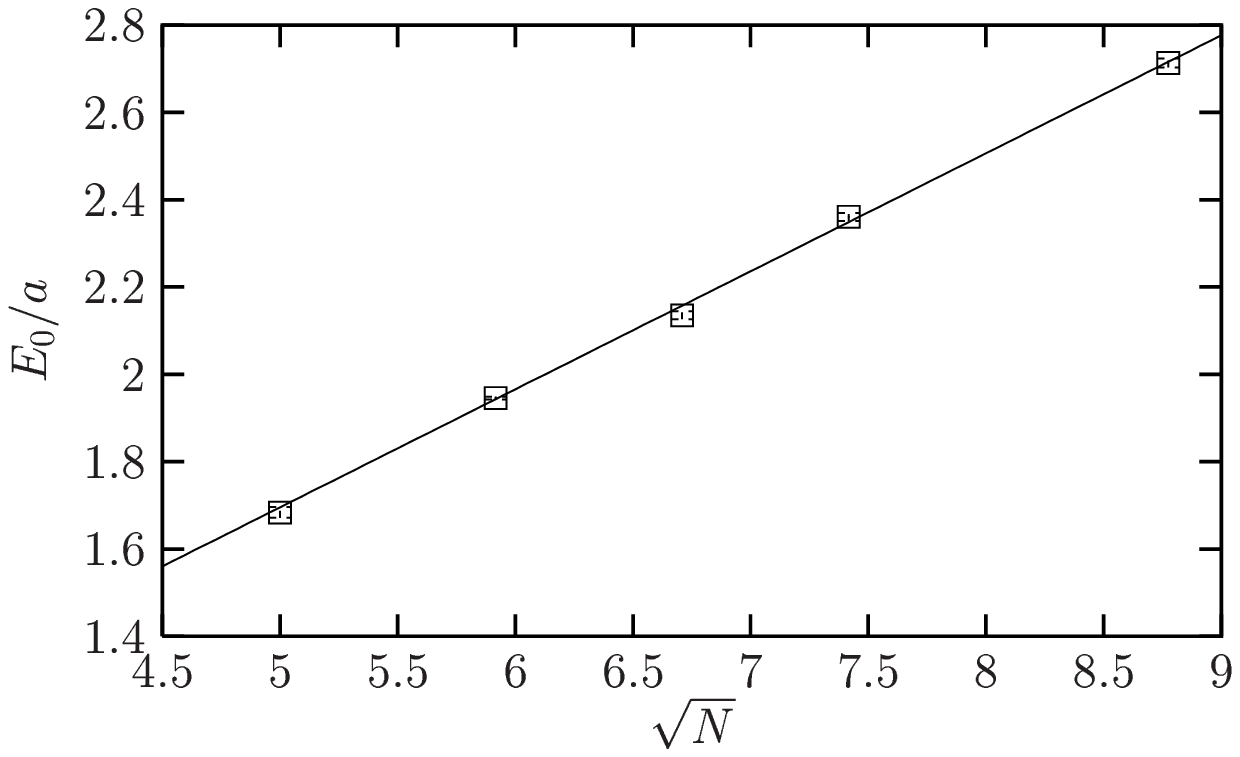,width=.55\linewidth,clip=}}
  \caption{The rest energy in physical units as a function of
    $\sqrt{N} \propto 1/a$. We clearly recognize a linear IR
    divergence in the continuum limit.\label{E0scal}}}
}

\section{A Nambu-Goldstone mode in the striped phase}\label{section8}

In the striped phase the translational symmetry is spontaneously
broken.  According to the Goldstone Theorem, we expect the emergence
of a zero-energy mode in the continuum theory. Generalities about
Nambu-Goldstone (NG) bosons in NC field theory are discussed in
refs.~\cite{NCGB}.

On the lattice the NG mode acquires a small energy since the
translational symmetry is discretized.\footnote{In the striped phase,
  the rotational symmetry is spontaneously broken as well. That
  symmetry is also explicitly broken by the boundary condition, unlike
  the translational symmetry.  The breaking by the boundary condition
  may be irrelevant when the wave length of the stripes is much
  smaller than the spatial extent.  However, the wave length has to be
  much larger than the lattice spacing in order to neglect the effects
  of discretization.  This makes it very difficult to study the
  corresponding NG mode by simulations, and we do not attempt it in
  this work.}  In this section we present numerical results from the
striped phase on this issue.

When dealing with the spontaneous breakdown of a symmetry, one has to
define the vacuum expectation values (VEVs) with some care to make
them meaningful.  In figure~\ref{spacorfig1} (below, right), for
instance, we have rotated each configuration before the measurements
so that the stripe pattern is oriented in one particular direction.
More precisely we have rotated each configuration such that the
momentum $\vec{p}$ which gives the maximum in $M(k)$ --- defined in
eq.~(\ref{Mkdef}) --- points to a particular direction.

The standard VEV $\langle \phi (\vec{x},t) \rangle$ vanishes all over
the phase diagram. Alternatively we now introduce an \emph{oriented
  VEV}, which is based on configurations which are all rotated and
shifted before the measurements.  In this section we restrict our
attention to the two-stripe patterns parallel to an axis.  We define
the oriented VEV by a rotation so that the stripes are vertical to the
$x_2$-axis, and in addition by shifts on the rotated configurations,
\begin{equation}  \label{phiprime}
\phi '(\vec{x},t) = \phi (x_1 , x_2 - \sigma ,t)_{\rm rotated} \,,
\end{equation} 
which maximize the overlap with a characteristic profile.

\FIGURE[t]{\centerline{\epsfig{file=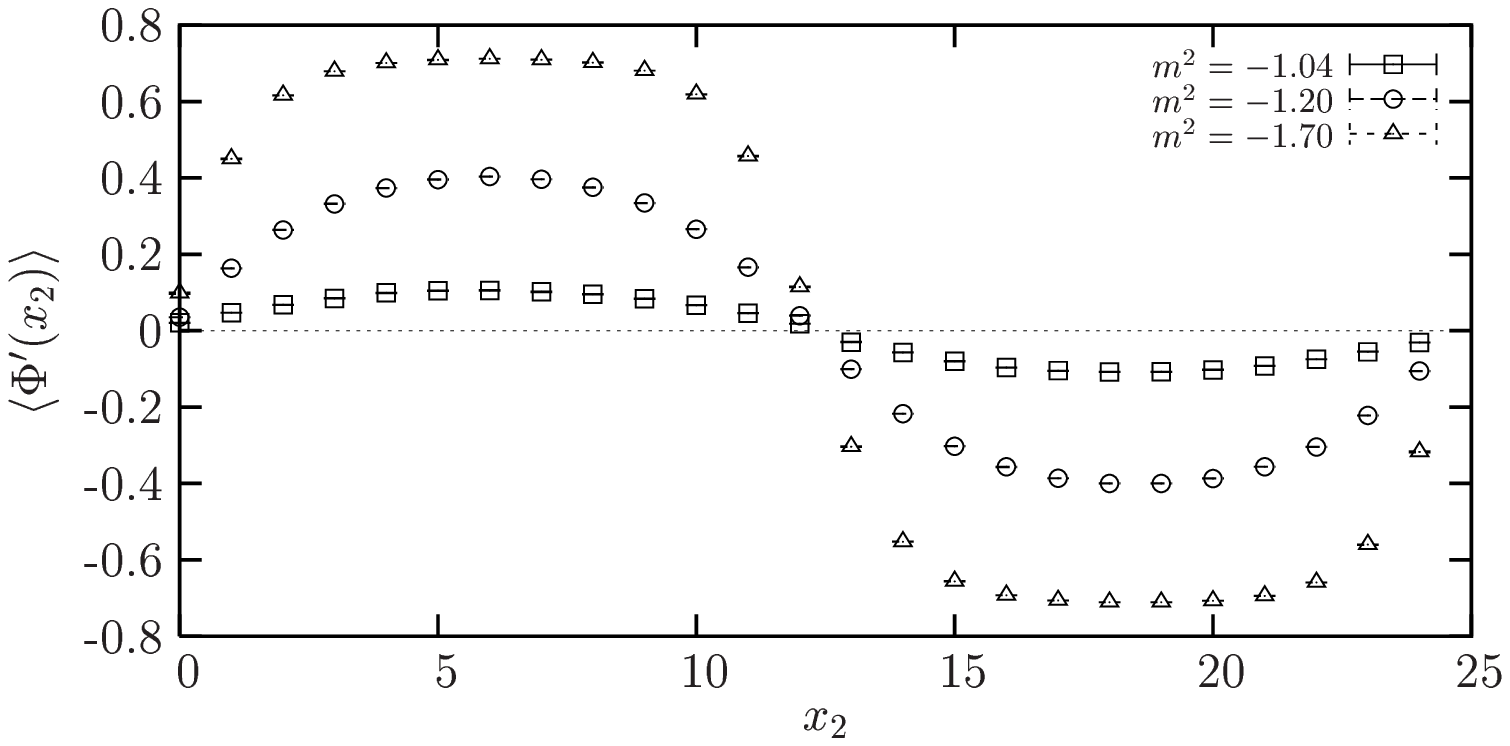,width=.55\linewidth,angle=0,clip=}}
  \caption{The oriented VEV $\langle \Phi '(x_2 ) \rangle =
    \frac{1}{NT} \sum_{x_1 ,t} \langle \phi '(\vec{x},t) \rangle$
    against $x_2$ for $\lambda =2$ and $m^2 = -1.04, -1.2, -1.7$.
    Rotations and shifts are carried out before taking the ensemble
    average, as we described in the text.\label{stripe}}}

For simplicity we use the step profile ${\rm sign} (N/2 - x_2 )$, so
we determine the shift $\sigma$ by maximizing the overlap
$\sum_{\vec{x},t} \phi ' (\vec{x},t) \, {\rm sign} (N/2 - x_2 )$.  In
figure~\ref{stripe} we show the results for the oriented VEV
\begin{equation}
\langle \Phi '(x_2 ) \rangle = \frac{1}{NT} \sum_{x_1 ,t} \langle \phi
'(\vec{x},t) \rangle \,.
\end{equation}
The feature is very similar to figure~\ref{profile} (where we
considered single configurations), and it demonstrates that the
translational symmetry in the $x_2$-direction is broken spontaneously.

Performing a Fourier transformation, we find that $ \frac{1}{NT}
\sum_{t} \langle \tilde{\phi} (\vec{p},t) \rangle$ is non-zero for
$\vec{p}= (0, \frac{2\pi}{N} )$.  Deeper in the striped phase we also
have condensations for $\vec{p}= (0, \frac{2\pi}{N} n_{2})$ with
$n_{2} = 3,5,\dots$, which corresponds to the deformation of the
leading sine pattern as we have seen in figure~\ref{order-k}.  Note
that the pattern is characterized by anti-periodicity in the
$x_2$-direction with the period $N/2$, and a constant behavior in the
$x_1$-direction.  All the modes consistent with this periodicity may
condense. Due to the condensation of non-zero momentum modes, the
momentum is no longer conserved in the striped phase.

In the continuum we can identify the NG mode in the standard manner.
We introduce a field $\phi '(x,t)$ in analogy to eq.~(\ref{phiprime}),
i.e.\ oriented such that the dominant stripes are vertical to the
$x_2$ axis, and shifted properly.  Let us then decompose this field
$\phi '(x,t)$ into the VEV and the fluctuation as
\begin{equation}
\phi '(\vec{x},t) = \langle \phi '(\vec{x},t) \rangle + \delta \phi
'(\vec{x},t ) \,.
\end{equation}
In the case of the assumed stripe pattern, the VEV is not invariant
under the shift in the $x_2$ direction.  The corresponding NG mode is
given by the infinitesimal translation in the $x_2$ direction as
\begin{equation}
\delta \phi '(\vec{x},t) \propto \frac{\partial}{\partial x_2}\langle
\phi '(\vec{x},t) \rangle \,.
\end{equation}
Thus we find that the NG mode has the same Fourier components as the
VEV $\langle \phi '(\vec{x},t) \rangle$ itself.

In order to reveal the existence of the NG mode, we repeat the
analysis which we performed before to study the dispersion relation in
the disordered phase.  Let us define the connected two-point function
\begin{equation}
G(\vec{p},\vec{q};\tau)_{c} := \frac{1}{N^2} \left[ \frac{1}{T} \sum_t
  \langle \tilde \phi ' (\vec{p} , t ) ^{*} \tilde \phi ' (\vec{q} ,
  t+ \tau ) \rangle - \left( \frac{1}{T} \sum_t \langle \tilde \phi
  '(\vec{p} , t ) ^{*} \rangle \right) \left( \frac{1}{T} \sum_t
  \langle \tilde \phi '(\vec{q} , t ) \rangle \right) \right] .
\label{2pt_gen}
\end{equation}
Unlike the behavior in the disordered phase, the two-point function
can take non-zero values also for $\vec{p} \neq \vec{q}$ because of
the non-conservation of the momentum. From the exponential decay with
respect to $\tau$, we can extract the energy of the intermediate state
which couples to the operators considered, as we have done in
section~\ref{section6} in the disordered phase.  When the two-stripe
pattern is formed, the subtraction of the disconnected part is needed
for $\vec{p}= (0, \frac{2\pi}{N} n_{2}^{(p)})$ and $\vec{q}= (0,
\frac{2\pi}{N} n_{2}^{(q)})$, where $n_{2}^{(p)}$ and $n_{2}^{(q)}$
are odd integers.  It is precisely this case where the quasi NG mode
is expected to appear.

In figure~\ref{energyN21N25} we plot the energies extracted from the
two-point function $G(\vec{p},\vec{p};\tau)_{c}$ for some low momentum
modes against the bare mass squared.  In order to see the $N$
dependence, we compare the results for $N=21$ and $N=25$.

{\renewcommand\belowcaptionskip{-1em}
\FIGURE[t]{\epsfig{file=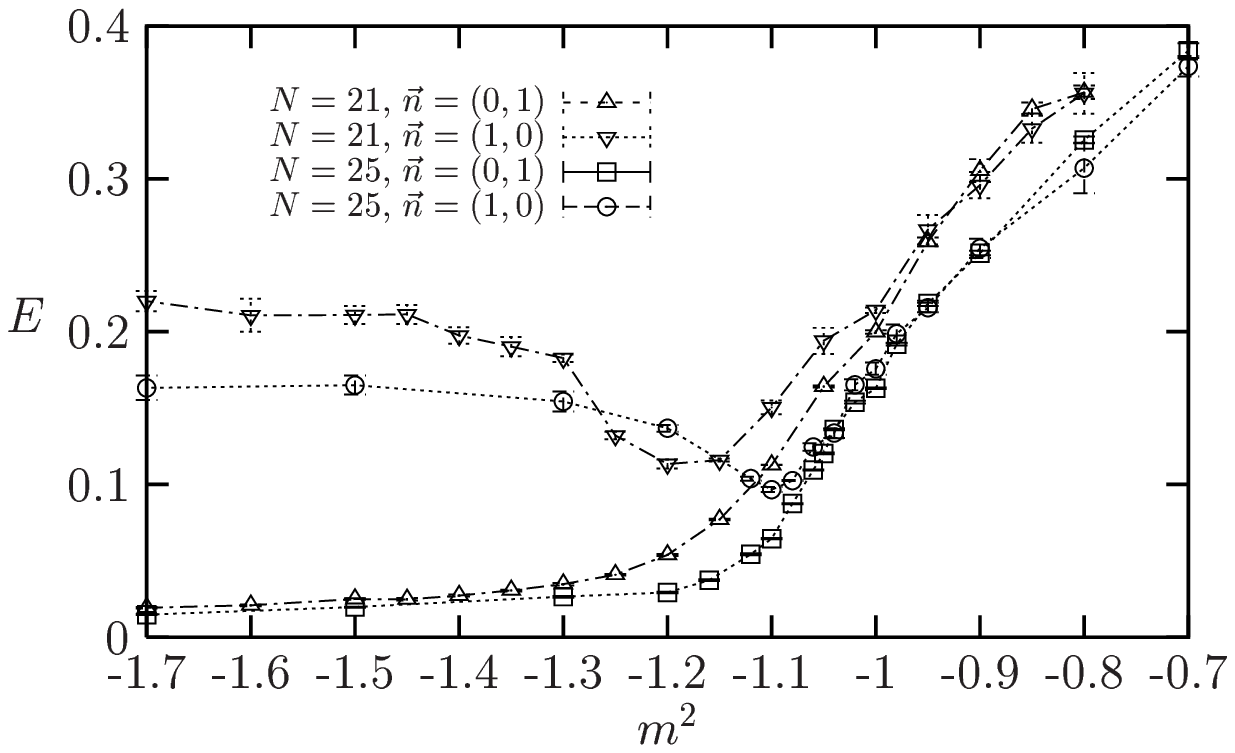,width=.54\linewidth,clip=}
  \epsfig{file=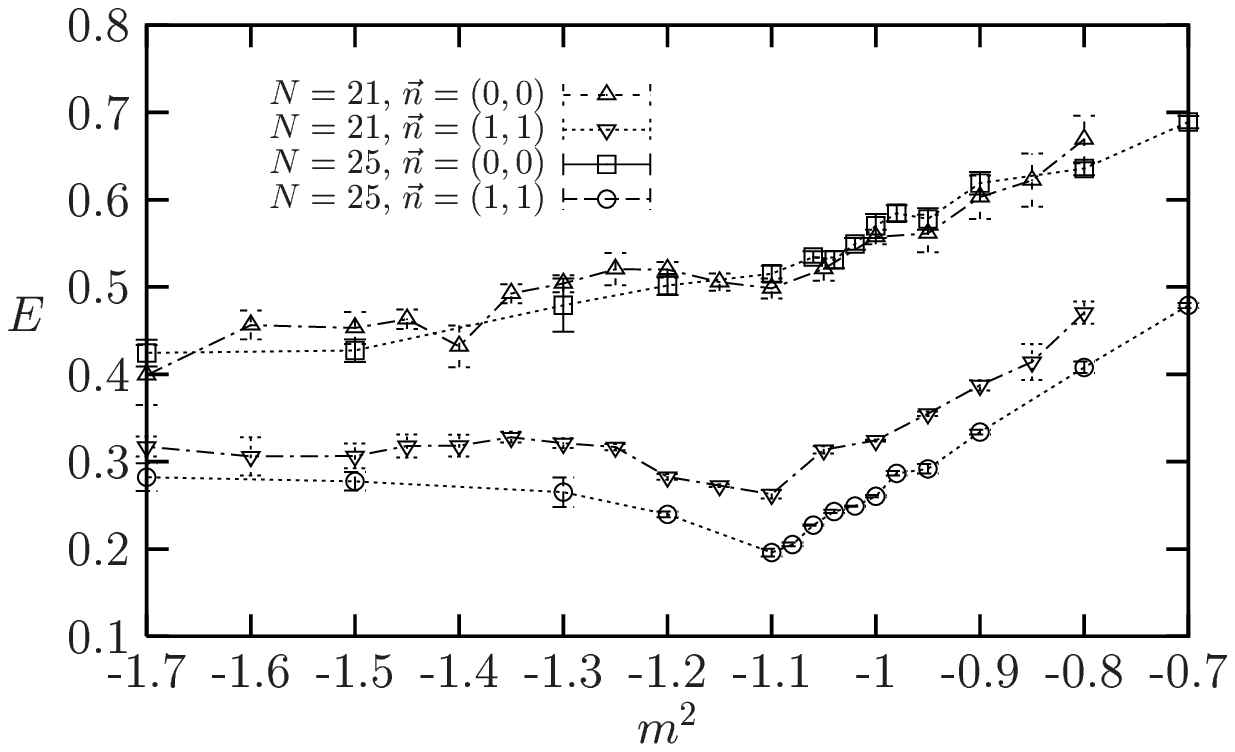,width=.54\linewidth,clip=}%
  \caption{The energies extracted from the connected two-point
    function, against the bare mass squared for $N=21$ and $N=25$ at
    $\lambda = 2$. We show above the results for the modes $(0,
    \frac{2\pi}{N} )$ and $(\frac{2\pi}{N} , 0 )$, and below the
    results for the modes $(0,0)$ and $(\frac{2\pi}{N},
    \frac{2\pi}{N})$ (with $\vec p = 2 \pi \vec n
    /N$).\label{energyN21N25}}}
}

The energy extracted from the $(\frac{2\pi}{N} , 0 )$ mode and the
$(0, \frac{2\pi}{N} )$ mode coincide in the disordered phase, but they
differ in the striped phase because the rotational invariance is
broken.  The energy of the $(\frac{2\pi}{N} , 0 )$ mode and the
$(\frac{2\pi}{N}, \frac{2\pi}{N})$ mode have a minimum near the
critical point.  The minimum values decrease for larger $N$, and they
seem to approach zero, having a cusp-like dip.  Thus these modes are
expected to have finite dimensionful energy in the continuum limit.

The energy of the $(0, \frac{2\pi}{N} )$ mode is small in the striped
phase, and it becomes even smaller as $N$ increases.  At larger $N$
the energy seems to approach zero not only at the critical point, but
all the way in the striped phase.  This behavior suggests the
existence of the quasi NG mode in the striped phase, whose tiny
non-zero energy is due to the discretization of the translational
symmetry.  Our results also support that the discretization effects
are governed by the ratio of the lattice spacing to the wave length of
the stripe pattern, which goes to zero at large $N$ in the striped
phase.

The energy extracted from the zero mode $(0,0)$ does not have a
minimum near the critical point and it does not decrease as $N$
increases.  This is consistent with our observation in the disordered
phase that the dispersion relation has an infrared singularity, and
the $(0,0)$ mode, in particular, has as energy which diverges linearly
with the cutoff scale as shown in figure~\ref{E0scal}.

\vspace{-1pt minus 4pt}

\section{Conclusions}\label{section9}
\vspace{-1pt minus 2pt}

We have presented a comprehensive non-perturbative study of the
$\lambda \phi^4$ model in a three dimensional space, with two
non-commutative spatial coordinates and a commutative Euclidean
time. The non-commutativity tensor is constant. This system is lattice
discretized and then mapped onto a hermitean matrix model. This
mapping is of great advantage for our numerical simulations, which
would be extremely tedious to run directly on the lattice because of
the star product.

\pagebreak[3]

\looseness=1We introduced a set of order parameters which are able to detect the
uniform order as well as various types of stripe patterns.  Measuring
numerically the expectation values of these order parameters, we could
explore the explicit phase diagram of the lattice discretized model,
see figure~\ref{phasedia}.  Its structure is consistent with an
earlier qualitative conjecture by Gubser and Sondhi: at strongly
negative $m^2$ some order emerges, which is uniform at small $\lambda$
(as in the commutative case), but which is dominated by stripe
patterns at larger $\lambda$. The latter is unknown in the commutative
$\lambda \phi^4$ model; we denoted it as the striped phase. The phase
transition between the disordered phase and the ordered regime seems
to be of second order, both, for the uniform order and for the striped
order.

We then studied spatial correlation functions, which decay in an
irregular way due to the non-commutative spatial geometry. On the
other hand, the temporal correlation functions decay
exponentially. That property allowed us to evaluate the dispersion
relation of the scalar field, in particular in the disordered phase
close to the ordering transition. In the vicinity of the uniform phase
we find the usual linear dispersion, whereas the vicinity of the
striped phase corresponds to the case of an energy minimum at a finite
momentum. In fact, the stripe formation means a condensation of such
modes of lower energy than the zero momentum mode.

\looseness=1If the momentum is not that small, we enter --- also in the vicinity
of stripes --- into a linear regime of the dispersion relation. From
the extrapolation of this linear behavior we obtained an effective
mass, which allowed us to identify a ``physical'' (i.e.\ dimensionful)
lattice spacing $a$ (in the planar limit).  This provides a
prescription how to take the limit to the NC $\lambda \phi^4$ model in
the continuum: we take the double scaling limit, which keeps the
product $N a^2$ constant, where $N$ is the size of the lattice resp.\
the matrices. Thus we fix the non-commutativity parameter
$\vartheta$. Approaching this limit, which is at the same time an
infinite volume limit, we found stable dispersion relations.  In
particular, the shift of the energy minimum to a finite momentum
survives this limit, which shows that the striped phase is indeed
there to stay in the continuum. This limit completes the confirmation
of the conjecture by Gubser and Sondhi.  Moreover, the convergence of
the dispersion in the double scaling limit shows that this model can
be renormalized non-perturbatively.

The stripe formation implies the spontaneous breaking of translation
and rotation invariance in the spatial plane.  The Nambu-Goldstone
mode of the broken translation symmetry can be seen approximately from
the numerical data.  Interestingly, this symmetry breaking also occurs
in the two dimensional model, see ref.~\cite{AC} and
appendix~\ref{2dmod}.  This is possible because this model does not
obey the assumptions for the proof of the Mermin-Wagner Theorem.

The energy of the zero mode itself (rest energy) diverges in the limit
of zero lattice spacing and infinite volume. This confirms that the
UV/IR mixing is a fundamental property of NC field theories, which is
also manifest beyond perturbation theory.  Diagrammatically, this
mixing effect is driven by the non-planar diagrams, which are not
suppressed in the double scaling limit (in contrast to the planar
limit). This mixing is the basis for the different types of long
ranged orders that occur in this model.

As an \emph{outlook}, we are now elaborating a dispersion relation for
the photon (or, strictly speaking, for the corresponding ``glueball'')
in a four dimensional NC space~\cite{prep}. We have studied earlier
QED in a NC plane~\cite{2dU1}, where we already saw striking
manifestations of UV/IR mixing.  We expect such effects also in four
dimensions, in particular a $\Theta$-deformed dispersion relation for
the photon.  This dispersion has been studied before to one loop in
perturbation theory~\cite{photon}, though with an uncontrolled
behavior of higher orders.  In contrast, explicit non-perturbative
results would allow us to establish bounds on the magnitude of the
non-commutativity in nature.

For instance, blazars (highly active galactic nuclei) are assumed to
emit bursts of photons simultaneously, which cover a broad range of
energies, see e.g.\ ref.~\cite{Aha}.  Experimentalists are already
trying to detect a relative delay of these photons upon their arrival
on earth, depending on the frequency~\cite{galax}.  Such experimental
efforts will be intensified in the near future.  For instance, the
GLAST project~\cite{GLAST} is scheduled to be launched in 2006 and to
monitor gamma rays from $20\,{\rm MeV}$ up to $1\,{\rm TeV}$.  If we
arrive at results for the $\Theta$-deformed photon dispersion ---
analogous to figure~\ref{dispcont} --- they could then be confronted
with such experimental data~\cite{Camel}.

\vspace{-1pt minus 6pt}

\acknowledgments

\vspace{-1pt minus 4pt}

It is our pleasure to thank Jan Ambj\o rn, Simon Catterall, Hikaru
Kawai, Dieter L\"{u}st and Richard Szabo for useful comments.  The
simulations for this study were performed on the PC clusters at
Humboldt Universit\"{a}t and Freie Universit\"{a}t, as well as the IBM
machine at the Konrad Zuse Zentrum, all of them in Berlin.

The work of J.N.\ was supported in part by Grant-in-Aid for Scientific
Research (No.\ 14740163) from the Ministry of Education, Culture,
Sports, Science and Technology.

\vspace{-1pt minus 6pt}

\appendix

\section{The mapping between the lattice and the matrix formulation}\label{section10}
\label{maplatmat}

\vspace{-1pt minus 4pt}

To describe the connection between the lattice formulation and the
matrix model that we actually simulated, we first go back to the
continuum and introduce the terminology of Weyl operators.

We start from a scalar field $\phi (x)$ in the (commutative) euclidean
space $\R ^d$, which falls off fast enough for the Fourier integral
\begin{equation}
\tilde \phi (p) = \int d^d x \, \phi (x) e^{-ix_{\mu}p_{\mu}}
\end{equation}
to converge. The corresponding NC space is characterized by the
coordinate operators $\hat x_{\mu}$, which obey the non-commutativity
relation~(\ref{NC1}). We recall that we are dealing with commutative
momenta in the NC space. Now we introduce the \emph{Weyl operators}
\begin{equation}
\hat W [ \phi ] = \frac{1}{(2\pi )^{d}} \int d^d p \, \tilde \phi (p)
\ e^{i \hat x_{\mu} p_{\mu}} \,,
\end{equation}
which build a NC, associative algebra. In coordinate space, the map
between the scalar fields and the corresponding Weyl operators reads
\begin{equation}
\hat W [ \phi ] = \int d^d x \, \phi (x) \hat \Delta (x) \,, \qquad
\hat \Delta (x) = \frac{1}{(2\pi )^{d}} \int d^d p \, e^{i (\hat
  x_{\mu}-x_{\mu}) p_{\mu}} \,,
\end{equation}
where $\hat \Delta (x)$ is a hermitean operator.

\pagebreak[3]

Derivatives of the Weyl operators are defined by an anti-hermitean
operator $\hat \partial_{\mu}$, which obeys
\begin{equation}
[ \hat \partial_{\mu}, \hat x_{\nu} ] = \delta _{\mu \nu} \quad
\rightarrow \quad [ \hat \partial_{\mu},\hat \Delta (x)] =-
\partial_{\mu} \hat \Delta (x) \,.
\end{equation}
This implies the desired property
\begin{equation}
[ \hat \partial_{\mu} , \hat W [\phi ] \, ] 
= \hat W [ \partial_{\mu}\phi ] \,.
\end{equation}
Hence a translation can be expressed by unitary operators $\exp
(v_{\mu} \hat \partial_{\mu})$, (c.f.\ eqs.~(\ref{shiftop})
and~(\ref{con2})),
\begin{equation}
\hat \Delta (x+v) = e^{v_{\mu}\hat \partial_{\mu}} \hat \Delta (x)
e^{-v_{\mu}\hat \partial_{\mu}} \,.
\end{equation}
Accordingly a trace on the algebra of Weyl operators is translation
invariant, and it is given explicitly by
\begin{equation}
 \Tr \, \hat W [\phi ] = \int d^d x \, \phi (x) \,, \quad \Tr \, \hat
 \Delta (x) = 1 \,.
\end{equation}

If $\Theta $ is invertible --- which requires the dimension $d$ of the
NC space to be even --- it can be shown that \cite{Szabo}
\begin{equation}
 \Tr \left( \hat \Delta (x) \hat \Delta (y) \right) =
 \delta^{(d)}(x-y) \,.
\end{equation}
Thus we arrive at the inverse map
\begin{equation}
\phi (x) = \Tr \left( \hat W [\phi ] \, \hat \Delta (x) \right).
\end{equation}
Based on this inverse map, we obtain the product of Weyl operators as
\begin{equation}
\hat W [\phi ] \, \hat W [\psi ] = \hat W [\phi \star \psi ] \,,
\end{equation}
which justifies the use of the star product in eq.~(\ref{stern}) and
below.

Let us now focus on $d=2$ and reduce the plane to a periodic $N\times
N$ lattice. Then also the Weyl operators are reduced to $N\times N$
matrices, for instance
\begin{equation}
\hat \Delta (x) = \sum_{n_{1},n_{2}=1}^{N} \hat Z_{1}^{n_{1}} \hat
Z_{2}^{n_{2}} \exp \left( - \frac{2\pi i}{N} ( n_{1} n_{2} + n_{\mu}
x_{\mu}) \right),
\end{equation}
where the operators $\hat Z_{j}$ are defined in eq.~(\ref{Zop}) (here
we set $a=1$).

This provides a map between a lattice field $\phi$ and the Weyl
matrices, which is analogous to the map from continuum fields to Weyl
operators,
\begin{equation}
\hat \phi = \frac{1}{N^{2}} \sum_{x} \phi (x) \, \hat \Delta (x) \,,
\qquad \phi (x) = \frac{1}{N} \Tr \, \Big( \hat \phi \, \hat \Delta
(x) \Big) \,.
\end{equation}
The lattice field in momentum space,
\begin{equation}
\tilde \phi (n) = \frac{1}{N} \sum_{x} \phi (x) e^{-2\pi i n_{\mu}
  x_{\mu}/N} \,,
\end{equation}
is related to the Weyl matrix as
$$
\hat \phi = \frac{1}{N} \sum_{n} \tilde \phi (n) \hat J(n) \,,\qquad
\tilde \phi (n) = \frac{1}{N}  \Tr \, 
\left( \hat \phi \, \hat J^{\dagger}(n) \right),  
$$
where 
\begin{equation}
\hat J(n) = \hat Z_{1}^{n_{1}} \hat Z_{2}^{n_{2}} \, e^{-2\pi i n_{1}
  n_{2}/N} \,.
\end{equation}

\section{The non-commutative $\lambda \phi^{4}$ model in $d=2$}\label{section11}
\label{2dmod}

In this appendix we want to address the case of only two NC spatial
coordinates, i.e.\ we omit the time direction now (resp.\ we reduce it
to one point).

In this context, we would like to repeat that the formation of stripes
necessarily implies the spontaneous breaking of translational and
rotational invariance. For this reason, a striped phase was originally
not expected in $d=2$~\cite{GuSo}.

However, a numerical study by Ambj\o rn and Catterall revealed that
the striped phase does exist even in the 2d model~\cite{AC}.  They
also mapped the lattice formulation onto a matrix model and simulated
$N=39$ at a fixed value of $\lambda$, probing one line for
$m^{2}$. They saw the two stripe pattern along with more complicated
multi-stripe patterns. They pointed out that this does not contradict
the Mermin-Wagner Theorem~\cite{MW}, because the derivations of the
latter are all based on assumptions like locality, which are not
realized in this case.  Intuitively one may wonder why the
short-ranged non-commutativity can induce a long-range order. However,
keeping in mind that we found UV/IR mixing to be a fundamental
property also beyond perturbation theory, this does not seem
surprising any longer.

We verified the results reported in ref.~\cite{AC} for $d=2$, and
extended also that investigation to a full exploration of the phase
diagram.  First we run the simulation with the same algorithm as in
ref.~\cite{AC}: in a Metropolis step the matrix $\hat \phi$ may be
replaced by $\hat \phi + \varepsilon \hat \eta$, where $\hat \eta$ is
a hermitean random matrix and $\varepsilon$ is a small parameter which
is tuned for an acceptance rate around 50 \%.  After about 500 steps
we saw a rich structure of patterns in the striped phase, in agreement
with ref.~\cite{AC}.  Some typical multi-stripe patterns that appeared
in this way are shown in figure~\ref{snap2d}.  However, when we extend
the history to $O(10^{6}) \dots O(10^{7})$ steps only the two stripe
patterns survive, as in the three dimensional case.

\FIGURE[b]{\epsfig{file=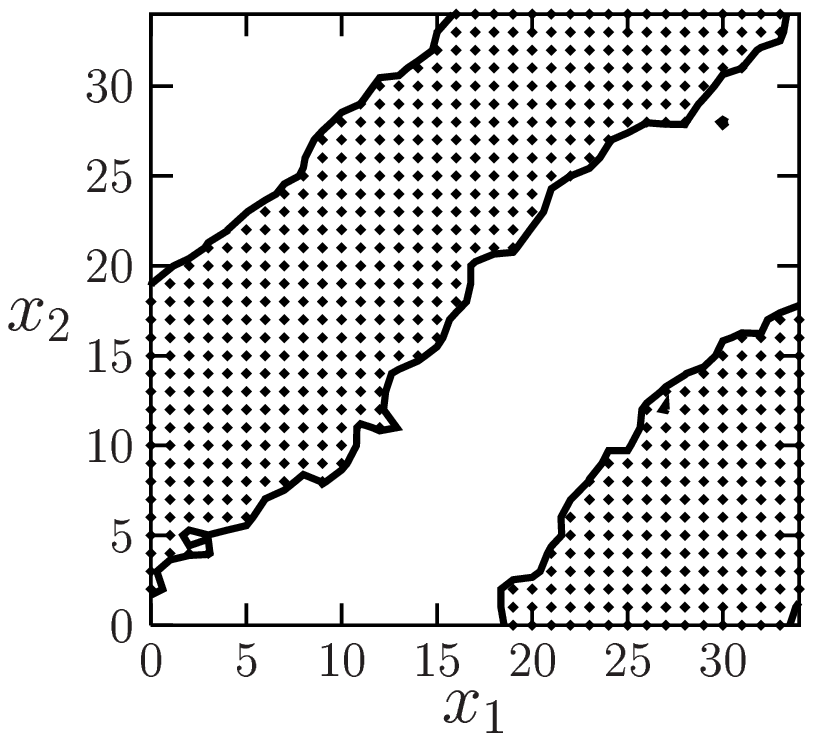,width=.3\linewidth,clip=}
  \epsfig{file=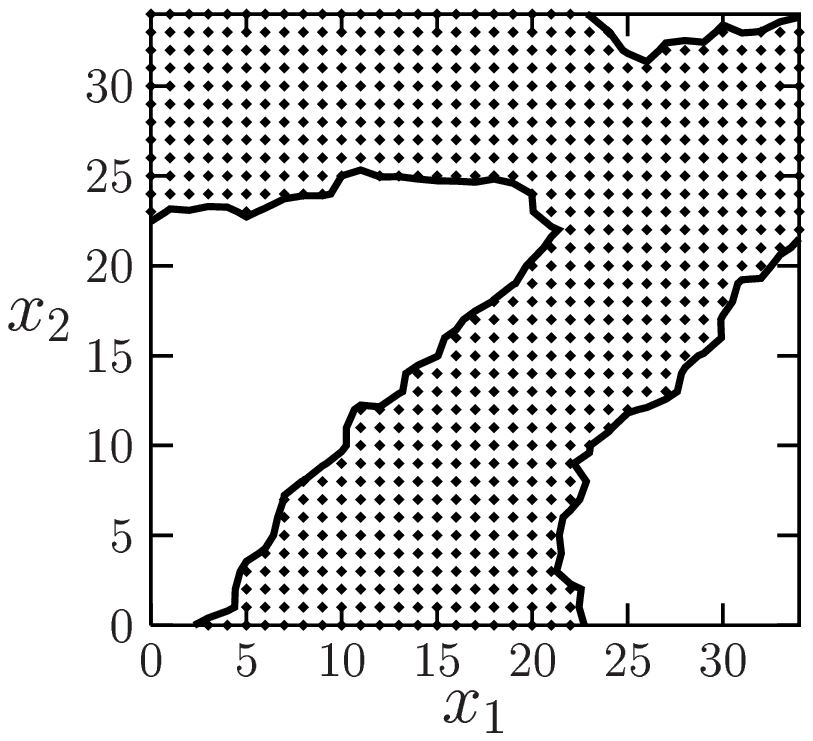,width=.3\linewidth,clip=}
  \epsfig{file=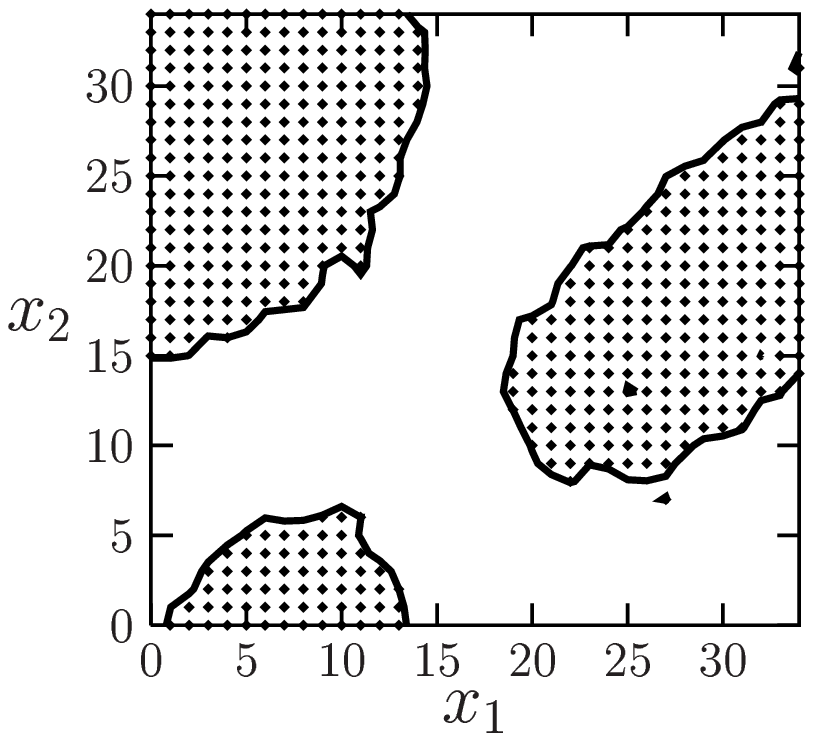,width=.3\linewidth,clip=}
  \caption{Three snapshots of meta-stable multi-stripe patterns in the
    2d model, at $N=35$, $N^{3/2} m^2 \simeq - 213 $ and $N^{2}\lambda
    =350$, in the striped phase.\label{snap2d}}}

We then checked this result with the more powerful algorithm that we
also used in the 3d case. The difference is that now the matrices
$\hat \phi$ are updated by changing only pairs of conjugate matrix
elements $\hat \phi_{ij} = \hat \phi_{ji}^{*}$.  This takes more
steps, but the corresponding parameter $\varepsilon$ can be taken much
larger. In total this results in a reduction of the thermalization
time by a factor $\sim 1/N$. For further numerical details we refer to
appendix~\ref{simu}.

The result was that we could see now with less steps the behavior
described above, namely that in this magnitude of $N$ only two stripes
parallel to one axis are ultimately stable.  Of course, also the
variety of meta-stable patterns found first by Ambj\o rn and Catterall
is interesting in itself and their observation complements the
description of the 2d model.

The feature of the phase diagram is similar to the 3d case, but as a
qualitative difference the mass axis has to be scaled as $N^{3/2}
m^{2}$ (rather than $N^{2} m^{2}$).  Then we find again an accurate
transition line between the disordered phase and the ordered regime,
and a somewhat broad but stable transition region between the uniform
phase and the striped phase, see figure~\ref{phasedia2d}.

\FIGURE[t]{\epsfig{file=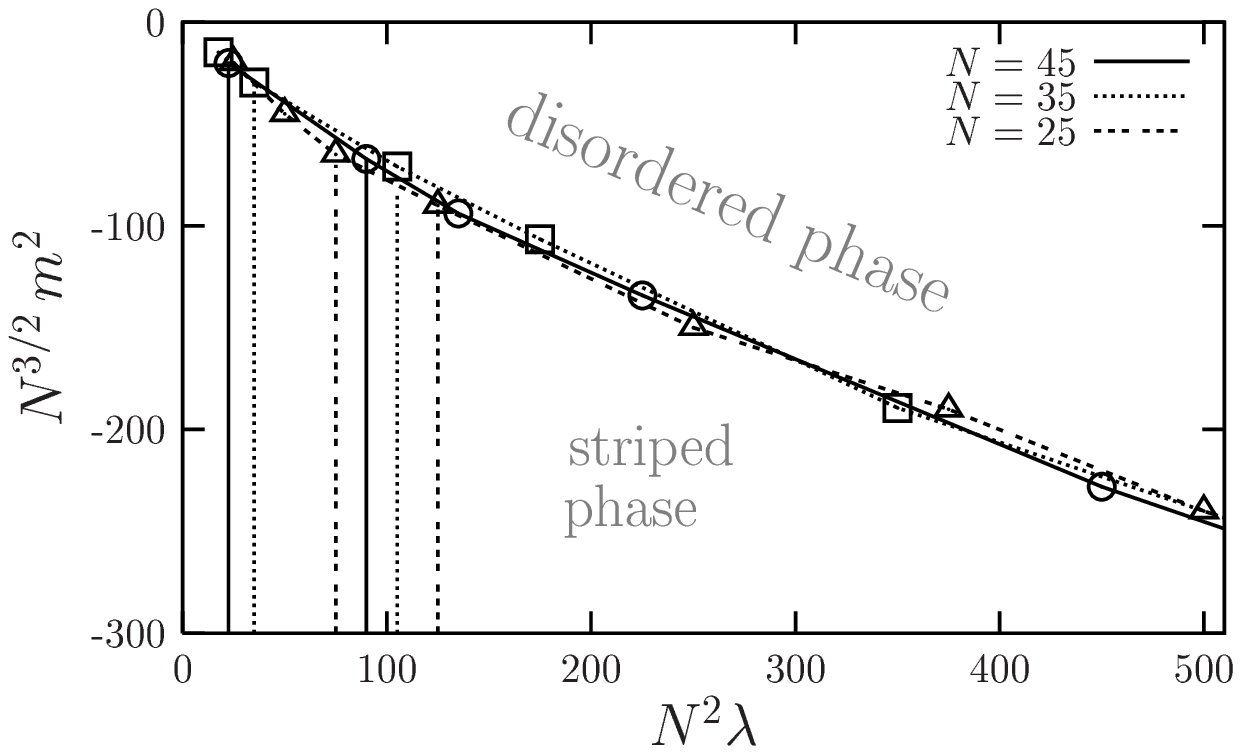,width=.7\linewidth,clip=}
  \caption{The phase diagram of the 2d NC $\lambda \phi^{4}$ model,
    identified on a lattice of size $N^{2}$. Note that the vertical
    axis has to be scaled differently from the 3d case shown in figure
    \protect\ref{phasedia}.\label{phasedia2d}}}

Finally we would like to add that the 2d $\lambda \phi^4$ model has
recently been studied numerically on a ``fuzzy sphere'', which also
represents a NC manifold. Ref.~\cite{XM} observed a ``matrix phase'',
which is likely to coincide with the striped phase.  For another
numerical study of a matrix model involving the fuzzy sphere, see
ref.~\cite{ABNN}.

\section{The simulations}\label{section12}
\label{simu}

Our Monte Carlo simulations were performed with the standard Metropolis
algorithm. The update steps of a configuration run over the single
matrix elements $\hat \phi_{ij}(t)$ $(i \geq j)$ which may be modified as
\begin{equation}
\hat \phi_{ij}(t) \to \hat \phi_{ij}(t) + \varepsilon \eta \,, \qquad
\hat \phi_{ji}(t) \to \hat \phi_{ji}(t) + \varepsilon \eta^{*} \,,
\qquad \varepsilon \in \R_{+} \,.
\end{equation}
The matrix elements $\hat \phi_{ij}(t) = \hat \phi_{ji}(t)^{*}$ of the
initial hot configurations, and the variable $\eta$ are complex random
numbers, where the real and the imaginary part have a flat probability
density in $[ -0.5, 0.5]$.  The parameter $\varepsilon$, which
controls the step size, is set in the beginning to $N^{-1/4}$. It is
then adapted dynamically as follows: after updating the complete
lattice configuration, it is increased resp.\ decreased by 20 \% if
the Metropolis acceptance rate is larger than 0.6, resp.\ less then
0.3. Thus the acceptance rate is pushed into this interval, and the
parameter $\varepsilon$ typically stabilizes around one order of
magnitude below its starting value.

We compared this updating procedure to an alternative one which
updates the whole matrix $\hat \phi (t)$ at once, c.f.\
appendix~\ref{2dmod}.  We found the method described above to be
superior by far with respect to the thermalization time and to the
autocorrelation.

For our method, the autocorrelation time ranged from about 10 to 90
update steps, depending on the point in the phase diagram.  To keep
this effect under control, we proceeded as follows: for each
simulation we performed six completely independent hot starts.  Then
we thermalized over about 500 to 1000 update steps. After this we
measured the observables on configurations which were separated by 100
steps each time.

The statistical error was evaluated with the jack-knife and the
binning method. In both cases we varied the bin size and we took
finally the over-all maximum as a careful estimate for the error bar,
which is shown in the plots. For the bulk of the points in the phase
diagram we arrived at small errors already with a modest statistics.
To compute the 2-point functions of the order parameter, however, we
had to handle in particular the region close to the phase transitions,
which required a much larger number of measurements.  Our statistics
is given in table~\ref{tabstat}.

\TABLE[t]{\begin{tabular}{|c|c|c|c|c|c|c|}
\hline
$N$ & & 15 & 25 & 35 & 45 & 55 \\ 
\hline
\multirow{2}{45mm}{number of configurations}
 & phase diagram & 300 & 180 & 60 & 30 & --- \\
 \cline{2-7}
 & 2-point function & 10000 & 10000 & 5000 & 3000 & 2700 \\
\hline
\end{tabular}%
\caption{Our statistics at the different values of $N=T$
in the measurements for the phase diagram 
in figure \protect\ref{phasedia}, and for the connected
2-point function $\langle M^{2}(k) \rangle_{c}$
(examples are shown in figure \protect\ref{order2-fig}).\label{tabstat}}}

As a test for the correctness of our code, we expanded the expectation
value of the action analytically to the first order in $\lambda$. The
result is in excellent agreement with our numerical data up to
$\lambda \approx 0.5$.  For the corresponding plot, and for further
details about the simulation, we refer to ref.~\cite{Diss}.

\end{document}